%% file: TW-Oct-14-1431.tex
\documentclass[journal]{IEEEtran}
\IEEEoverridecommandlockouts
\usepackage[latin1]{inputenc}
\usepackage{amsfonts}
\usepackage{amsmath}
\usepackage{mathrsfs}
\usepackage{balance}
\usepackage{cite}
\usepackage{subfigure}
\usepackage{hyperref}

\usepackage{flushend}

\usepackage{listings}
\usepackage{graphicx}

    \usepackage{graphicx,subfigure,float}
\usepackage{color}
\usepackage{units}
\usepackage{nicefrac}
\usepackage[printonlyused]{acronym}
\usepackage{pstricks, pst-plot, pst-grad, pstricks-add, pst-node, pstricks-add}
\usepackage[section] {placeins}

\newcommand{\Prob}[0]{\mathbb{P}}
\newcommand{\Exp}[0]{\mathbb{E}}

\newcommand{\EulerGamma}[0]{\gamma_e}

\newcommand{\etal}{\textit{et al.}}

\newtheorem{theorem}{Theorem}

\newmuskip\pFqmuskip

\newcommand*\pFq[6][8]{%
  \begingroup 
  \pFqmuskip=#1mu\relax
  \mathcode`\,=\string"8000
  \begingroup\lccode`\~=`\,
  \lowercase{\endgroup\let~}\pFqcomma
  {}_{#2}F_{#3}{\left[\genfrac..{0pt}{}{#4}{#5};#6\right]}%
  \endgroup
}
\newcommand{\pFqcomma}{\mskip\pFqmuskip}

\acrodef{AWGN}[AWGN]{additive white Gaussian noise}
\acrodef{FEC}[FEC]{forward error correction}
\acrodef{MCS}[MCS]{modulation and coding scheme}
\acrodef{PDF}[PDF]{probability density function}
\acrodef{CDF}[CDF]{cumulative density function}
\acrodef{CRT}[CRT]{complexity-rate tradeoff}
\acrodef{RAN}[RAN]{radio access network}
\acrodef{RRH}[RRH]{remote radio head}
\acrodef{SNR}[SNR]{signal-to-noise ratio}
\acrodef{SINR}[SINR]{signal-to-interference-and-noise ratio}
\acrodef{C-RAN}[C-RAN]{centralized radio access network} 
\acrodef{CB}[CB]{code block}
\acrodef{TB}[TB]{transport block}
\acrodef{CBLER}[CBLER]{code block error rate}
\acrodef{TBLER}[TBLER]{transport block error rate}
\acrodef{VM}[VM]{virtual machine}
\acrodef{RAP}[RAP]{radio access point}
\acrodef{RB}[RB]{resource block}
\acrodef{FLOPS}[FLOPS]{floating-point operations per second}
\acrodef{PPP}[PPP]{Poisson point process}
\acrodef{LP}[LP]{local processing}
\acrodef{MRS}[MRS]{max-rate selection}
\acrodef{CAS}[CAS]{computationally aware selection}
\acrodef{BS}[BS]{base station}
\acrodef{IT}[IT]{information technology}
\acrodef{MIMO}[MIMO]{multiple-input and multiple-output}
\acrodef{BS}[BS]{base-station}
\acrodef{LDPC}[LDPC]{low-density parity check}
\acrodef{pmf}[pmf]{probability mass function}
\acrodef{RE}[RE]{resource element}
\acrodef{CCI}[CCI]{co-channel interference}
\acrodef{SINR}[SINR]{signal-to-interference and noise ratio}

\newrgbcolor{darkgreen}{0.2 0.6 0.2}
\newrgbcolor{darkred}{0.6 0.2 0.2}
\newrgbcolor{darkblue}{0.2 0.2 0.6}

\makeatletter
  \def\DrawDot[#1]{ moveto gsave #1 stroke grestore gsave .8 .8 scale 1 setgray #1 fill grestore }
  \def\CirclePath{ DS DS scale  0.9 0.9 scale currentpoint 1.5 0 360 arc fill }
  \@namedef{psds@o}{ /Dot { \DrawDot[\CirclePath] } bind def }
  \def\TrianglePath{ DS DS scale 0.55 0.55 scale 0 2.5 rmoveto 2.3 -4 rlineto -4.6 0 rlineto 2.3 4 rlineto closepath }
  \@namedef{psds@triangle}{ /Dot { \DrawDot[\TrianglePath] } bind def }
  \def\DiamondPath{ DS DS scale 0.55 0.55 scale 0 2.5 rmoveto 1.5 -2.5 rlineto -1.5 -2.5 rlineto -1.5 2.5 rlineto closepath }
  \@namedef{psds@diamond}{ /Dot { \DrawDot[\DiamondPath] } bind def }
  \def\SquarePath{ DS DS scale 0.55 0.55 scale -2 -2 rmoveto 4 0 rlineto 0 4 rlineto -4 0 rlineto closepath }
  \@namedef{psds@square}{ /Dot { \DrawDot[\SquarePath] } bind def }
\makeatother

\begin{document}
\title{The Complexity-Rate Tradeoff \\ of Centralized Radio Access Networks}
\author{Peter Rost,~\IEEEmembership{Senior~Member,~IEEE},
Salvatore Talarico,~\IEEEmembership{Student Member,~IEEE},
and Matthew C. Valenti~\IEEEmembership{Senior~Member,~IEEE}.
\thanks{Manuscript received Oct. 6, 2014; revised Mar. 27, 2015; accepted June 7, 2015.  Date of publication XXX. XX, 2015. Date of current version XXX. XX, 2015.  The associate editor coordinating the review of this paper and approving it for publication was A. Kwasinski.}
\thanks{P. Rost was formerly with NEC Laboratories Europe, Heidelberg, Germany, and is now with Nokia Networks, Munich, Germany (email: Peter.Rost@ieee.org).}
\thanks{ S.~Talarico and M.~C.~Valenti are with West Virginia University, Morgantown, WV, U.S.A. (email:Salvatore.Talarico81@gmail.com;valenti@ieee.org).}
\thanks{The research leading to these results has received partly funding from the European Union Seventh Framework Programme (FP7/2007-2013) under grant agreement n\textordmasculine~317941 (www.ict-ijoin.eu).  The authors would like to acknowledge the contributions of their colleagues in iJOIN, although the views expressed are those of the authors and do not necessarily represent the project.}
\thanks{Digital Object Identifier 10.1109/TWC.2015.1XXXXXXX}
}

\date{}
\maketitle

\begin{abstract}
In a centralized \ac{RAN}, the signals from multiple \acp{RAP} are processed centrally in a data center.  Centralized \ac{RAN} enables advanced interference coordination strategies while leveraging the elastic provisioning of data processing resources.  It is particularly well suited for dense deployments, such as within a large building where the \acp{RAP} are connected via fibre and many cells are underutilized. This paper considers the computational requirements of centralized \ac{RAN} with the goal of illuminating the benefits of pooling computational resources.   A new analytical framework is proposed for quantifying the computational load associated with the centralized processing of uplink signals in the presence of block Rayleigh fading, distance-dependent path-loss, and fractional power control.  Several new performance metrics are defined, including computational outage probability, outage complexity, computational gain, computational diversity, and the complexity-rate tradeoff.
The validity of the analytical framework is confirmed by comparing it numerically with a simulator compliant with the 3GPP LTE standard.
Using the developed metrics, it is shown that centralizing the computing resources provides a higher net throughput per computational resource as compared to local processing.
\end{abstract}

\IEEEpeerreviewmaketitle

\thispagestyle{empty}

\section{Introduction}
\begin{figure}
      \centering
      \scalebox{0.8}{\small{\input{cloud.architecture.tex}}
      }
      \caption{Typical centralized RAN architecture}
      \label{fig:system.model:cloud.architecture}
      \vspace{-0.50 cm}
    \end{figure}
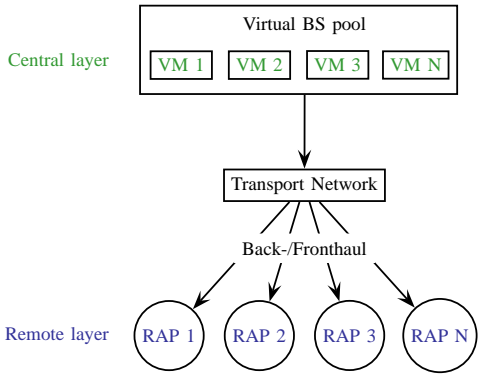

  Compared with today's networks, future mobile networks will need to cope with a dramatic increase in the density and demand of users.
 Several novel technologies have been proposed to meet the needs of future systems, including
massive \acs{MIMO}, millimeter wave signaling, very densely deployed small-cell networks, and
  centralized \acfp{RAN} \cite{Wang.ChinaComm.2010,Wuebben.etal.SPM.2014}.
  The centralized \ac{RAN} concept is illustrated in Fig. \ref{fig:system.model:cloud.architecture}. A centralized \ac{RAN} system consists of multiple \acfp{RAP}, which
  perform only simple processing of the input signal, including sampling and optionally performing an FFT operation, a fronthaul or backhaul connection, a transport network, and a central data center. The data center manages a virtual \ac{BS} pool composed of a scalable number of virtual machines, each representing one or more \acp{BS} and performing the majority of the baseband processing.

  To date, research on centralized \ac{RAN} has focused on the \ac{RAN} functional split, the applicability of joint processing, system performance, and fronthaul requirements.
  Recently, the implementation on cloud-computing platforms attracted more interest \cite{Checko.etal.CST.2014}: cloud-computing platforms allow resources to be virtualized,
  improve resource  elasticity, and allow mobile networks to be implemented in software.
  However, the application of centralized \ac{RAN} to cloud-computing platforms (also referred to as Cloud-RAN) requires a better understanding
  of the interplay between available computing resources and the required communications performance. As we will show, the amount of computational resources and the degree to which they are centralized have a direct impact on
  how the communication performance is effected by the communication channel properties and resource allocation policies.

  \subsection{Data processing complexity in mobile networks}
    Each mobile network is characterized by a well-defined set of functionality, timing requirements, and protocols. This imposes very precise
    requirements on the operation of each \ac{RAP}, including data processing requirements in order to maintain real-time properties.
    For instance, 3GPP LTE defines a set of \acp{MCS}, each with a prescribed number of information bits per \ac{TB}.  Since the choice of \acp{MCS} depends on the \ac{SNR} and the computational requirements depend on the number of received information bits, there is an inherent relationship between the channel quality and the computational requirements.  Furthermore, 3GPP LTE specifies strict processing deadlines that are necessary to guarantee given latency constraints and prevent unnecessary hybrid-ARQ retransmissions.
    Due to randomness in the channel and user demand, the data processing requirements are typically found by resorting to extensive simulation studies.

   Because the local processing capability of each small cell is anticipated to be far less than that of a macro cell, the dimensioning of data
    processing resources needs to be revisited. Furthermore, very dense networks are subject to dramatic temporal and spatial traffic fluctuations
    such that many small cells may be strongly underutilized. It is therefore not economically viable to equip a small cell based on peak data processing
    requirements, yet under-dimensioning the computational resources limits its capabilities.
    This dilemma motivates the concept of centralized \ac{RAN}, which enables computational load balancing across multiple \acp{BS} to avoid peak-pro\-vi\-sio\-ning. The goal is to dimension the computational assets of the network to exploit the \emph{computational diversity} present in a pool of resources, while still guaranteeing the communication service requirements of each individual cell.   The utilization of processing resources should be maximized for a given service guarantee.
    These goals require a framework that enables the prediction of the required data processing resources under a given performance constraint.



  \subsection{Related work}
  The major benefits of centralized \ac{RAN} have been unveiled in \cite{Wang.ChinaComm.2010}.  Key among these benefits is computational load balancing, which exploits the variable number of users processed within the data center (a consequence of the spatial and temporal traffic fluctuations) and variations of the processing requirements per user (due to the channel variability).
    In \cite{Bhaumik.Chandrabose.Jataprolu.Kumar.Muralidhar.Polakos.Srinivasan.Woo.MobiCom.2012}, Bhaumik \etal~use
    the the OpenAir \cite{OpenAir.2014} LTE implementation and real-world measurements to predict that centralized \ac{RAN} can save up to $\unit[22]{\%}$ of the data processing resources. Furthermore, the paper considers statistical guarantees that sufficient computational resources are provided with very high probability. However, these results do not characterize the source of fluctuations, they do not allow the quantification of the centralized \ac{RAN} gains depending on the variation of channel parameters, and they do not allow for active shaping of the data processing load.
    In \cite{Valenti.Talarico.Rost.GC.2014}, the impact of the decoder complexity on the data processing requirements is evaluated numerically. Furthermore, the impact of
    computational constraints is investigated on the overall system performance, and it is shown that taking these constraints into account by the scheduler allows the system to recover much of the performance loss caused by imposing a computational budget. However, the work was largely based on simulation, and does not provide a formal analytical framework for the evaluation of centralized \ac{RAN} and its computational requirements.

    Implementation options for centralized \ac{RAN} have been investigated in \cite{Rost.etal.ComMag.2014}, which considers
    flexible centralization over heterogeneous backhaul with possibly high latency. In \cite{Wuebben.etal.SPM.2014}, the signal processing requirements of this
    system are investigated. Among others, different centralization options are analyzed with respect to their required backhaul capacity, latency, and challenges for
    the signal processing. A preferred functional split is to centralize the forward error correction coding and decoding (possibly on commodity hardware), while all other lower layer functionalities
    are executed at the \ac{RAP}. In \cite{Zhu.etal.ACMCF.2011}, a virtual \ac{BS} pool based on
    commodity hardware and Linux operating system is investigated. It is shown that such a system can provide the required real time and throughput guarantees.

    Evaluating the complexity requirements of a centralized \ac{RAN} system requires an analysis of the corresponding baseband processing.  However, the evaluation of computational complexity in mobile networks has been primarily focused on the power consumption of receivers.
    So far, mainly energy-limited devices have been investigated where complexity and power consumption are of particular interest, e.g. in \cite{Troesch.etal.ICUWB.2007}
    where ultra-wideband receivers have been analyzed. The receiver complexity is affected by different components such as the decoding
    and detection process as well as filtering process.
    Hong and Stark estimated in \cite{Hong.Stark.ECS.1999} the power consumption
    and decoding performance of prominent decoders for codes such as turbo, Hadamard, block, and convolutional codes.
    In \cite{Zimmermann.PhD.2007}, a thorough analysis of the sphere decoding and detection has been performed with particular
    focus on complexity reduction in multiple antenna systems. Similarly, \cite{Pusane.etal.ISIT.2004} analyzed \ac{LDPC} codes and possible complexity reductions. The filter complexity of the receiver has been investigated in \cite{Mahesh.Vinod.PIMRC.2006} in the context of software-defined radio. In
    software-defined radio, it is important to reduce complexity particularly at the lower layers.
    At higher layers, scheduling incurs a major part of the complexity, which has been  formally
    analyzed in \cite{Moscibroda.Wattenhofer.Infocom.2006}, which derives expressions quantifying the complexity required to schedule a successful transmission.  In \cite{Elias.Jalden.ISIT.2010}, Elias and Jalden investigate in  the trade-off between rate, reliability (expressed by outage probability), and complexity in the high-\ac{SNR}
    regime.

    Complexity is equally important from an infrastructure point of view. In \cite{Andrews.Dinitz.Infocom.2009}, the network capacity is investigated using
    game theory and distributed algorithms, where the complexity is distributed across multiple nodes.
    In \cite{Howard.Schlegel.Iniewski.EURASIP.2006}, Howard \etal~provide a formal model for quantifying power consumption and derive the critical distance at which the transmit
    power consumption equals the decoder power consumption. This critical distance has been quantified for particular decoder implementations. In \cite{Grover.Woyach.Sahai.JSAC.2011}, Grover \etal~derive a comprehensive framework for analytically modeling the power consumption of a decoder and introduce an optimization problem whose main objective is to minimize the system-wide power consumption including both transmission and decoding power.
    The optimal rate allocation to minimize the system-wide power consumption is strictly lower than Shannon capacity.  In this paper, we leverage the model proposed in \cite{Grover.Woyach.Sahai.JSAC.2011} to evaluate the data processing requirements of a centralized \ac{RAN} system.

  \subsection{Contribution and Outline}
    In this paper, we introduce a framework for modeling, analyzing the performance, and determining the data processing requirements of centralized \ac{RAN} systems, which applies for both macro and small cells.  The model is used to quantify the computational load associated with the centralized processing of uplink signals in the presence of block Rayleigh fading, distance-dependent path-loss, and fractional power control.  Several new performance metrics are defined, including computational outage probability, outage complexity, computational gain, computational diversity, and the complexity-rate tradeoff.
The validity of the analytical framework is confirmed by comparing it numerically with a simulator compliant with the 3GPP LTE standard.
Using the developed metrics,  the advantages of centralized \ac{RAN} over conventional (distributed) \ac{RAN} are illuminated.

    The remainder of this paper is organized as follows.
    Section \ref{sec:system.complexity.model} introduces the system model, presents the complexity model, and defines the new performance metrics.
    In Section \ref{sec:complexity.framework}, an analytical framework for computational complexity is introduced, culminating in tractable expressions of the proposed complexity metrics suitable for cellular uplinks characterized by block Rayleigh fading and fractional power control. Section \ref{sec:numerical.verification} performs a numerical verification of the proposed framework. Section \ref{sec:results} presents and discusses results obtained with our framework. Finally, the paper concludes in Section \ref{sec:conclusions}.

\section{Complexity model and Performance metrics}\label{sec:system.complexity.model}

Consider a {\em \ac{RAP} group} containing $N_\mathsf{c}$ \emph{\acp{RAP}}, where each \ac{RAP} corresponds to a single \ac{BS} and whose signals are jointly processed in a virtual \ac{BS} pool. Let $\gamma$ indicate the instantaneous \ac{SNR} for a particular link and $\overline{\gamma}=\mathbb E [\gamma]$ indicate the average \ac{SNR}.  In  the  following, it  is  assumed  that  the  channel  gain  remains  fixed for the duration of one \ac{TB}, but varies independently from \ac{TB} to \ac{TB}, which corresponds to a \emph{block-fading} model.  For each \ac{TB}, the channel is conditionally subject to \ac{AWGN}.
When the \ac{SNR} of a channel is $\gamma$, its capacity is $\log_2(1 + \gamma)$ per channel use \cite{Shannon.1948}, which provides an upper bound on the rate.  In practice, the rate is selected for a given $\gamma$ from a set of \acp{MCS} in such a way that a constraint on outage probability is satisfied.  Because no finite-length code satisfies the capacity limit with equality, the rate is selected to be below the corresponding capacity limit according to some margin.

\subsection{Link Adaptation}

In order to perform link adaptation, the transmitter can choose from $N_R$ different \acp{MCS}, with $R_k$ denoting the rate of the $k^{th}$ \ac{MCS}, $k \in \{1,...,N_R\}$.
As in LTE, we assume that each \ac{MCS} corresponds to setting the rate of a turbo code and selecting from among three modulations: QPSK, 16-QAM, and 64-QAM.   See \cite{3GPP.TS.36.211, 3GPP.TS.36.212} for details of how the turbo-code rate and modulation format is defined for each MCS.
Per the standard, the $k^{th}$ \ac{MCS} segments each \ac{TB} into $C_k$ \acp{CB}, each of which conveys $D_{k}$ information bits.
As we advocate later, the selection of the \ac{MCS} should take into account the complexity of the turbo decoder, and in particular, the number of expected iterations required to meet the outage constraint is the key factor in making the selection.

Suppose that the turbo decoder is run with a certain number of maximum iterations $L_\mathsf{max}$.  Let $\gamma^R_{k}$ indicate the minimum \ac{SNR} for which the $k^{th}$ \ac{MCS} satisfies the outage constraint after the $L_\mathsf{max}$-th iteration, on average.
The value of $\gamma^R_{k}$ for each \ac{MCS} can be obtained as follows.  Simulations are used to obtain \ac{CBLER} curves for each possible \ac{MCS} and for an arbitrarily large number of maximum iterations (here, set to $L_\mathsf{max} = 8$). For the $k^{th}$ \ac{MCS}, $\gamma^R_{k}$ is selected to be the value of \ac{SNR} for which the \ac{TBLER} satisfies a particular  constraint for the channel outage $\hat\epsilon_\text{channel}$ (a transport block error occurs when any of the code blocks in a transport block fails).

It is well known that the complexity of a turbo decoder depends on how much the channel \ac{SNR} exceeds the capacity limit.  If a turbo code of rate $R_k$ is transmitted over a channel whose capacity is sufficiently higher than $R_k$, then typically only a few iterations are required.  Thus, to manage complexity, an \ac{SNR} \emph{margin} $\Delta \gamma$ may be used.  With such a margin, the \ac{SNR} must be at least $\Delta \gamma$ above $\gamma^R_{k}$ in order to select the $k^{th}$ \ac{MCS}.  The rate selection for channel \ac{SNR} $\gamma$ and \ac{SNR} margin $\Delta \gamma$ can be represented by the function
\begin{eqnarray}
   r\left( \gamma, \Delta \gamma\right)
   & = &
   \begin{cases}
       R_1 & \mbox{if $\frac{\gamma}{\Delta \gamma} \leq  \gamma_1^R $ } \\
       R_k & \mbox{if $\gamma_k^R   < \frac{\gamma}{\Delta \gamma} \leq  \gamma_{k+1}^R $ } \\
       R_{N_R}  & \mbox{if $\gamma_{N_R}^R  < \frac{\gamma}{\Delta \gamma}. $ } \\
   \end{cases}
   \label{rate function}
\end{eqnarray}

\subsection{Complexity Model}\label{sec:outage.network:complexity.model}



The computational effort required to decode a turbo code
is linear in the number of iterations and in the number of
(information) bits. Thus a reasonable metric for computational
effort  is  the {\em bit-iteration}. For a given SNR $\gamma$ and SNR margin $\Delta\gamma$, let
the random variable $L_{r}(\gamma, \Delta\gamma)$ be the number of iterations required to decode a particular \ac{CB},
and $\text{E}\left[L_{r}(\gamma,\Delta\gamma)\right]$ be the average number of iterations required to decode a \ac{CB}. The expected decoding complexity (averaged over the number of iterations required per \ac{CB}) expressed in bit-iterations per channel use (pcu) can be evaluated as
\begin{equation}
\mathcal{C}(\gamma,\Delta\gamma)=
\frac{ D_{k} C_{k} \text{E}\left[L_{r}(\gamma,\Delta\gamma)\right]}{S_\mathsf{re}}
      \label{eq:average.outage.network:300}
\end{equation}
where $S_\mathsf{re}$ is the number of channel uses or \acp{RE} required to convey the \ac{TB}.  The expected number of decoding iterations can be found using the same set of simulation curves used to determine the MCS thresholds $\gamma_k^R$.  For a given rate $r\left( \gamma, \Delta \gamma\right)$, the simulation results will show the CBLER for each iteration up to the $L_\mathsf{max}$-th iteration.  These CBLER curves can be interpreted as the \emph{\ac{pmf}} of the number of iterations at $\gamma$, and from the pmf coefficients, the average number of iterations is computed.

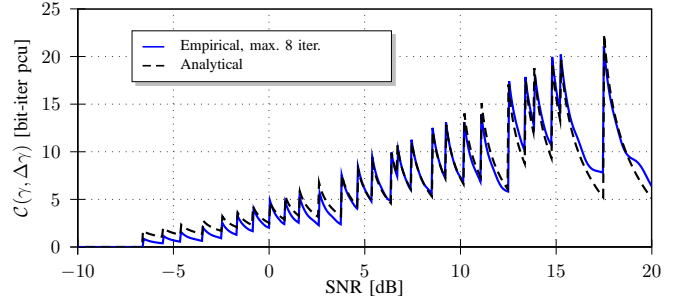
\begin{figure}
\vspace{-0.25 cm}
      \centering
      \scalebox{.72}{\input{Complexity_SNR_theory.tex}}
      \caption{Complexity as a function of the SNR obtained both through simulations (blue) and based on (\ref{eq:outage.network:200}) (black curve), when the maximum number of iterations used to decode a \ac{CB} is limited to eight.}
      \label{fig:outage.network:complexity_in_snr_theory}
      \vspace{-0.50 cm}
\end{figure}

As an example, we have computed the complexity using (\ref{eq:average.outage.network:300}) for an example system.  For this and all of the examples in this paper, the choice of MCS assumes the use of the normal cyclic prefix and a 10 MHz bandwidth corresponding to 50 \acp{RB}.  However, since up to 5 \acp{RB} must be reserved for the physical uplink control channel and the number of \acp{RB} must be a multiple of 2, 3, or 5 \cite{3GPP.TS.36.211, 3GPP.TS.36.212}, the transmitted signal occupies 45 \acp{RB}.  Since there are 12 information-bearing SC-FDMA symbols per sub-frame and 12 subcarriers per \ac{RB}, it follows that $S_\mathsf{re} = 45 \times 12 \times 12 = 6480$.  The MCS selection is implemented according to (\ref{rate function}) with the threshold found for $L_\mathsf{max} = 8$ iterations, an outage constraint $\hat\epsilon_\text{channel}=0.1$, and no excess margin ($\Delta \gamma = 0$). This empirically determined complexity is shown in Fig. \ref{fig:outage.network:complexity_in_snr_theory} by the blue curve, while the black curve shows the complexity predicted by the complexity model that will be discussed next.


To enable a deeper analysis of complexity issues, it is essential to build an accurate yet wieldy complexity model.  The groundwork for such a model is provided by \cite{Grover.Woyach.Sahai.JSAC.2011}, which accurately predicts the power consumption of a decoder for a given code rate, coding scheme, channel, and decoder type. According to \cite[Eq. (4) and (9)]{Grover.Woyach.Sahai.JSAC.2011}, the expected decoding complexity  under a given instantaneous \ac{SNR} $\gamma$ and \ac{SNR} margin $\Delta \gamma$ can be modeled as
\begin{eqnarray}
      \mathcal{C}(\gamma, \Delta\gamma) \hspace{-0.2cm}& = & \hspace{-0.2cm}\frac{r\left(\gamma, \Delta\gamma  \right)}{\log_2\left(\zeta - 1\right)} \log_2\left[\frac{-\log_{10}(\hat\epsilon_\text{channel})}{ l^2\left( \gamma, \Delta\gamma \right) K' }\frac{\zeta - 2}{\zeta} + \frac{2}{\zeta}\right] \label{eq:outage.network:201} \nonumber \\
      \hspace{-0.2cm} & \approx & \hspace{-0.2cm} \frac{r\left(\gamma, \Delta\gamma  \right)}{\log_2\left(\zeta - 1\right)} \log_2\left[\frac{-\log_{10}(\hat\epsilon_\text{channel})}{l^2\left( \gamma, \Delta\gamma \right) K'}\frac{\zeta - 2}{\zeta}\right] \nonumber \\
      \hspace{-0.2cm} & = & \hspace{-0.2cm} \frac{r\left(\gamma, \Delta\gamma \right)}{\log_2\left(\zeta - 1\right)}\left[\log_2\left(\frac{\zeta - 2}{K(\hat\epsilon_\text{channel})\zeta}\right) \right. \nonumber \\
      \hspace{-0.2cm} & &  \hspace{-0.2cm}- 2\log_2 \left[ l\left( \gamma, \Delta\gamma \right) \right]  \bigg]
      \label{eq:outage.network:200}
\end{eqnarray}
where $\zeta$ is a parameter of the model related to the connectivity of the decoder when represented as a graph,
\begin{eqnarray}
      l\left( \gamma, \Delta\gamma \right)& = & \log_2(1 + \gamma) - r\left(\gamma, \Delta\gamma  \right), \label{eq:l_function} \\
      K(\hat\epsilon_\text{channel})& = &-\frac{K'}{\log_{10}(\hat\epsilon_\text{channel})},
\end{eqnarray}
$K'$ is a parameter of the model, and $\hat\epsilon_\text{channel}$ is a constraint on the channel outage probability.  For the sake of simplicity $K(\hat\epsilon_\text{channel})$ is denoted by $K$ in the remainder of the text.

When evaluating (\ref{eq:outage.network:200}), the function $r(\gamma, \Delta \gamma)$ is defined by (\ref{rate function}).
The relationship between each threshold $\gamma_k^R$  and the corresponding rate $R_k$ in (\ref{rate function}) can be modeled as
\begin{eqnarray}
  R_k
  & = &
  \log_2\left( 1+ \frac{\gamma^{R}_{k}}{\nu } \right )
\end{eqnarray}
where $\nu$ is a parameter that models the gap between the capacity at $\gamma^{R}_{k}$ and the SNR for the actual code to meet the performance objective at rate $R_k$.   The values of $(\zeta, K', \nu )$ could be found for each MCS by statistically fitting the empirically observed complexity to the complexity predicted by the model.  However, it is more convenient to have values for these parameters that are common to all the MCSs.




 The values of $(\zeta, K', \nu )$ that provide the best fit to the empirical complexity (blue curve) of Fig. \ref{fig:outage.network:complexity_in_snr_theory} are $\zeta= 6, K' = 0.2,$ and $\nu = 0.2$ dB.  Using these values with (\ref{eq:outage.network:200}) provides the predicted complexity shown by the black curve in Fig. \ref{fig:outage.network:complexity_in_snr_theory}.
 As can be seen, these two curves are quite similar, demonstrating the effectiveness of the model.  We note that a more accurate model could be achieved by determining a different set of $(\zeta, K', \nu )$ for each \ac{MCS}, but this significantly complicates the analysis, as will be subsequently evident. Fig. \ref{fig:outage.network:complexity_in_snr_theory} further shows a highly volatile behavior, which results from the link-adaptation process. When the system operates at the same \ac{MCS} (rate), a decoder requires fewer iterations as the \ac{SNR} increases. However when link adaptation is performed,
 the system requires more complexity each time it moves towards an higher \ac{MCS}, since it operates closer to capacity, which is the cause of the peaky behavior in Fig. \ref{fig:outage.network:complexity_in_snr_theory}.

  \subsection{Performance Metrics}\label{sec:system.model:definitions}
    In this subsection, several performance metrics are defined, which are used throughout the paper to evaluate the performance of a centralized \ac{RAN} system and highlight its benefits over a distributed implementation.


    \subsubsection{Computational outage probability}
      Fig. \ref{fig:outage.network:complexity_in_snr_theory} shows that the required computational complexity to decode a \ac{TB} may differ significantly depending
      on the \ac{SNR}. Furthermore, when the system is distributed, each \ac{BS} will be provisioned with a limited amount of computational resources $C_{\text{max}}$.
      By contrast, in a centralized system the  virtual  \ac{BS}  pool   will  be  equipped  with  a  total limited amount of computational resources, which scales linearly with the number of \acp{BS}. If the total required computational effort  exceeds  the computational resources before  a  decoding  deadline  is reached,  a {\em computational outage}  will occur.  From the operator's perspective, a computational outage is no different than a channel outage, because in both cases the information is not made available to the end user.

      More formally, the computational outage probability is defined by
      \begin{equation}
	 \epsilon_\text{comp}(N_\text{c}, C_{\text{max}}) = \Prob\left[
	     \sum_{i=1}^{N_\text{c}} \mathcal{C}(\gamma_i, k) > N_\text{c} C_{\text{max}}
	 \right], \label{eq:outage.network:500}
      \end{equation}
      where $\gamma_i$ denotes the instantaneous \ac{SNR} for the $i$-th \ac{RAP}.
      When a computational outage occurs, the centralized processor is unable to finish the decoding of at least one uplink user.  However, when this condition occurs it is still possible that some of the uplink transport blocks can be decoded, even though at least one of the blocks has failed.

\subsubsection{Outage complexity}\label{outage_complexity}
The outage complexity $\mathcal{C}_\text{out}\left(\hat{\epsilon},N_{\text{c}}\right)$ is the minimum amount of computational resources required to satisfy an outage constraint on the per-cell computational outage probability $\hat\epsilon_{\text{comp}}\left( N_\text{c}\right)$. More formally, the outage complexity is defined as
\begin{equation}
\mathcal{C}_\text{out}(\hat\epsilon, N_\text{c}) = \text{arg}\min_{C_{\text{max}}}\Big\{ \epsilon_\text{comp}(N_\text{c}, C_{\text{max}}) \leq \hat\epsilon_{\text{comp}}\left( N_\text{c}\right) \Big\}.\label{eq:outage.complexity}
      \end{equation}
The constraint of the computational outage probability of the system $\hat{\epsilon}_\text{comp}(N_\text{c})$ is normalized to a per-cell basis. Assuming
independent processing with a non-priority-based scheduler, the constraint for a centralized system is related to the constraint for a distributed system with value $\hat\epsilon_{\text{comp}}\left( 1\right)=\hat\epsilon $ by
      \begin{equation}
	\hat\epsilon_\text{comp}(N_\text{c}) = 1 - \left(1 - \hat\epsilon \right)^{N_\text{c}}.\label{eq:outage.network:510}
      \end{equation}

    \subsubsection{Computational gain and diversity}
      Using the previous definition of computational outage, the centralization gain of a centralized \ac{RAN} mobile network is determined. In particular, the impact on the performance by centralizing communication-processing resources can be measured in two ways: 1) by considering the reduction of the required
      computational resources; 2) by examining the improved outage behavior.

The {\em computational gain} quantifies the reduction of computational resources made possible by the centralization of processing.  When $N_\text{c}$ \acp{RAP} are jointly processed, the computational gain is
      \begin{eqnarray}
	g_\text{comp}\left( N_\text{c}\right)
	& = &
\frac{N_\text{c}\mathcal{C}_\text{out}(\hat\epsilon, 1)}{\mathcal{C}_\text{out}(\hat\epsilon, N_\text{c})}
      \end{eqnarray}
The asymptotic computation gain is found by letting the number of \acp{RAP} go to infinity,
      \begin{equation}
	g_\text{comp}^{\infty} = \lim\limits_{N_\text{c}\rightarrow\infty} g_\text{comp}\left( N_\text{c}\right).
	\label{eq:outage.network:520}
      \end{equation}

An alternative metric is the {\em computational diversity}, which quantifies the rate of computational improvement as more \acp{RAP} are jointly processed.  This is found by determining the rate at which the computational outage probability is improved as a function of $N_\text{c}$, i.\,e., the computational diversity is
      \begin{equation}
	d_\text{comp}\left( N_\text{c}\right) = -\frac{\partial \log_{10}\epsilon_\text{comp}(N_\text{c},C_{\text{max}})}{\partial N_\text{c}}.
	\label{eq:outage.network:530}
      \end{equation}
This is the downward slope of the curve that relates computational outage probability to the number of jointly processed \acp{RAP}.  The larger this negative slope, the  faster the maximum computational gain (\ref{eq:outage.network:520}) is achieved.
       In a mobile network, the computational diversity provides an indication for the required number of
      centralized \acp{BS} per data-center. 

We note that the slope is at its maximum magnitude when $N_\text{c}=1$, and thus for the remainder of the paper when we discuss computational diversity, we evaluate  the derivative of (\ref{eq:outage.network:530}) at $N_\text{c}=1$.  Under this condition, we can drop the dependence on $N_\text{c}$ and simply express $d_\text{comp}\left( N_\text{c}\right)$ by $d_\text{comp}$.

\subsubsection{Average achievable rate}
Given a fixed average \ac{SNR} $\overline{\gamma}$, and a fixed \ac{SNR} margin $\Delta\gamma$, the rate associated to a given user depends upon the \ac{MCS} selected based on the channel quality. Due to the random effect of fading, the code rate selected per user is aleatory. One of the conventional ways to measure system performance involves evaluating the average code rate. Given the distribution of the instantaneous \ac{SNR} $\gamma$, the expected code rate can be computed as follows
\begin{eqnarray}
    \hspace{-0.5 cm} \mathbb{E}_\gamma \left\{R\left( \Delta\gamma \right)\right\} \hspace{-0.2 cm} & = & \hspace{-0.2 cm} \sum\limits_{k = 1}^{N_R} \Prob\left\{\gamma^R_{k} \leq \gamma < \gamma^R_{k+1}\right\} r\left( \gamma^R_{k}, \Delta\gamma\right) \nonumber \\
       \hspace{-0.5 cm}  \hspace{-0.2 cm} & =&  \hspace{-0.2 cm} \sum\limits_{k = 1}^{N_R} \left[ F_\gamma(\gamma^R_{k+1}) - F_\gamma(\gamma^R_{k}) \right] r\left( \gamma^R_{k}, \Delta\gamma\right)
         \label{eq:AverageAchievableRate}
\end{eqnarray}
where $F_\gamma(\gamma)$ is the \ac{CDF} of the instantaneous \ac{SNR} $\gamma$.

\subsubsection{Complexity-Rate Tradeoff}
Focusing only on outage complexity ignores the influence of rate.  An alternative metric should consider the relationship between the rate and the \ac{SNR} margin $\Delta\gamma$.   As the margin $\Delta\gamma$ increases, the complexity decreases, but so does the rate since the system operates farther away from capacity. Therefore, the \emph{\ac{CRT}} can be quantified, and it measures how much additional complexity is required in order to further improve the rate. In particular, the complexity-rate tradeoff can be evaluated as
\begin{equation}
	 t_\text{comp}(N_\text{c}, \Delta\gamma) = \hspace{-0.2cm} \lim\limits_{\Delta\gamma'\rightarrow \Delta\gamma} \hspace{-0.2cm}
	  \frac{\partial \mathbb{E}_\gamma\left\{R\left(\Delta\gamma' \right)\right\}}{\partial \Delta\gamma'}
\left[\frac{\partial \mathcal{C}_\text{out}(\hat\epsilon, N_\text{c})}{\partial \Delta\gamma' }\right]^{-1} \hspace{-0.5cm}. \hspace{-0.1cm}
	\label{eq:outage.network:550}
\end{equation}

\section{A Framework for Complexity Analysis}\label{sec:complexity.framework}
In this section, we provide an analytical framework for computing the previously introduced performance metrics.  We begin in Section \ref{sec:complexity.framework:constant.pl} by assuming that the user has a fixed \emph{average} SNR $\overline{\gamma}$, which is true when the uplink is fully power controlled.  We continue in Section \ref{sec:complexity.framework:with.pl} by considering the use of partial power control with a user that is randomly positioned in the cell.  For ease of exposition, we start with $N_\text{c} = 1$ and extend the results in Section \ref{sec:complexity.framework:centralized} to the case of arbitrary $N_\text{c}$.

  \subsection{Local Processing with Rayleigh Fading}\label{sec:complexity.framework:constant.pl}
Consider a single user with fixed average \ac{SNR} $\overline{\gamma}$, and that this user is processed locally ($N_\text{c} = 1$).  The instantaneous value of $\gamma$ is a random variable, which, for the sake of tractability, we assume is an exponential random variable corresponding to the Rayleigh fading model\footnote{We note that other distributions for $\gamma$ may be used, more accurately modeling the effects of interference, shadowing, and signal propagation, but this comes at the cost of reduced tractability.  Most such distributions will require a simulation or numerical integration    to compute the performance metrics.}. The \ac{PDF} of such a $\gamma$ is given by:
 \begin{equation}
 f_{\gamma}\left( \gamma\right)=\frac{1}{\overline{\gamma}} \exp\left( -\frac{\gamma}{\overline{\gamma}}\right), \gamma \geq  0.
	\label{eq:pdf_gamma}
\end{equation}

\subsubsection{Expected complexity}
\begin{theorem}
The expected value of the decoding complexity conditioned on the average \ac{SNR} $\overline{\gamma}$ can be evaluated by
    \begin{multline}
      \mathbb{E}_{\gamma| \overline\gamma}\{\mathcal{C}_i\} \approx \sum\limits_{k = 1}^{N_R}\frac{r\left(\gamma^R_{k}, \Delta\gamma\right)}{\log_2\left(\zeta - 1\right)} \times \\
      \left[ I_1\left(\gamma^R_{k+1},\gamma^R_{k}\right)+I_2\left(\gamma^R_{k+1},\gamma^R_{k}\right)\right] \label{eq:ExpextedComplexity}
\end{multline}
    where the functions $I_1\left( \cdot,\cdot\right)$ and $I_2\left( \cdot,\cdot\right)$ are given respectively by  (\ref{eq:T1}) and (\ref{eq:T2_Proof3}).
\begin{IEEEproof}
The derivation is given in Appendix \ref{App:Expectation}.
\end{IEEEproof}
\end{theorem}

\subsubsection{Variance of the complexity}
\begin{theorem}
The variance of the computational complexity conditioned on the average \ac{SNR} $\overline{\gamma}$ can be evaluated by
\begin{multline}
      \text{Var}_{\gamma| \overline\gamma}\left\{\mathcal{C}_i\right\} \approx \sum\limits_{k = 1}^{N_R} \left[\frac{2 r\left(\gamma^R_{k}, \Delta\gamma \right)}{\log_2\left(\zeta - 1\right)}\right]^2 \times \\
      I_4\left(\gamma^R_{k+1},\gamma^R_{k}\right) -\mathbb{E}_{\gamma| \overline\gamma}\{\mathcal{C}_i\}^2  \label{eq:VarianceComplexity}
\end{multline}
where $\mathbb{E}_{\gamma| \overline\gamma}\{\mathcal{C}_i\}$ is evaluated using (\ref{eq:ExpextedComplexity}) and  $I_4\left(\cdot,\cdot\right)$ is given by (\ref{eq:lastT4}).

\begin{IEEEproof}
  The derivation is given in the Appendix \ref{App:Variance}.
\end{IEEEproof}
\end{theorem}

\subsubsection{Outage complexity}
In the case of an isolated user processed locally, the outage complexity in a distributed system is the value $\mathcal{C}_\text{out}$ that satisfies the following expression
\begin{equation}
      \Prob\{\mathcal{C}_i \geq \mathcal{C}_\text{out}(\hat\epsilon, 1)\} \leq \hat\epsilon
\end{equation}
which can be evaluated once the \ac{CDF} of the complexity is known.

 The \ac{CDF} of the complexity can be written as follows
\begin{eqnarray}
      F_{\mathcal{C}}(\mathcal{C}_\text{thr}) & = & \Prob\left\{\mathcal{C}_i \leq \mathcal{C}_\text{thr} \right\} \label{eq:CDFComplexity}
\end{eqnarray}
where $\mathcal{C}_\text{thr}$ is a complexity threshold chosen based on the quality of the channel and the computational resources available. Using the notion that the complexity monotonically decreases in $\gamma$ in the interval $[\gamma^R_{k}; \gamma^R_{k+1}]$, (\ref{eq:CDFComplexity}) can be replaced by
    \begin{equation}
      F_{\mathcal{C}}(\mathcal{C}_\text{thr})
	 =  \frac{1}{\Prob\left\{\gamma > \gamma^R_{1}\right\}} \sum\limits_{k = 1}^{N_R}\Prob\left\{\gamma_{k, \text{min}} \leq \gamma < \gamma^R_{k+1}\right\}  \label{eq:CDFComplexity2}
    \end{equation}
where
\begin{eqnarray}
      \gamma_{k, \text{min}} & = & \max\left[\gamma^R_{k}, \min\left(\gamma^R_{k+1}, 2^{\Delta R + r\left(\gamma, \Delta\gamma \right)} - 1\right)\right] \label{eq:MinGamma} \\
      \Delta R & = & 2^{-\displaystyle\mathcal{C}_\text{thr} \displaystyle\nicefrac{\log_2(\zeta - 1)}{\left[2 r\left(\gamma, \Delta\gamma\right) \right]}} \sqrt{\frac{\zeta - 2}{K\zeta}}. \label{eq:DELTA}
\end{eqnarray}

Finally, (\ref{eq:CDFComplexity2}) can be written as follows
    \begin{equation}
      F_{\mathcal{C}}(\mathcal{C}_\text{thr})
	=  \frac{1}{1 - F_\gamma\left(\gamma^R_{1}\right)} \sum\limits_{k = 1}^{N_R}\left[F_\gamma\left(\gamma^R_{k+1}\right) - F_\gamma\left(\gamma_{k, \text{min}}\right)\right] \label{eq:CDFComplexity3} \hspace{-0.2cm}
    \end{equation}
where $F_\gamma(\cdot)$ is the \ac{CDF} of the instantaneous \ac{SNR} with \ac{PDF} given by (\ref{eq:pdf_gamma}).

%

\subsubsection{Example}\label{sec:complexity.framework:constant.pl-Example}
    \begin{figure*}
      \vspace{-0.25 cm}
      \centering
      \subfigure[Expected complexity]{\input{Complexity_scaling_NR.tex}}
      \subfigure[Variance of the complexity]{\input{Complexity_scaling_Var_NR.tex}}
      \subfigure[Outage complexity when $\hat\epsilon =0.1$]{\input{Complexity_outage_NR.tex}}
      \subfigure[Expected achievable rate]{\input{Rate_scaling_NR.tex}}
      \caption{Expected complexity, variance of the complexity, outage capacity, and expected achievable rate as function of the number of MCS schemes used, for three different values of $\Delta\gamma$ and block Rayleigh fading. The average \ac{SNR} is  $\overline\gamma=\unit[10]{dB}$. The constraint on the computational outage is $\hat\epsilon = 0.1$. Solid lines are obtained analytically, while the dots are obtained through simulations.}
      \label{fig:network:complexity_scaling}
      \vspace{-0.5 cm}
    \end{figure*}
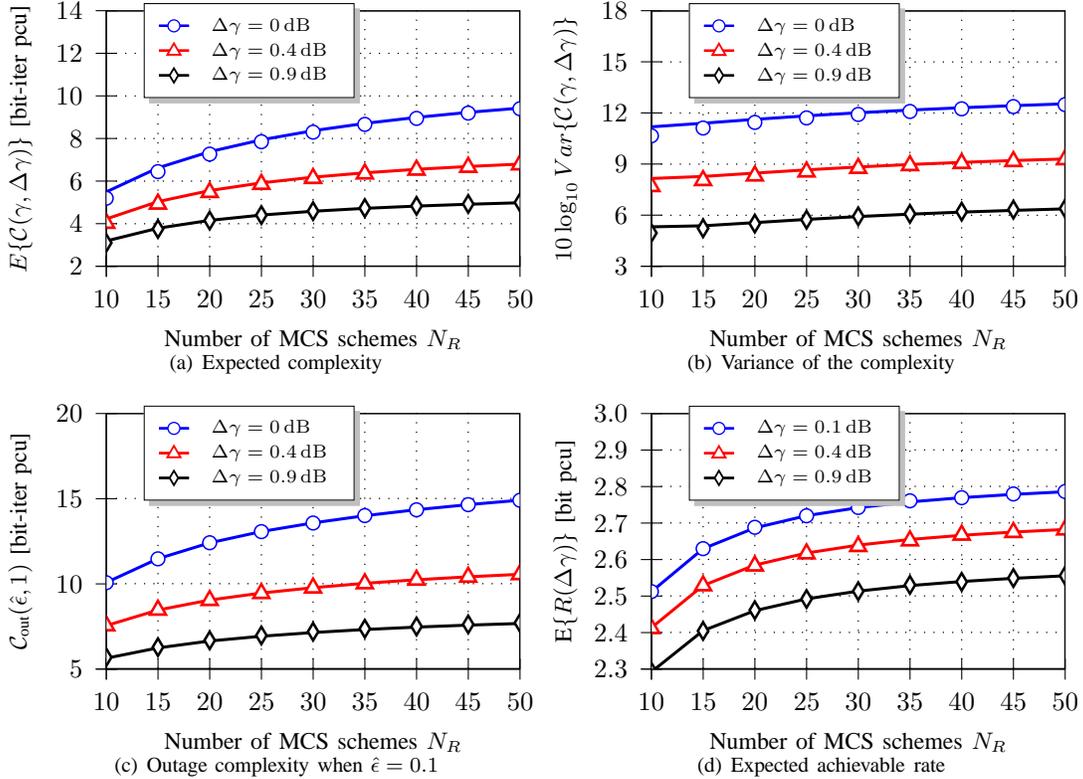

    Fig. \ref{fig:network:complexity_scaling} shows a comparison between analytical (solid lines) and simulation (dots) results, in terms of: a) expected complexity; b) variance of the complexity; c) outage complexity when the constraint on the computational outage is set to be $\hat\epsilon= 0.1$; and d) expected achievable-rate. The simulations are obtained as follows. During each trial, the instantaneous \ac{SNR} $\gamma_i$ is determined for each uplink by drawing an exponentially distributed random variable. For the given SNR, the rate is determined  according to the function $r(\gamma_i, \Delta \gamma)$.  Next, the complexity $C\left(\gamma_i,\Delta\gamma\right)$ is determined using (\ref{eq:outage.network:200}). Once a sufficient number of trials is run, the average and variance of $C\left(\gamma_i,\Delta\gamma\right)$ are found along with the average achievable rate, while $\mathcal{C}_\text{out}(\hat\epsilon,1)$ is evaluated such that (\ref{eq:outage.complexity}) is satisfied for $N_\text{c}=1$, given a specific constraint on both the computational and channel outage. For the following results, one million trials are run for each of $N_{R}$ \ac{MCS} schemes and for several representative values of \ac{SNR} margin $\Delta\gamma$.  In doing this experiment, the value of $N_{R}$ is varied in an effort to gain some insight into the required number of \acp{MCS} schemes.

    All four sub-figures show both the dependency on the number of MCS schemes $N_{R}$ and the \ac{SNR} margin $\Delta\gamma$.  The number of \ac{MCS} schemes $N_R$ determines how well the capacity curve is sampled, while the \ac{SNR} margin $\Delta\gamma$ provides a direct indication of how close the code and decoder operate to Shannon capacity. This \ac{SNR} margin decreases when more turbo-decoder iterations are allowed. The value of $\gamma^R_{k}$ for each \ac{MCS} is equally-spaced with the values of its adjacent \acp{MCS}. The first and the last \ac{MCS} have assigned values, respectively $\gamma^R_{1} = \unit[-6.4]{dB}$ and $\gamma^R_{N_R} = \unit[17.6]{dB}$ (based on 3GPP LTE). The average \ac{SNR} is fixed to $\overline\gamma=\unit[10]{dB}$.
    Fig. \ref{fig:network:complexity_scaling} shows a good agreement between the analysis and the simulations  for all values of $\Delta\gamma$ and $N_R$. However, as the number of \ac{MCS} schemes decreases,
    the mismatch between the analysis and simulations increases for the variance of the complexity and this is imputable to the non-linearity of $\log(1 + x)$ and the linearization method adopted to derive
    (\ref{eq:VarianceComplexity}).

  \subsection{Effect of Path Loss and Power Control}\label{sec:complexity.framework:with.pl}
Now consider the case that the user is subject to a fractional power control policy and is located randomly within the cell.  For the sake of tractability, we assume a circular cell of unit radius and that the user is placed according to a uniform distribution. A more complicated model for the placement of the user may be used, such as one that accounts for the actual shape of the Voronoi region.  While more accurate, this comes at the cost of tractability and will typically require simulation or numerical integration. Let $Y_i$ indicate the $i^{th}$ \ac{BS} and its location. Each \ac{BS} $Y_i$ serves only one user, which is located at $X_i$ at distance $|Y_i - X_i|$ from the \ac{BS}.
Under a fractional power control policy, $X_i$ will transmit using power
\begin{equation}
 P_i = P_0|Y_i - X_i|^{s\eta}
\end{equation}
where $P_0$ is a reference power (typically taken to be the power received at unit distance from the transmitter), $\eta$ is the path-loss exponent, and $s, 0 \leq s \leq 1,$ is the \emph{compensation factor} for fractional power control. The average \ac{SNR} received by $Y_i$ for the $i^{th}$ mobile is given by
\begin{equation}
\overline\gamma_i = \overline\gamma_{\text{ud}}|Y_i - X_i|^{-\eta(1-s)} \label{eq:instantaneousSNR}
\end{equation}
where $\overline\gamma_{\text{ud}}$
is the average reference \ac{SNR}, measured assuming unit-distance transmission.

 In this case, it is not possible to determine a closed-form expression or a suitable approximation for both the expected and the variance of the complexity. However, knowing that the \ac{PDF} of $r_i=|X_i - Y_i|$ is
  \begin{eqnarray}
f_{r_i}(\omega)
& = &
\begin{cases}
  2\omega & \mbox{, $ 0 \leq  \omega\leq 1$} \\
  0 & \mbox{, otherwise}
\end{cases} \label{eq:pdf_radius}
\end{eqnarray}
  the expected complexity $\Exp\{\mathcal{C}_i\}$ and variance $\text{Var}\{\mathcal{C}_i\}$ can be obtained by performing a numerical integration over $r$ of (\ref{eq:ExpextedComplexity}) for the expected complexity and (\ref{eq:VarianceComplexity}) for the complexity variance.

\begin{theorem}
 Under this scenario, the \ac{CDF} of the instantaneous \ac{SNR} can be computed in closed form and it yields
    \begin{multline}
      F_\gamma(\gamma) = 1-\frac{2\left(\frac{\gamma}{\gamma_\text{ud}}\right)^{-2/[\eta(1 - s)]}}{\eta(1-s)} \times \\
     \left[ \Gamma\left(\frac{2}{\eta(1-s)}, \frac{\gamma}{\gamma_\text{ud}}\right) - \Gamma\left(\frac{2}{\eta(1-s)}, 0\right)\right] \label{eq:pathlossSNR}
    \end{multline}
where $\Gamma(\cdot, \cdot)$ is the incomplete gamma function \cite{Abramowitz:1965}.
\begin{IEEEproof}
 The derivation is given in the Appendix \ref{App:CDF_SNR}.
\end{IEEEproof}
\end{theorem}

The outage complexity can be determined in this case analytically, since
the \ac{CDF} of the decoding complexity can be evaluated by using (\ref{eq:pathlossSNR}) into (\ref{eq:CDFComplexity3}). Even if (\ref{eq:MinGamma}) cannot be expressed in closed form in this case, well-known algorithms such as bisection, Newton, or gradient-search algorithm can be applied. Furthermore, the average achievable rate can be determined by substituting (\ref{eq:pathlossSNR}) into (\ref{eq:AverageAchievableRate}).

  \subsection{Centralized Processing}\label{sec:complexity.framework:centralized}
    Using the previously derived expressions for the expectation and the variance of the complexity, in this section the analytical framework is expanded in order to be able to compute through an approximation the maximum decoding complexity when $N_\text{c}$ \acp{BS} are centralized. More specifically, the central limit theorem (Lindeberg/Levy) \cite{Bronstein.2000} is used, which yields
    \begin{equation}
      \mathcal{C}_\text{out}(\hat\epsilon, N_\text{c}) =
	 \sqrt{\frac{\text{Var}\{\mathcal{C}_i\}}{N_\text{c}}}\underbrace{\sqrt{2}Q^{-1}\left(2(1 - \hat\epsilon)^{N_\text{c}} - 1\right)}_{\text{inv. norm. CDF}} + \mathbb{E}\{\mathcal{C}_i\}
	\label{eq:outage.network:computation.diversity:400}
    \end{equation}
    where $Q^{-1}(\cdot)$ is the inverse error function, which has the following series representation \cite{Abramowitz:1965}
    \begin{eqnarray}
    Q^{-1}(x) =  \sum_{k=0}^\infty\frac{v_k}{2k+1}\left(\frac{\sqrt{\pi}}{2}x\right )^{2k+1} \label{eq:InverseErrorFunction}
    \end{eqnarray}
    where $v_0$=1 and
    \begin{eqnarray}
    v_k=\sum_{m=0}^{k-1}\frac{v_m v_{k-1-m}}{(m+1)(2m+1)}.
    \end{eqnarray}

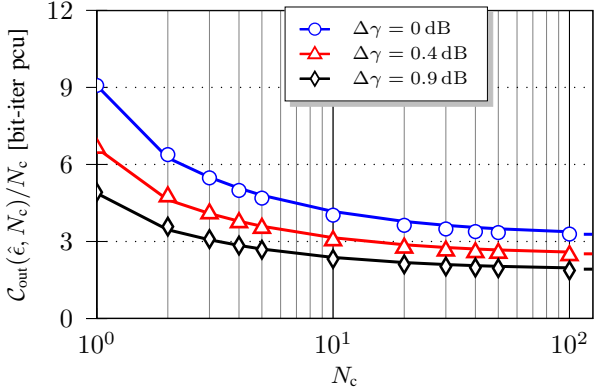
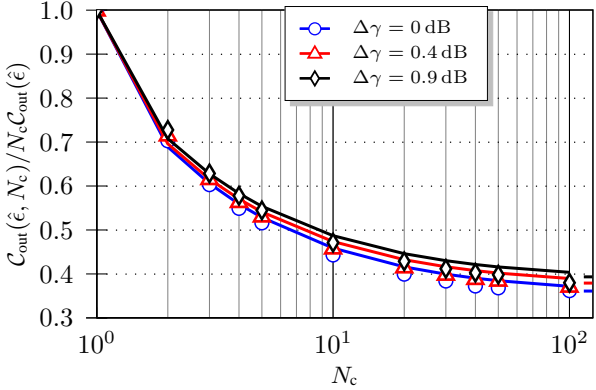
\begin{figure}[tb]
      \centering
      \subfigure[Absolute outage complexity]{\input{Complexity_outage_network_Nusers_with_pl.tex}}
      \subfigure[Relative outage  complexity]{\input{Complexity_outage_network_Nusers_rel_with_pl.tex}}
      \caption{Outage  complexity to ensure the per-cell outage constraint $\hat\epsilon= 0.1$. Solid lines are obtained through analysis, while the dots are the results of simulations. The notches on the right side of each sub-figure show the behavior as $N_\text{c} \rightarrow \infty$.}
      \label{fig:network:complexity_scaling_nusers_with_path_loss}
      \vspace{-0.25 cm}
\end{figure}

    The two sub-figures of Fig. \ref{fig:network:complexity_scaling_nusers_with_path_loss} show respectively the absolute and the relative outage complexity which ensure the per-cell outage constraint $\hat\epsilon= 0.1$. The normalization of the relative outage complexity is done over the decentralized case. In both sub-figures the dependency to the number of centralized \acp{BS} and the \ac{SNR} margin $\Delta\gamma$ is shown. As expected, the normalized outage complexity decreases as the number of centralized \acp{BS} or the \ac{SNR} margin increases. Fig. \ref{fig:network:complexity_scaling_nusers_with_path_loss} shows curves obtained both analytically (solid lines) and through simulations (dots) using the 3GPP LTE parametrization introduced in Section \ref{sec:outage.network:complexity.model}. The simulated results are obtained as follows. At each trial, the instantaneous \ac{SNR} $\gamma_i$ is determined by drawing a random variable exponentially distributed. Once the correct MCS is found, the complexity $C_\text{norm}$ is determined using (\ref{eq:outage.network:200}). Once a sufficient number of trials is run, $\mathcal{C}_\text{out}(\hat\epsilon, N_\text{c})$ is evaluated such that (\ref{eq:outage.network:500}) is satisfied, given a specific constraint on both the computational and channel outage. For the following results, hundred-thousand trials are run for each number of centralized \acp{BS} and for each \ac{SNR} margin $\Delta\gamma$ used.
     The path-loss exponent is set to $\eta = 2$, the average \ac{SNR} at unit distance is $\gamma_\text{ud} = \unit[0]{dB}$ and the compensation factor is fixed to $s = 0.1$.
     The notches on the right side of each sub-figure show the behavior as $N_\text{c} \rightarrow \infty$.
     Both sub-figures show a good match between analysis and simulations. Furthermore, the right sub-figure emphasizes the benefit of using a centralized \ac{RAN} rather than a distributed system.

\section{Numerical verification}\label{sec:numerical.verification}
In order to evaluate the validity of the analytical framework derived in Sec. \ref{sec:complexity.framework}, in this section numerical results are provided based on the 3GPP LTE standard and a 3GPP LTE decoder. The numerical results are compared against those obtained through the previously derived framework. LTE uses adaptive modulation and coding based on turbo codes with overall 27 distinct \acp{MCS} ($N_{R}=27$). Each MCS is identified by an MCS index, $I_\mathsf{mcs} = \{0,...,26\}$, and is characterized by a different combination of code rate and modulation format \cite{3GPP.TS.36.211, 3GPP.TS.36.212}.  Three kinds of modulation are used: QPSK ($0\leq I_\mathsf{mcs} \leq 10$), 16-QAM ($11\leq I_\mathsf{mcs} \leq 20$), and 64-QAM ($21\leq I_\mathsf{mcs} \leq 26$).  When a \ac{TB} is larger than 6144 bits, it is segmented into multiple \acp{CB}.  Each CB is separately turbo encoded, and all $C_{k}$ \acp{CB} in the \ac{TB} must be correctly decoded for the TB to be correct. Let $\epsilon_\mathsf{cb}$ be the probability that a CB is in a channel outage.  It follows that $\epsilon_\mathsf{channel} = 1-(1-\epsilon_\mathsf{cb})^{C_{k}}$ is the probability that the TB is in a channel outage.

\begin{figure}[tb]
      \centering
      \subfigure[Rayleigh fading]{\input{AbsComplexity.tex}}
      \subfigure[Distance dependent path-loss model]{\input{AbsComplexity_PathLoss.tex}}
      \caption{Outage complexity to ensure per-cell outage constraint $\hat\epsilon = 0.1$ as function of the number of RAPs, whose signals are centrally processed. Solid lines are evaluated analytically, while dots are obtained through simulations using one million trials. The notches on the right side of each sub-figure show the behavior as $N_\text{c} \rightarrow \infty$.}
      \label{fig:ComplexityNOICI}
      \vspace{-0.5 cm}
\end{figure}
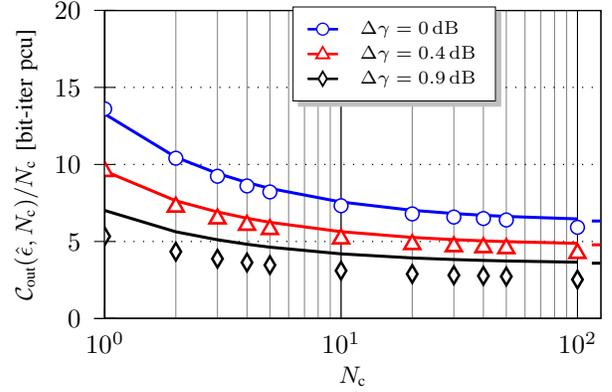
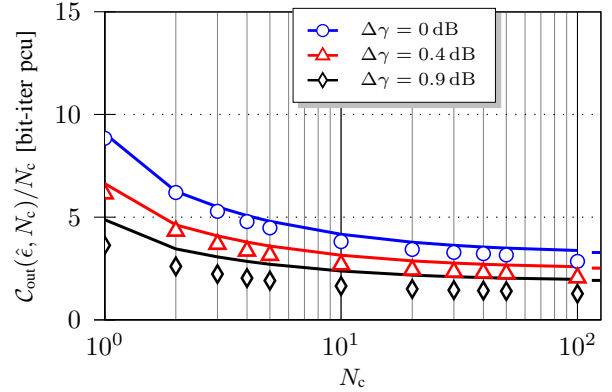

For a given value of SNR $\gamma_i$, the MCS is selected to satisfy $\epsilon_\mathsf{channel} \leq \hat{\epsilon}_\mathsf{channel}$ where $\hat{\epsilon}_\text{channel}= 0.1$, which is a typical value for an LTE network. When complexity is a concern, an \ac{SNR} margin $\Delta\gamma$ can be applied, which corresponds to the case when
 the MCS is selected such that the outage constraint is met after a specific number of decoder iterations.

Assume all the available resources per cell are allocated to a single users. Furthermore, a full-buffer scenario is assumed: each of the $N_\text{c}$ \acp{RAP}, whose signals are processed at the data center, always serves a user over the complete bandwidth. However, 3GPP LTE allows for multi-user opportunistic scheduling such that one \ac{TB} may carry information from more than one user. When resources are shared among multiple users
more diversity and therefore a lower outage complexity per cell can be achieved compared to the case when all the resources are monopolized by a single user.
Therefore, this assumption represents a worst-case scenario regarding computational complexity, since it does not allow for any computational diversity within a given cell.

A block Rayleigh fading channel is considered, although in practice a system experiences more channel diversity due to multi-path propagation, and the computational complexity varies from user to user and from \ac{CB} to \ac{CB}. This scenario, in an extreme case, leads the users to experience a sufficient channel diversity for a close-to-\ac{AWGN} channel, which results in a lower computational gain and diversity.

The outage complexity $\mathcal{C}_\text{out}(\hat\epsilon, N_\text{c})$ is obtained through a simulation campaign.  During each trial, the SNR $\gamma_i$ is drawn from an exponential distribution when we do not consider the path loss, or from a distribution whose \ac{CDF} is given by (\ref{eq:pathlossSNR}) when the path loss is taken into consideration. The MCS scheme for a given $\gamma_i$ is determined according to the value of $\Delta \gamma$ chosen.
Each of the $C_k$ \acp{CB} in the \ac{TB} is marked as being in a channel outage with probability $\epsilon_\mathsf{cb}$, which can be precomputed.  If any of the CB are in outage, then the entire TB is declared to be in an outage.  If the $r^{th}$ CB is in an outage, then $L_{r}\left(\gamma, \Delta\gamma\right) = L_{\mathsf{max}}$, which is the maximum number of attempted iterations.  Otherwise, $L_{r}\left(\gamma, \Delta\gamma\right)$ is determined by drawing a random variable distributed according to the pdf of $L_{r}\left(\gamma, \Delta\gamma\right)$, which can be precomputed by tracking the error-rate as a function of the number of iterations. Once the decoding complexity is computed, finally the outage complexity is evaluated by (\ref{eq:outage.complexity}).
The outage complexity is obtained by using one million trials when evaluated through simulations.

Fig. \ref{fig:ComplexityNOICI} shows the outage complexity to ensure per-cell outage constraint $\hat\epsilon= 0.1$ as function of the number of RAPs, whose signals are centrally processed. Results in the left sub-figure consider a scenario characterized by a Rayleigh fading channel with $\overline{\gamma}=\unit[10]{dB}$, while in the right sub-figure a distance dependent path loss model is taken into account, as described in Sec. \ref{sec:complexity.framework:with.pl}. For both models, three values of $\Delta\gamma$ are considered. For the right sub-figure, the path loss exponent is fixed to $\eta=2$, the compensation factor is $s=0.1$ and the average SNR at unit distance is set to $\gamma_{\text{ud}}=\unit[0]{dB}$. In Fig. \ref{fig:ComplexityNOICI}, the solid lines are evaluated analytically as derived in Section \ref{sec:complexity.framework:constant.pl} and \ref{sec:complexity.framework:with.pl}, while the dots are obtained through simulations using the methodology described above. The notches on the right side of each sub-figure show the behavior as $N_\text{c} \rightarrow \infty$.  Fig. \ref{fig:ComplexityNOICI} shows a good agreement for both models between the simulated and the analytical results, emphasizing the validity of the proposed analytical framework.

In Section \ref{sec:outage.network:complexity.model}, the complexity is assumed to be deterministic for a fixed \ac{SNR}, while in an actual implementation this would be a random variable with expectation $\mathcal{C}(\gamma, \Delta\gamma)$. This additional source of randomness is not considered in the analytical framework, and it is one of the reason for the slight mismatch between the empirical and analytical results. Another reason for the slight mismatch is that a lower-bound on the complexity is not applied: an actual decoder always performs at least one iteration which determines the minimum required complexity.
If $\mathcal{C}(\gamma, \Delta\gamma)$ is  a random variable, the computational variance would increase. This would consequently cause an increment of the computational gain for a centralized \ac{RAN}. Furthermore, if the lower bound is applied, it causes a slight increment of the expected complexity, even if the variance might decrease, which lead also in this case to a higher computational gain.

\section{Results}\label{sec:results}
In this section, the analytical framework developed in Section \ref{sec:complexity.framework} is applied to perform an evaluation of a centralized system using some of the metrics introduced in Section \ref{sec:system.model:definitions}.
In the following, the path-loss exponent is fixed to $\eta = 2$, the compensation factor is $s = 0.1$, the per-cell constraint on the computational outage is set to $\hat\epsilon= 0.1$, and the average \ac{SNR} at unit distance is $\gamma_\text{ud} = \unit[0]{dB}$.

\begin{figure}[t]
	\centering
	\subfigure[Gain as function of $N_\text{c}$]{\input{Complexity_gain_Nusers.tex}}
	\subfigure[Gain as function of  $\hat\epsilon$]{\input{Complexity_gain_epsilon.tex}}
	\caption{Complexity gain for $\gamma_\text{ud}=\unit[0]{dB}$, $\eta=2$, $s = 0.1$. The notches on the right side of the first sub-figure show the behavior as $N_\text{c} \rightarrow \infty$.}
	\label{fig:results:complexity_gain}
\vspace{-0.5 cm}
\end{figure}
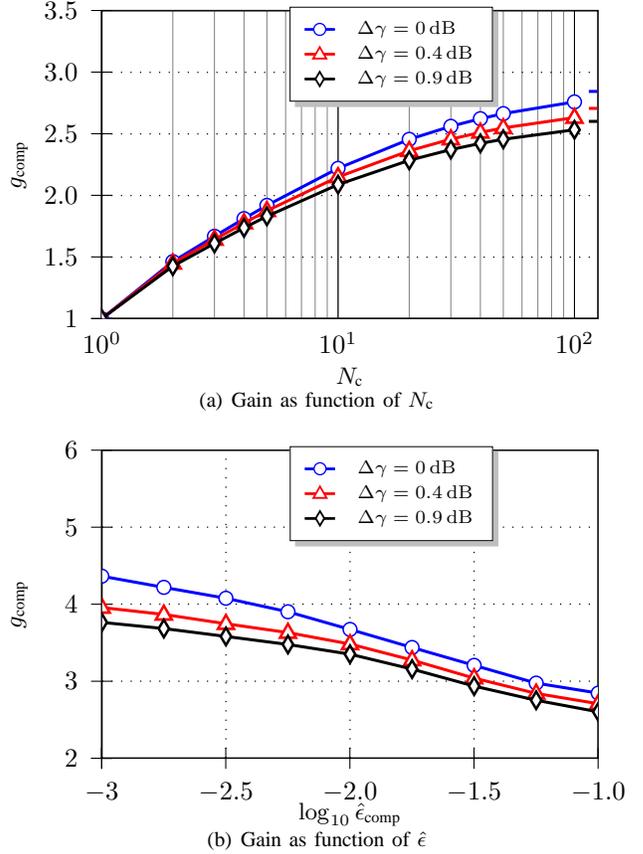


\subsection{Computational Gain}
At first, the computational gain $g_\text{comp}$ and its dependence on the number of centralized \acp{BS} and on the \ac{SNR} margin $\Delta\gamma$ is considered. The first sub-figure shows the computational gain as
function of the number of centralized \acp{BS}. The second sub-figure shows the computational gain as function of the target computational outage probability.

Fig. \ref{fig:results:complexity_gain}(a) shows that the computational gain increases sharply for small values of $N_\text{c}$, before it becomes flat for larger values of $N_\text{c}$. Furthermore, it shows a high computational gain compared to a distributed system, emphasizing the benefit of a centralized solution. The notches on the right side of the sub-figure show the behavior as $N_\text{c} \rightarrow \infty$.

Fig. \ref{fig:results:complexity_gain}(b) shows that the computational gain increases as the target computational outage probability decreases, since more computational resources are required to make sure that also peak-requirements are satisfied. This is a consequence of the fact that the complexity scales with $-\log_{10}(\hat\epsilon)$ in (\ref{eq:outage.network:201}).
In this case, centralization provides more benefits as diversity effects can be exploited to load balance between the individual \acp{BS}.
Furthermore, the computational gain differs more significantly for different \ac{SNR} offsets as the target computational outage probability decreases. This effect is mainly due to the fact that at lower target outage probability it is more challenging to operate close to capacity because the decoder would be required to have a steeper error rate curve.

\subsection{Computational Diversity}
    \begin{figure}
      \centering
      \input{Computational_diversity.tex}
      \caption{Computational diversity as function of the target computational outage probability.}
      \label{fig:results:computational_diversity}
      \vspace{-0.5 cm}
    \end{figure}
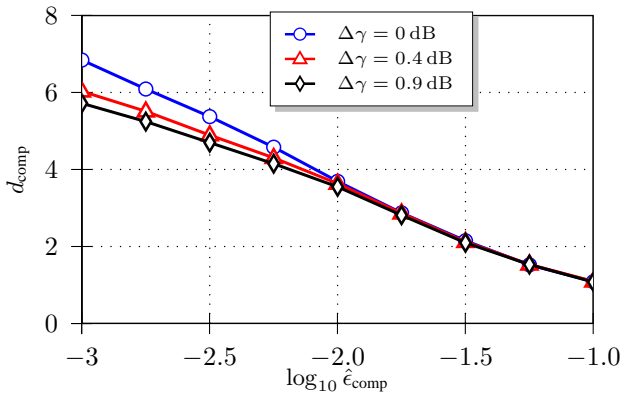
    Fig. \ref{fig:results:computational_diversity} shows the computational diversity as function of the target computational outage probability. For each \ac{SNR} margin $\Delta\gamma$ and target computational outage probability, the computational diversity is computed by numerically evaluating the derivative in (\ref{eq:outage.network:530}), once the computational outage probability is obtained from (\ref{eq:outage.network:computation.diversity:400}). The computational diversity, similarly to the computational gain, decreases by increasing the target computational outage probability. This is due to the over-provisioning of resources at lower target computational outage probability. Furthermore, the computational diversity differs significantly for different \ac{SNR} offsets as the target computational outage probability decreases.


    \begin{figure}
      \centering
      \input{Complexity_rate_tradeoff.tex}
      \caption{Complexity-rate tradeoff (CRT) as a function of $N_\text{c}$. The notches on the right side of the figure show the behavior as $N_\text{c} \rightarrow \infty$. }
      \label{fig:results:complexity_rate_tradeoff}
      \vspace{-0.5 cm}
    \end{figure}
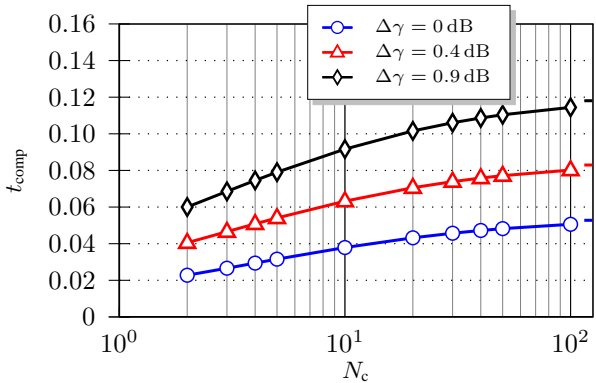

  \subsection{Complexity-Rate Tradeoff}

 Fig. \ref{fig:results:complexity_rate_tradeoff} shows the \ac{CRT} as function of the number of centralized \acp{BS}, which quantifies the increase of the achievable rate as computational resources are added. Fig. \ref{fig:results:complexity_rate_tradeoff} shows also the dependence of the \ac{CRT} on the \ac{SNR} margin offset.  The \ac{CRT} defined in (\ref{eq:outage.network:550}) is the ratio between the slope of the average rate and the slope of the outage complexity both computed at a given \ac{SNR} margin. In Fig. \ref{fig:results:complexity_rate_tradeoff}, the two slopes are individually evaluated numerically using (\ref{eq:AverageAchievableRate}) and (\ref{eq:outage.network:computation.diversity:400}) for each value of $N_\text{c}$ and \ac{SNR} margin $\Delta\gamma$. The notches on the right side of the figure show the behavior as $N_\text{c} \rightarrow \infty$.

  Fig. \ref{fig:results:complexity_rate_tradeoff} shows that the \ac{CRT} increases as the margin offset is increased and this is due to the fact that the decoder operates farther away from capacity and therefore does not cause high peak complexity. Hence, additional complexity is beneficial in order to gain additional achievable rate. In other words, the closer the decoder operates to capacity, the more resources must be invested to increase the achievable rate. Fig. \ref{fig:results:complexity_rate_tradeoff} points out that an optimum point to operate a system exists. Assume that a minimum average achievable rate is required. In this case, the operator may use a less complex decoder, which reduces the required computational resources (and therefore reduces costs for \ac{IT} infrastructures). However, a minimum computational complexity should be maintained since the penalty in terms of achievable rate may otherwise be significant. In addition, due to the lower per-cell achievable rate, more \acp{BS} must be deployed, which implies higher deployment costs. Depending on the costs for \ac{IT} infrastructure and mobile network infrastructure, an optimal operating point can be determined where deployment costs are minimized.

\section{Conclusions}\label{sec:conclusions}

Computational limitations in mobile wireless networks have a significant impact on the network performance.  Such  computational  limitations are of particular interest to small-cell networks, where \acp{RAP} are computationally limited due to economic constraints. The computational constraints are also of fundamental importance to the design of centralized-\ac{RAN} architectures.

In this paper a new analytical framework has been developed to evaluate the data processing requirements for a centralized-\ac{RAN}. Several performance metrics have been introduced to help quantifying the benefit of such a network compared to a conventional \ac{RAN}: \emph{computational outage probability}, \emph{computational gain}, \emph{computational diversity} and \emph{complexity-rate tradeoff}. The novel concept of \emph{computational outage} helps to quantify the complexity-throughput tradeoff in centralized RAN platforms.  Such a quantification allows for specifying computational requirements for centralized RAN platforms in a meaningful way.  Furthermore, insights into the \emph{computational gain} and \emph{computational diversity} offered by centralized processing are provided.
Finally, the \emph{complexity-rate tradeoff} is used to measure  the  additional
complexity needed to further improve the achievable rate.

The analytical framework has been verified by comparing it numerically with a large-scale system level simulator compliant with the 3GPP LTE standard. It has been shown that the analytical framework provides a good match with simulations despite simplifying assumptions which are necessary for the tractability of the derivations.

The analysis in this paper does not account explicitly for channel outage. For instance, the required computational resources may be higher in the case of a channel outage, since in this case the maximum allowed number of iterations needs to be used. In a more realistic scenario, the computational outage and channel outage can both occur, but only the system outage is relevant for the operator. We leave it as future work to include channel outage into our analytical framework.
Furthermore the effect of co-channel interference could be taken into account, e.\,g., by using tools from stochastic geometry, e.\,g. \cite{haenggi:2012} and bounds or asymptotical expressions could be derived for the novel performance metrics introduced here.


\appendices
\section{} \label{App:Expectation}
This section provides details leading to (\ref{eq:ExpextedComplexity}). By knowing the
\ac{PDF} of the instantaneous \ac{SNR} $\gamma$ provided by (\ref{eq:pdf_gamma}) and by using the complexity model in Section \ref{sec:outage.network:complexity.model}, the expected decoding complexity conditioned on the average \ac{SNR} $\overline\gamma$ can be evaluated as follows
\begin{eqnarray}
      \mathbb{E}_{\gamma_i, \Delta\gamma | \overline\gamma}\{\mathcal{C}_i\} & =& \int\limits_{0}^{\infty} \mathcal{C}_i(\gamma, \Delta\gamma) f_{\gamma_i}(\gamma) d\gamma \nonumber \\
      & =& \sum\limits_{k = 1}^{N_R} \int\limits_{\gamma^R_{k}}^{\gamma^R_{k+1}} \mathcal{C}_i(\gamma, \Delta\gamma) f_{\gamma_i}(\gamma) d\gamma \label{eq:ExpextedComplexityProof1}
\end{eqnarray}
By substituting (\ref{eq:pdf_gamma}) and (\ref{eq:outage.network:200}) into (\ref{eq:ExpextedComplexityProof1}), the integral in (\ref{eq:ExpextedComplexityProof1}) becomes
\begin{eqnarray}
\int\limits_{\gamma^R_{k}}^{\gamma^R_{k+1}} \mathcal{C}_i(\gamma, \Delta\gamma) f_{\gamma_i}(\gamma) d\gamma &=& \frac{r\left(\gamma^R_{k}, \Delta\gamma\right)}{\log_2\left(\zeta-1\right)}  \left[ I_1\left(\gamma^R_{k+1},\gamma^R_{k}\right) + \right. \nonumber \\ & & \left. I_2\left(\gamma^R_{k+1},\gamma^R_{k}\right)\right].
	\label{eq:ExpextedComplexityProof2}
\end{eqnarray}
where
\begin{eqnarray}
I_{1}\left(x_1,x_2\right) &=& \log_2 \left(\frac{\zeta - 2}{K\zeta}\right) \int\limits_{x_2}^{x_1}
	  \frac{1}{\overline\gamma} \exp\left(-\frac{\gamma} {\overline\gamma}\right)  d\gamma. \nonumber \\
 &=& - \log_2 \left(\frac{\zeta - 2}{K\zeta}\right) \exp\left( -\frac{\gamma} {\overline\gamma}\right)\biggr\rvert_{x_2}^{x_1}
 \label{eq:T1}
\end{eqnarray}
\begin{eqnarray}
I_{2}\left(x_1,x_2\right) &=& \frac{- 2}{\overline\gamma}  \int\limits_{x_2}^{x_1} \log_2\left[l_k(\gamma) \right]
	   \exp\left(-\frac{\gamma} {\overline\gamma}\right)  d\gamma.	
 \label{eq:T2}
\end{eqnarray}
with
\begin{eqnarray}
 l_k(\gamma)=\log_2\left( 1+\gamma\right)-r\left(\gamma^R_{k}, \Delta\gamma\right) \label{eq:LK}.
\end{eqnarray}
Eq. (\ref{eq:T2}) cannot be solved in closed form. However, a piece-wise linearization of (\ref{eq:LK}) can be done, which leads to
\begin{eqnarray}
l_k(\gamma) \approx c_k(\gamma)= a_k \gamma +b_k \label{eq:LKlinearized}
\end{eqnarray}
where
\begin{eqnarray}
      a_k & = & \frac{l_k(\gamma^R_{k+1}) - l_k(\gamma^R_{k})}{\gamma^R_{k+1} - \gamma^R_{k}} \label{eq:AK}\\
      b_k & = & l_k(\gamma^R_{k}) - \frac{\gamma^R_{k}}{\gamma^r_{k+1} - \gamma^R_{k}}\left[l_k(\gamma^R_{k+1}) - l_k(\gamma^R_{k})\right].
      \label{eq:BK}
\end{eqnarray}
Substituting (\ref{eq:LKlinearized}) into (\ref{eq:T2}) and by integrating by parts yields
\begin{eqnarray}
I_{2}\left(x_1,x_2\right) \approx 2 \log_2\left[ c_k(\gamma) \right] \exp\left( -\frac{\gamma} {\overline\gamma}\right)\biggr\rvert_{x_2}^{x_1} - I_{3}\left(x_1,x_2\right)
 \label{eq:T2_Proof2}
\end{eqnarray}
where
\begin{eqnarray}
I_{3}\left(x_1,x_2\right) &=& 2  \int\limits_{x_2}^{x_1} \left\{ \frac{\partial}{\partial \gamma} \log_2\left[l_k(\gamma) \right]\right\}
	   \exp\left(-\frac{\gamma} {\overline\gamma}\right)  d\gamma.	 \nonumber\\
&=& \frac{2}{\log\left( 2\right)}  \int\limits_{x_2}^{x_1} \frac{a_k}{a_k \gamma +b_k}
	   \exp\left(-\frac{\gamma} {\overline\gamma}\right)  d\gamma.
 \label{eq:T3}
\end{eqnarray}
After few algebraic manipulations and using the change of variable $\displaystyle \gamma= \frac{a_k t + b_k}{a_k \overline\gamma}$, (\ref{eq:T3}) becomes
\begin{eqnarray}
I_{3}\left(x_1,x_2\right) &=& \frac{2}{\log\left( 2\right)} \exp\left( \frac{b_k}{a_k \overline\gamma}\right)  \left[ \int\limits_{-\tau(x_1)}^{\infty}  \frac{1}{t}\exp(-t) dt - \right. \nonumber \\ & & \left.\int\limits_{-\tau(x_2)}^{\infty} \frac{1}{t}\exp(-t)   dt \right]
 \label{eq:T3_Proof1}
\end{eqnarray}
where $\displaystyle \tau(x)=\frac{a_k x + b_k}{a_k \overline\gamma}$.

The integrals in (\ref{eq:T3_Proof1}) are exponential integrals, defined as (i.e, \cite{Abramowitz:1965})
\begin{eqnarray}
    \text{E}_1(x)=\int_{x}^{\infty}\frac{e^{-t}}t\,dt.\, \label{eq:ExponentialFunction}
\end{eqnarray}
By substituting (\ref{eq:ExponentialFunction}) into (\ref{eq:T3_Proof1}), and (\ref{eq:T3_Proof1}) into (\ref{eq:T2_Proof2}) yields
\begin{eqnarray}
I_{2}\left(x_1,x_2\right) &\approx& 2 \log_2\left[ c_k(\gamma) \right] \exp\left( -\frac{\gamma} {\overline\gamma}\right)\biggr\rvert_{x_2}^{x_1} + \nonumber \\
& &
\frac{2}{\log\left( 2\right)} \exp\left( \frac{b_k}{a_k \overline\gamma}\right) \text{E}_1\left[\tau(x)\right] \biggr\rvert_{x_2}^{x_1}
 \label{eq:T2_Proof3}
\end{eqnarray}
Finally, an approximated closed form expression for the expected decoding complexity conditioned over the average \ac{SNR} is obtained by substituting (\ref{eq:ExpextedComplexityProof2}), (\ref{eq:T1}), and (\ref{eq:T2_Proof3}) into (\ref{eq:ExpextedComplexityProof1}).

\section{}\label{App:Variance}
This section provides details leading to (\ref{eq:VarianceComplexity}). The variance of the decoding complexity conditioned over the average \ac{SNR} $\overline\gamma$ is
\begin{eqnarray}
      \text{Var}_{\gamma_i, \Delta\gamma | \overline\gamma}\left\{\mathcal{C}_i\right\} & = & \mathbb{E}_{\gamma_i, \Delta\gamma | \overline\gamma}
    \left\{\left[\mathcal{C}_i - \mathbb{E}_{\gamma_i, \Delta\gamma | \overline\gamma}\left\{\mathcal{C}_i\right\}\right]^2\right\} \nonumber \\
    & = & \mathbb{E}_{\gamma_i, \Delta\gamma | \overline\gamma}
    \left\{\mathcal{C}_i^2\right\} - \mathbb{E}_{\gamma_i, \Delta\gamma | \overline\gamma}\left\{\mathcal{C}_i\right\}^2 \label{eq:VC}
\end{eqnarray}
where $\mathbb{E}_{\gamma_i, \Delta\gamma | \overline\gamma}\left\{\mathcal{C}_i\right\}$ is given by
(\ref{eq:ExpextedComplexity}) and
\begin{eqnarray}
\mathbb{E}_{\gamma_i, \Delta\gamma | \overline\gamma}
    \left\{\mathcal{C}_i^2\right\} & = & \int\limits_{0}^{\infty} \mathcal{C}_i^2(\gamma, \Delta\gamma) f_{\gamma_i}(\gamma) d\gamma \nonumber \\
    & = & \sum\limits_{k = 1}^{N_R} \int\limits_{\gamma^R_{k}}^{\gamma^R_{k+1}} \mathcal{C}_i^2(\gamma, \Delta\gamma) f_{\gamma_i}(\gamma) d\gamma. \label{eq:VarianceComplexityProof1}
\end{eqnarray}
By substituting (\ref{eq:pdf_gamma}) and (\ref{eq:outage.network:200}) into (\ref{eq:VarianceComplexityProof1}) yields
\begin{eqnarray}
\mathbb{E}_{\gamma_i, \Delta\gamma | \overline\gamma}
    \left\{\mathcal{C}_i^2\right\} =  \sum\limits_{k = 1}^{N_R} \left[\frac{2 r\left(\gamma^R_{k}, \Delta\gamma \right)}{\log_2\left(\zeta - 1\right)}\right]^2  I_4\left(\gamma^R_{k+1},\gamma^R_{k}\right) \label{eq:VarianceComplexityProof2}
\end{eqnarray}
where
\begin{eqnarray}
   I_4\left(x_1,x_2\right) \hspace{-0.2 cm}& = & \hspace{-0.2 cm} \frac{1}{\overline\gamma} \int\limits_{x_2}^{x_1}
	  \left\{\log_2\left[\frac{\log_2\left(1+\gamma\right)-r\left(\gamma, \Delta\gamma\right) }{G}\right]\right\}^2 \times \nonumber \\
\hspace{-0.2 cm} & &\hspace{-0.2 cm}	  \exp\left(-\frac{\gamma}{\overline\gamma}\right) d\gamma
\label{eq:VarianceComplexityProof3}
\end{eqnarray}
and $G=\sqrt{\frac{\zeta - 2}{K\zeta}}$.

The integral in (\ref{eq:VarianceComplexityProof3}) cannot be solved in closed form. However, a piece-wise linearization of (\ref{eq:LK}) can be done, and (\ref{eq:VarianceComplexityProof3}) becomes
\begin{eqnarray}
   I_4\left(x_1,x_2\right)  \approx  \frac{1}{\overline\gamma} \int\limits_{x_2}^{x_1}
	  \log_2^2\left(\frac{a_k}{G} \gamma +\frac{b_k}{G} \right)
	  \exp\left(-\frac{\gamma}{\overline\gamma}\right) d\gamma.
\label{eq:VarianceComplexityProof4}
\end{eqnarray}
By integrating (\ref{eq:VarianceComplexityProof4}) by parts yields
\begin{eqnarray}
   I_4\left(x_1,x_2\right)  &\approx&  -\log_2^2\left( \frac{a_k}{G} \gamma +\frac{b_k}{G} \right) \exp\left( -\frac{\gamma}{\overline\gamma}\right)\biggr\rvert_{x_2}^{x_1}  + \nonumber \\
   & &  I_5\left(x_1,x_2\right)
\label{eq:lastT4b}
\end{eqnarray}
where
\begin{eqnarray}
   I_5\left(x_1,x_2\right) \hspace{-0.2cm}&=& \hspace{-0.2cm}\int\limits_{x_2}^{x_1}
	  \exp\left(-\frac{\gamma}{\overline\gamma}\right) \frac{\partial}{\partial \gamma}\log_2^2\left(\frac{a_k}{G} \gamma +\frac{b_k}{G} \right)
 d\gamma. \nonumber \\
 \hspace{-0.2cm} &=&\hspace{-0.2cm} \frac{2}{\left[\log\left(2\right)\right]^2}\int\limits_{x_2}^{x_1}
	  \exp\left(-\frac{\gamma}{\overline\gamma}\right) \frac{ a_k}{ a_k \gamma + b_k} \times \nonumber \\
\hspace{-0.2cm} & & \hspace{-0.2cm}\log\left(\frac{a_k}{G} \gamma +\frac{b_k}{G} \right) d\gamma.
 \label{eq:VarianceComplexityProof6}
\end{eqnarray}
After making the substitution $t=a_k \gamma+b_k$ into (\ref{eq:VarianceComplexityProof6}) and by defining the function $h\left(x\right)= a_k x + b_k$ yields
\begin{eqnarray}
   I_5\left(x_1,x_2\right) \hspace{-0.2 cm} &=& \hspace{-0.2 cm}\frac{2 }{\left[\log\left(2\right)\right]^2}\exp\left(\frac{b_k}{a_k\overline\gamma}\right)\nonumber \\
    \hspace{-0.2 cm} & & \hspace{-0.2 cm}
   \int\limits_{h\left(x_2\right)}^{h\left(x_1\right)}
	  \frac{1}{ t} \exp\left(-\frac{t}{ a_k \overline\gamma}\right)  \log\left(\frac{t}{G} \right) dt.
 \label{eq:VarianceComplexityProof7}
\end{eqnarray}
By integrating again by parts, (\ref{eq:VarianceComplexityProof7}) becomes
\begin{eqnarray}
   I_5\left(x_1,x_2\right) \hspace{-0.2 cm}&=& \hspace{-0.2 cm} \frac{2 \exp\left(\frac{b_k}{a_k\overline\gamma}\right)}{\left[\log\left(2\right)\right]^2}
   \left[ -\text{E}_1\left( \frac{t}{a_k \overline\gamma}\right) \log\left( \frac{t}{G} \right)\biggr\rvert_{{h\left(x_2\right)}}^{{h\left(x_1\right)}} + \right. \nonumber \\
   \hspace{-0.2 cm} & & \hspace{-0.2 cm}
   I_6\left[h\left(x_1\right),h\left(x_2\right)\right] \Bigg]
 \label{eq:VarianceComplexityProof8}
\end{eqnarray}
where $\text{E}_1(\cdot)$ is an exponential integral given as (\ref{eq:ExponentialFunction}) and
\begin{eqnarray}
 I_6\left(x_1,x_2\right) &=& \int\limits_{x_2}^{x_1}  \frac{1}{t} \text{E}_1\left(\frac{t}{a_k \overline\gamma}\right) dt. \label{eq:T6}
\end{eqnarray}
The integral (\ref{eq:T6}) can be solved using the equality \cite[Eq. (18), Sec. 4.1]{Geller.1969},
\begin{eqnarray}
\int_{b}^{\infty} \frac{\text{E}_1\left( a x \right)}{x}  dx &= & \frac{1}{2} \left\{ \left[ \EulerGamma + \log\left( a b\right) \right]^2 + \zeta\left( 2\right) \right\} + \nonumber \\ & &  \sum_{n=1}^{\infty} \frac{\left(-a b \right)^n}{n! n^2}. \label{eq:GellerEdward}
\end{eqnarray}
where $\EulerGamma$ is the Euler-Mascheroni constant and $\zeta\left( \cdot\right)$ is the Riemann zeta function \cite{Abramowitz:1965}.
Furthermore, let use the identify
\begin{eqnarray}
\sum_{n=1}^{\infty} \frac{\left(-a x \right)^n}{n! n^2} =  -x   \pFq{3}{3}{1,1,1}{2,2,2}{x} \label{eq:IdentityGHF}
\end{eqnarray}
where, by using the Pochhammer symbol $(a)_0 = 1$ and $(a)_n = a(a+1)(a+2)...(a+n-1),n \geq 1$,
$_pF_q(a_1,\ldots,a_p;b_1,\ldots,b_q;z)$ is the generalized hypergeometric function defined as (i.e, \cite{Abramowitz:1965})
\begin{eqnarray}
\pFq{p}{q}{a_1,\ldots,a_p}{b_1,\ldots,b_q}{x}= \sum_{n=0}^\infty \frac{(a_1)_n\dots(a_p)_n}{(b_1)_n\dots(b_q)_n} \, \frac {x^n} {n!}
\end{eqnarray}
By applying (\ref{eq:GellerEdward}) and (\ref{eq:IdentityGHF}) into (\ref{eq:T6}) yields
\begin{eqnarray}
 I_6\left(x_1,x_2\right) &=& - \frac{1}{2} \left\{ \left[ \EulerGamma + \log\left( \frac{t}{a_k \overline\gamma}\right) \right]^2 + \zeta\left( 2\right) \right\}\biggr\rvert_{x_2}^{x_1} + \nonumber \\
 & &
 \frac{t}{a_k \overline\gamma}\pFq{3}{3}{1, 1, 1}{2, 2, 2}{-\frac{t}{a_k \overline\gamma}}\biggr\rvert_{x_2}^{x_1}. \label{eq:T6_2}
\end{eqnarray}
While the generalized hypergeometric function is itself an infinite series as well as the Riemann zeta function, they are widely known and they are implemented as a single function call in most mathematical programming languages, including Matlab.

Finally, by substituting (\ref{eq:VarianceComplexityProof8}) and (\ref{eq:T6_2}) into (\ref{eq:lastT4b}),
and after few manipulations
yields
\begin{eqnarray}
 \hspace{-0.3 cm} & & \hspace{-0.3 cm} I_4\left(x_1,x_2\right) \approx  \left\{ \frac{2 \exp\left(\frac{b_k}{a_k\overline\gamma}\right)}{\left[\log\left(2\right)\right]^2} \left[ \frac{t}{a_k \overline\gamma}\pFq{3}{3}{1, 1, 1}{2, 2, 2}{-\frac{t}{a_k \overline\gamma}}- \right. \right.
  \nonumber \\
   & & \hspace{0.6 cm}
  \text{E}_1\left( \frac{t}{a_k \overline\gamma}\right) \log\left( \frac{t}{G} \right) - \frac{1}{2} \left\{ \left[ \log\left( \frac{t}{a_k \overline\gamma}\right)+ \EulerGamma \right]^2 + \right.
 \nonumber \\
   & & \hspace{0.6 cm}
   \left.
  \zeta\left( 2\right) \biggr\} \right]-
  \left[\log_2\left( t \right)\right]^2 \exp\left( \frac{b_k-t}{a_k \overline\gamma}\right)
  \Biggr \} \Biggr\rvert_{{h\left(x_2\right)}}^{{h\left(x_1\right)}}
\label{eq:lastT4}
\end{eqnarray}



\section{} \label{App:CDF_SNR}
This section provides details leading to (\ref{eq:pathlossSNR}). The \ac{CDF} of the instantaneous \ac{SNR} $\gamma$ averaged over the spatial distribution can be obtained by
\begin{eqnarray}
      F_\gamma(\gamma) & = & \int F_{\gamma}\left(\gamma \big| \omega\right) f_{r_i}\left( \omega\right) \,{\rm d}\omega \label{eq:pathloss_Proof}
\end{eqnarray}
where $F_{\gamma}\left(\gamma \big| \omega\right)=1-\exp\left( - \frac{\gamma}{\overline\gamma} \right)$, $\gamma \geq 0$ and $\overline\gamma$ is given by (\ref{eq:instantaneousSNR}).
Substituting (\ref{eq:pdf_radius}) into (\ref{eq:pathloss_Proof}) yields
\begin{eqnarray}
      F_\gamma(\gamma) & = & \int_{0}^{1} \left[ 1 -\exp \left( - \frac{\gamma}{\gamma_\text{ud}} \omega^{\eta\left( 1-s\right)}  \right) \right] 2 \omega \,{\rm d}\omega \nonumber \\
      & = &  1 - 2 \int_{0}^{1} \exp \left( - \frac{\gamma}{\gamma_\text{ud}} \omega^{\eta\left( 1-s\right)}  \right) \omega \,{\rm d}\omega    \label{eq:pathloss_Proof2}
\end{eqnarray}
where $\eta$ is the path loss exponent, $s$ is the compensation factor for fractional power control, and $\gamma_\text{ud}$ is the average \ac{SNR} at unit distance away.

After few algebraic manipulations and using the change of variable $\displaystyle t=\frac{\gamma}{\gamma_\text{ud}} \omega^{\eta\left( 1-s\right)}$, (\ref{eq:pathloss_Proof2}) becomes
\begin{eqnarray}
      F_\gamma(\gamma) & = & 1 - \alpha \left( \frac{\gamma}{\gamma_\text{ud}}\right)^{-\alpha} \left[ \int_{\gamma/\gamma_\text{ud}}^{\infty} \exp \left( - t \right) t^{\alpha-1}\,{\rm d}t  - \right. \nonumber \\
       & & \left.
       \int_{0}^{\infty} \exp \left( - t \right) t^{\alpha-1}\,{\rm d}t\right]
           \label{eq:pathloss_Proof3}
\end{eqnarray}
where $\alpha=\frac{2}{\eta\left( 1-s\right)}$ and the integrals in (\ref{eq:pathloss_Proof3}) are gamma incomplete functions. Finally, by using the gamma incomplete functions in (\ref{eq:pathloss_Proof3}), (\ref{eq:pathlossSNR}) is obtained.

\bibliographystyle{IEEEtran}
\bibliography{IEEEfull,references}

\vspace{-0.65cm}

\begin{IEEEbiography}[{\includegraphics[width=1in, height=1.25in,clip,keepaspectratio]{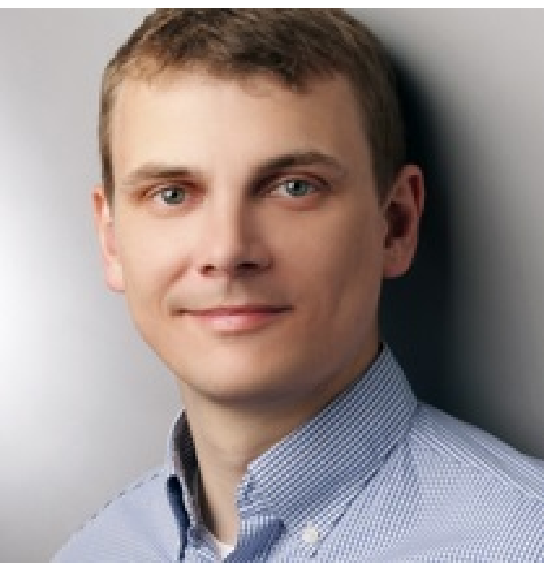}}]{Peter Rost} (S'06-M'10-SM'15) received his Ph.D. degree from Technische Universität Dresden, Dresden, Germany, in 2009 under supervision of Prof. G. Fettweis, and his M.Sc. degree from University of Stuttgart, Stuttgart, Germany, in 2005. Peter has been with Fraunhofer Institute for Beam and Material Technologies, Dresden, Germany; IBM Deutschland Entwicklung GmbH, Böblingen, Germany; and NEC Laboratories Europe, Heidelberg, Germany.
Since May 2015, Peter is member of the Radio Systems research group at Nokia Networks, Munich, Germany, where he contributes to the European H2020 projects 5G-NORMA and METIS-II, and works in business unit projects on 5G Architecture. Peter has been involved in several EU projects (e.g. FP7 iJOIN as Technical Manager), and standardization (e.g. 3GPP RAN2).
Currently, Peter serves as member of IEEE ComSoc GITC, IEEE Online GreenComm Steering Committee, and VDE ITG Expert Committee Information and Communication Theory. He is an Executive Editor of IEEE Transactions of Wireless Communications. Peter further supported several conferences, e.g. IEEE ICC 2009 and IEEE VTC Spring 2013 (TPC Chair). Peter published more than 40 scientific publications and he is author of multiple patents and patent applications.
\end{IEEEbiography}

\vspace{-1.05cm}

\begin{IEEEbiography}[{\includegraphics[width=1in,height=1.25in,clip,keepaspectratio]{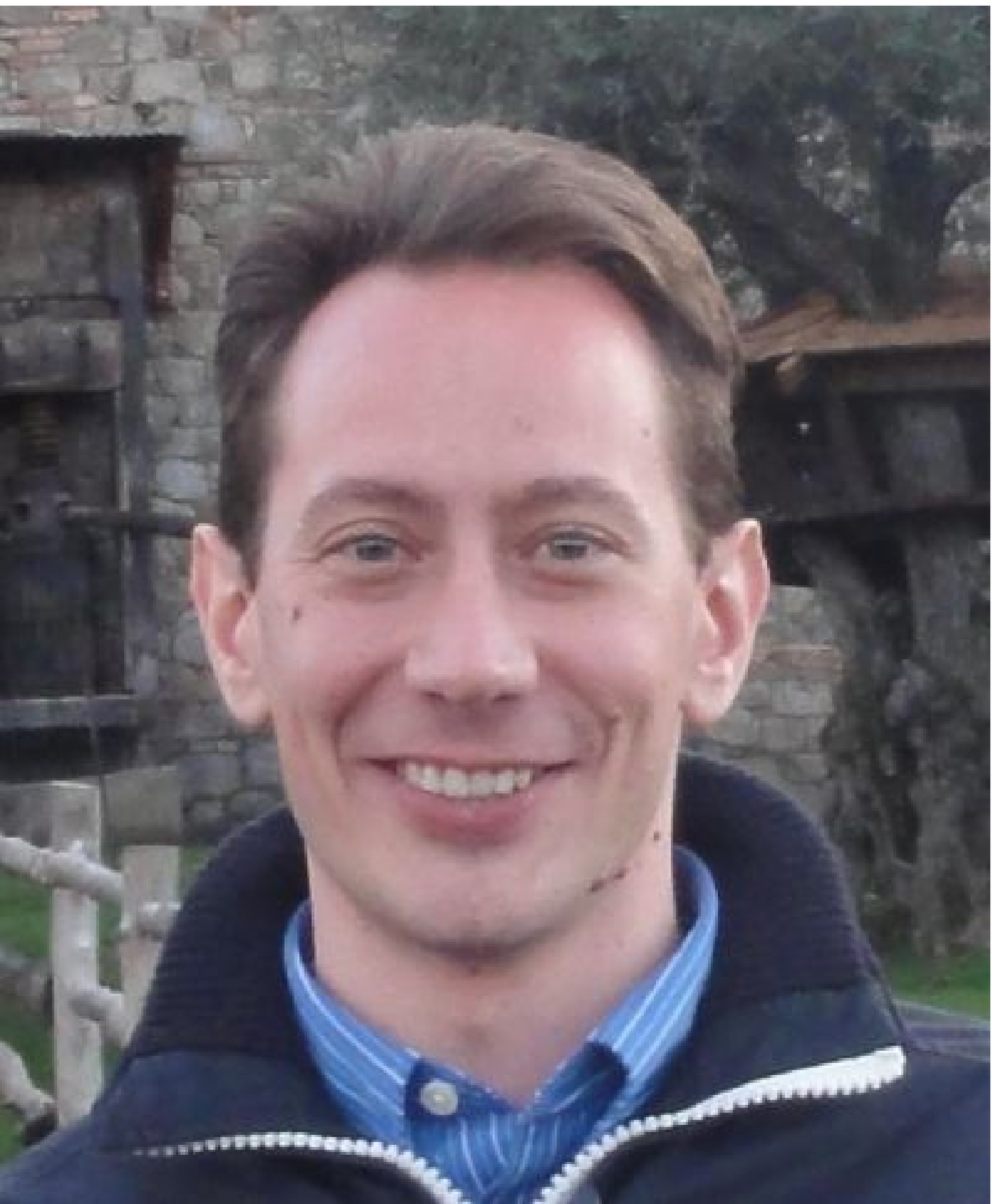}}]{Salvatore Talarico}
received the BSc and MEng degrees in electrical engineering from University of Pisa, Italy, in 2006 and 2007 respectively. From 2008 until 2010, he worked in the R$\&$D department of Screen Service Broadcasting Technologies (SSBT) as an RF System Engineer. He is currently a research assistant and a Ph.D. candidate in the Lane Department of Computer Science and Electrical Engineering at West Virginia University, Morgantown, WV. His research interests are in wireless communications, software defined radio and modeling, performance evaluation and optimization of ad-hoc and cellular networks.
\end{IEEEbiography}

\vspace{-0.85cm}

  \begin{IEEEbiography}[{\includegraphics[width=1in,height=1.25in,clip,keepaspectratio]{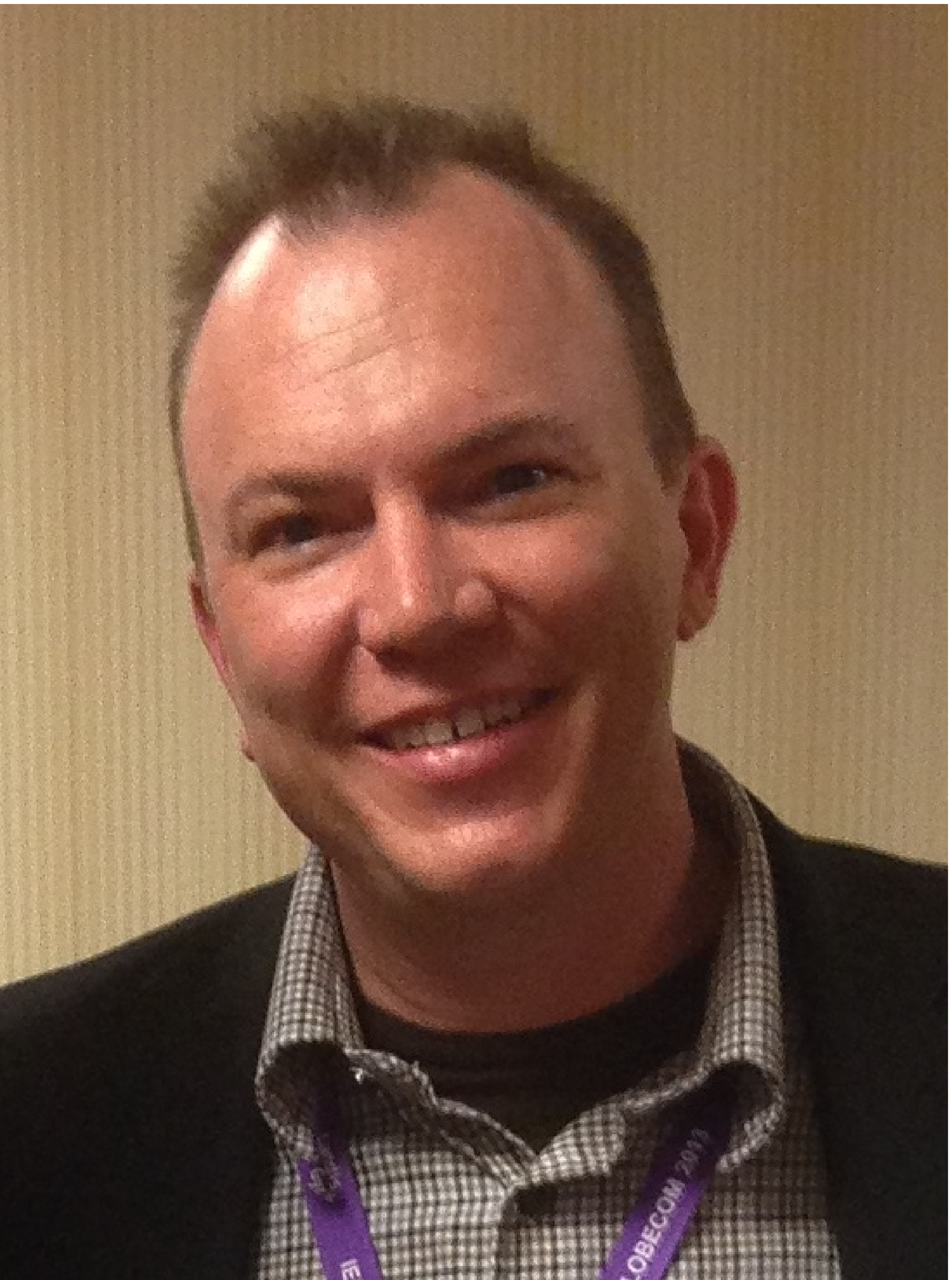}}]{Matthew C. Valenti}
is a Professor in the Lane Department of Computer Science and Electrical Engineering at West Virginia University and site director for the Center for Identification Technology Research (CITeR), an NSF Industry/University Cooperative Research Center (I/UCRC).  His research is in the area of wireless communications, including cellular networks, military communication systems, sensor networks, and coded modulation for satellite communications.   He is active in the organization of major conferences, including serving as the Technical Program Vice Chair for MILCOM 2015 and Globecom 2013, and as a track or symposium chair for MILCOM ('10,'12,'14), ICC ('09,'11), Globecom ('15), and VTC ('07).   He is an Executive Editor for IEEE Transactions on Wireless Communications, an Editor for IEEE Transactions on Communications, and the Chair of ComSoc's Communication Theory Technical Committee.
At WVU, he is a Faculty Senator and is engaged in curriculum development at all levels of the university.  Dr. Valenti is an alumnus of Virginia Tech, having received his BSEE in 1992 and Ph.D. in 1999, under the support of the Bradley Fellowship.  In addition, he received a MSEE from Johns Hopkins and worked as an Electronics Engineer at the US Naval Research Laboratory. He is registered as a Professional Engineer in the state of West Virginia.
\end{IEEEbiography}

\balance
\vfill

\end{document}

%% file: cloud.architecture.tex
\begingroup
\unitlength=1mm
\begin{picture}(79, 62)(0, 0)
  \psset{xunit=1mm, yunit=1mm, linewidth=0.3mm}
  \psset{arrowsize=3pt 4, arrowlength=1.4, arrowinset=.4}
  \rput(0, -3){%
    \rput(17, 0){%
      \cnodeput(10, 10){RRH1}{\textcolor{darkblue}{RAP 1}}%
      \cnodeput(25, 10){RRH2}{\textcolor{darkblue}{RAP 2}}%
      \cnodeput(40, 10){RRH3}{\textcolor{darkblue}{RAP 3}}%
      \cnodeput(55, 10){RRH4}{\textcolor{darkblue}{RAP N}}%
      \rput[r](0, 10){\textcolor{darkblue}{Remote layer}}%
      \rput[c](32.5, 35){\rnode{Transport}{\psframebox[linecolor=black]{Transport Network}}}%
      \rput(12, 55){\psframebox[linecolor=black]{\textcolor{darkgreen}{VM 1}}}%
      \rput(25, 55){\psframebox[linecolor=black]{\textcolor{darkgreen}{VM 2}}}%
      \rput(38, 55){\psframebox[linecolor=black]{\textcolor{darkgreen}{VM 3}}}%
      \rput(51, 55){\psframebox[linecolor=black]{\textcolor{darkgreen}{VM N}}}%
      \pnode(32.5, 50){VM}\psframe(5, 50)(58, 65)%
      \rput[t](32.5, 63){Virtual BS pool}%
      \rput[r](0, 55.5){\textcolor{darkgreen}{Central layer}}%
      \ncline{->}{VM}{Transport}%
      \ncline{->}{Transport}{RRH1}%
      \ncline{->}{Transport}{RRH2}%
      \ncline{->}{Transport}{RRH3}%
      \ncline{->}{Transport}{RRH4}%
      \rput(32.5, 24.5){\psframebox[linecolor=white, fillcolor=white, fillstyle=solid]{Back-/Fronthaul}}%
      }%
  }
\end{picture}
\endgroup

%% file: Complexity_SNR_theory.tex
\begingroup
\unitlength=1mm
\psset{xunit=3.53333mm, yunit=1.76000mm, linewidth=0.1mm}
\psset{arrowsize=2pt 3, arrowlength=1.4, arrowinset=.4}\psset{axesstyle=frame}
\begin{pspicture}(-13.96226, -9.09091)(20.00000, 25.00000)
\rput(-0.56604, -3.97727){%
\psaxes[subticks=0, labels=all, xsubticks=1, ysubticks=1, Ox=-10, Oy=0, Dx=5, Dy=5]{-}(-10.00000, 0.00000)(-10.00000, 0.00000)(20.00000, 25.00000)%
\multips(-5.00000, 0.00000)(5.00000, 0.0){5}{\psline[linecolor=black, linestyle=dotted, linewidth=0.2mm](0, 0)(0, 25.00000)}
\multips(-10.00000, 5.00000)(0, 5.00000){4}{\psline[linecolor=black, linestyle=dotted, linewidth=0.2mm](0, 0)(30.00000, 0)}
\rput[b](5.00000, -5.11364){SNR [dB]}
\rput[t]{90}(-13.39623, 12.50000){$\mathcal{C}(\gamma,\Delta\gamma)$ [bit-iter pcu]}
\psclip{\psframe(-10.00000, 0.00000)(20.00000, 25.00000)}
\psline[linecolor=blue, plotstyle=curve, linewidth=0.4mm, showpoints=false, linestyle=solid, linecolor=blue, dotstyle=triangle, dotscale=1.2 1.2, linewidth=0.4mm](-10.00000, 0.00000)(-9.95000, 0.00000)(-9.90000, 0.00000)(-9.85000, 0.00000)(-9.80000, 0.00000)(-9.75000, 0.00000)(-9.70000, 0.00000)(-9.65000, 0.00000)(-9.60000, 0.00000)(-9.55000, 0.00000)(-9.50000, 0.00000)(-9.45000, 0.00000)(-9.40000, 0.00000)(-9.35000, 0.00000)(-9.30000, 0.00000)(-9.25000, 0.00000)(-9.20000, 0.00000)(-9.15000, 0.00000)(-9.10000, 0.00000)(-9.05000, 0.00000)(-9.00000, 0.00000)(-8.95000, 0.00000)(-8.90000, 0.00000)(-8.85000, 0.00000)(-8.80000, 0.00000)(-8.75000, 0.00000)(-8.70000, 0.00000)(-8.65000, 0.00000)(-8.60000, 0.00000)(-8.55000, 0.00000)(-8.50000, 0.00000)(-8.45000, 0.00000)(-8.40000, 0.00000)(-8.35000, 0.00000)(-8.30000, 0.00000)(-8.25000, 0.00000)(-8.20000, 0.00000)(-8.15000, 0.00000)(-8.10000, 0.00000)(-8.05000, 0.00000)(-8.00000, 0.00000)(-7.95000, 0.00000)(-7.90000, 0.00000)(-7.85000, 0.00000)(-7.80000, 0.00000)(-7.75000, 0.00000)(-7.70000, 0.00000)(-7.65000, 0.00000)(-7.60000, 0.00000)(-7.55000, 0.00000)(-7.50000, 0.00000)(-7.45000, 0.00000)(-7.40000, 0.00000)(-7.35000, 0.00000)(-7.30000, 0.00000)(-7.25000, 0.00000)(-7.20000, 0.00000)(-7.15000, 0.00000)(-7.10000, 0.00000)(-7.05000, 0.00000)(-7.00000, 0.00000)(-6.95000, 0.00000)(-6.90000, 0.00000)(-6.85000, 0.00000)(-6.80000, 0.00000)(-6.75000, 0.00000)(-6.70000, 0.00000)(-6.65000, 0.00000)(-6.60000, 0.00000)(-6.55000, 0.86126)(-6.50000, 0.81200)(-6.45000, 0.76380)(-6.40000, 0.72414)(-6.35000, 0.68573)(-6.30000, 0.65393)(-6.25000, 0.62310)(-6.20000, 0.59783)(-6.15000, 0.57338)(-6.10000, 0.55244)(-6.05000, 0.53201)(-6.00000, 0.51395)(-5.95000, 0.49622)(-5.90000, 0.47960)(-5.85000, 0.46314)(-5.80000, 0.44842)(-5.75000, 0.43397)(-5.70000, 0.42155)(-5.65000, 0.40942)(-5.60000, 0.39983)(-5.55000, 0.39059)(-5.50000, 1.13484)(-5.45000, 1.06900)(-5.40000, 1.00590)(-5.35000, 0.95245)(-5.30000, 0.90187)(-5.25000, 0.85971)(-5.20000, 0.82005)(-5.15000, 0.78717)(-5.10000, 0.75631)(-5.05000, 0.72994)(-5.00000, 0.70492)(-4.95000, 0.68240)(-4.90000, 0.66063)(-4.85000, 0.63991)(-4.80000, 0.61949)(-4.75000, 0.59950)(-4.70000, 0.57963)(-4.65000, 0.56117)(-4.60000, 1.57202)(-4.55000, 1.47782)(-4.50000, 1.38447)(-4.45000, 1.30744)(-4.40000, 1.23121)(-4.35000, 1.17043)(-4.30000, 1.11029)(-4.25000, 1.06239)(-4.20000, 1.01503)(-4.15000, 0.97789)(-4.10000, 0.94115)(-4.05000, 0.91214)(-4.00000, 0.88332)(-3.95000, 0.85838)(-3.90000, 0.83348)(-3.85000, 0.80938)(-3.80000, 0.78521)(-3.75000, 0.75973)(-3.70000, 0.73433)(-3.65000, 0.71033)(-3.60000, 0.68649)(-3.55000, 0.66565)(-3.50000, 0.64508)(-3.45000, 2.03627)(-3.40000, 1.90855)(-3.35000, 1.79061)(-3.30000, 1.68601)(-3.25000, 1.59361)(-3.20000, 1.51114)(-3.15000, 1.43775)(-3.10000, 1.37215)(-3.05000, 1.31368)(-3.00000, 1.26190)(-2.95000, 1.21625)(-2.90000, 1.17641)(-2.85000, 1.14189)(-2.80000, 1.10990)(-2.75000, 1.08022)(-2.70000, 1.04992)(-2.65000, 1.01907)(-2.60000, 0.98695)(-2.55000, 0.95367)(-2.50000, 0.92140)(-2.45000, 2.56243)(-2.40000, 2.39226)(-2.35000, 2.24057)(-2.30000, 2.10312)(-2.25000, 1.98310)(-2.20000, 1.87650)(-2.15000, 1.78279)(-2.10000, 1.69902)(-2.05000, 1.62530)(-2.00000, 1.55933)(-1.95000, 1.50210)(-1.90000, 1.45162)(-1.85000, 1.40681)(-1.80000, 1.36634)(-1.75000, 1.32645)(-1.70000, 1.28700)(-1.65000, 3.30475)(-1.60000, 3.09585)(-1.55000, 2.89649)(-1.50000, 2.73283)(-1.45000, 2.57760)(-1.40000, 2.45393)(-1.35000, 2.33512)(-1.30000, 2.23452)(-1.25000, 2.13683)(-1.20000, 2.05003)(-1.15000, 1.96671)(-1.10000, 1.89643)(-1.05000, 1.83068)(-1.00000, 1.78189)(-0.95000, 1.73602)(-0.90000, 1.70109)(-0.85000, 1.66573)(-0.80000, 4.01320)(-0.75000, 3.76477)(-0.70000, 3.51996)(-0.65000, 3.32550)(-0.60000, 3.13437)(-0.55000, 2.98954)(-0.50000, 2.84680)(-0.45000, 2.73338)(-0.40000, 2.62061)(-0.35000, 2.51695)(-0.30000, 2.41401)(-0.25000, 2.32108)(-0.20000, 2.23000)(-0.15000, 2.16478)(-0.10000, 2.10130)(-0.05000, 2.06210)(0.00000, 2.02343)(0.05000, 4.86973)(0.10000, 4.55281)(0.15000, 4.24693)(0.20000, 3.99300)(0.25000, 3.75085)(0.30000, 3.56418)(0.35000, 3.38505)(0.40000, 3.24140)(0.45000, 3.10131)(0.50000, 2.97791)(0.55000, 2.85602)(0.60000, 2.74118)(0.65000, 2.63123)(0.70000, 2.54426)(0.75000, 2.46341)(0.80000, 2.41136)(0.85000, 4.74322)(0.90000, 4.44834)(0.95000, 4.16204)(1.00000, 3.93118)(1.05000, 3.71138)(1.10000, 3.53201)(1.15000, 3.36071)(1.20000, 3.21506)(1.25000, 3.07452)(1.30000, 2.96210)(1.35000, 2.85528)(1.40000, 2.77104)(1.45000, 2.69129)(1.50000, 2.62194)(1.55000, 2.55465)(1.60000, 5.26146)(1.65000, 4.90659)(1.70000, 4.61220)(1.75000, 4.33289)(1.80000, 4.10599)(1.85000, 3.89217)(1.90000, 3.71222)(1.95000, 3.54072)(2.00000, 3.39800)(2.05000, 3.26246)(2.10000, 3.15742)(2.15000, 3.05998)(2.20000, 2.97699)(2.25000, 2.89761)(2.30000, 2.81607)(2.35000, 2.73398)(2.40000, 2.64516)(2.45000, 2.55465)(2.50000, 2.46311)(2.55000, 2.37131)(2.60000, 2.29273)(2.65000, 5.66106)(2.70000, 5.31990)(2.75000, 4.97911)(2.80000, 4.71437)(2.85000, 4.44993)(2.90000, 4.24553)(2.95000, 4.04135)(3.00000, 3.88188)(3.05000, 3.72262)(3.10000, 3.60509)(3.15000, 3.48766)(3.20000, 3.39123)(3.25000, 3.29479)(3.30000, 3.19627)(3.35000, 3.09770)(3.40000, 2.98959)(3.45000, 2.88151)(3.50000, 2.77860)(3.55000, 2.67578)(3.60000, 2.59359)(3.65000, 2.51152)(3.70000, 2.45281)(3.75000, 2.39421)(3.80000, 7.56408)(3.85000, 7.14288)(3.90000, 6.76183)(3.95000, 6.41431)(4.00000, 6.11576)(4.05000, 5.84551)(4.10000, 5.61657)(4.15000, 5.40597)(4.20000, 5.22215)(4.25000, 5.04433)(4.30000, 4.87530)(4.35000, 4.70978)(4.40000, 4.54939)(4.45000, 4.39877)(4.50000, 4.26242)(4.55000, 4.14392)(4.60000, 4.05147)(4.65000, 8.45601)(4.70000, 8.01117)(4.75000, 7.58144)(4.80000, 7.22242)(4.85000, 6.87659)(4.90000, 6.59249)(4.95000, 6.31913)(5.00000, 6.09601)(5.05000, 5.87958)(5.10000, 5.69448)(5.15000, 5.50922)(5.20000, 5.32317)(5.25000, 5.14037)(5.30000, 4.97278)(5.35000, 4.81277)(5.40000, 9.44191)(5.45000, 8.93764)(5.50000, 8.43593)(5.55000, 8.03585)(5.60000, 7.63790)(5.65000, 7.32456)(5.70000, 7.01305)(5.75000, 6.77430)(5.80000, 6.53626)(5.85000, 6.32644)(5.90000, 6.11667)(5.95000, 5.90898)(6.00000, 5.70178)(6.05000, 5.51417)(6.10000, 5.32780)(6.15000, 5.19090)(6.20000, 5.05539)(6.25000, 4.97558)(6.30000, 4.89641)(6.35000, 4.84228)(6.40000, 9.92250)(6.45000, 9.35856)(6.50000, 8.89268)(6.55000, 8.44254)(6.60000, 8.07952)(6.65000, 7.72801)(6.70000, 7.44025)(6.75000, 10.16487)(6.80000, 9.61698)(6.85000, 9.11696)(6.90000, 8.68779)(6.95000, 8.29861)(7.00000, 7.96865)(7.05000, 7.65827)(7.10000, 7.37688)(7.15000, 7.10725)(7.20000, 6.85503)(7.25000, 6.61686)(7.30000, 6.39951)(7.35000, 6.20820)(7.40000, 6.05547)(7.45000, 11.26424)(7.50000, 10.62076)(7.55000, 10.07215)(7.60000, 9.56726)(7.65000, 9.14125)(7.70000, 8.74753)(7.75000, 8.41206)(7.80000, 8.08913)(7.85000, 7.78883)(7.90000, 7.50071)(7.95000, 7.23460)(8.00000, 6.98989)(8.05000, 6.78381)(8.10000, 6.60049)(8.15000, 6.45821)(8.20000, 6.32836)(8.25000, 6.22090)(8.30000, 6.11141)(8.35000, 5.99824)(8.40000, 5.87442)(8.45000, 5.73141)(8.50000, 5.57906)(8.55000, 12.51137)(8.60000, 11.82093)(8.65000, 11.16876)(8.70000, 10.62690)(8.75000, 10.12081)(8.80000, 9.69571)(8.85000, 9.29687)(8.90000, 8.94139)(8.95000, 8.59998)(9.00000, 8.28896)(9.05000, 7.98781)(9.10000, 7.73544)(9.15000, 7.49889)(9.20000, 7.31430)(9.25000, 13.11085)(9.30000, 12.38611)(9.35000, 11.67993)(9.40000, 11.12230)(9.45000, 10.58168)(9.50000, 10.13980)(9.55000, 9.70923)(9.60000, 9.33088)(9.65000, 8.95851)(9.70000, 8.64091)(9.75000, 8.32958)(9.80000, 8.08856)(9.85000, 7.85559)(9.90000, 7.67710)(9.95000, 7.50485)(10.00000, 7.34827)(10.05000, 7.19349)(10.10000, 7.02461)(10.15000, 6.85411)(10.20000, 13.09598)(10.25000, 12.36093)(10.30000, 11.68866)(10.35000, 11.11159)(10.40000, 10.58472)(10.45000, 10.12622)(10.50000, 9.70377)(10.55000, 9.33527)(10.60000, 8.99521)(10.65000, 8.70577)(10.70000, 8.44302)(10.75000, 8.20722)(10.80000, 7.98564)(10.85000, 7.76712)(10.90000, 7.55021)(10.95000, 7.32760)(11.00000, 7.10199)(11.05000, 6.87887)(11.10000, 13.92752)(11.15000, 13.04847)(11.20000, 12.36925)(11.25000, 11.69621)(11.30000, 11.15620)(11.35000, 10.62031)(11.40000, 10.19796)(11.45000, 9.77913)(11.50000, 9.44831)(11.55000, 9.12021)(11.60000, 8.84909)(11.65000, 8.57973)(11.70000, 8.32291)(11.75000, 8.06647)(11.80000, 7.80821)(11.85000, 7.54989)(11.90000, 7.30105)(11.95000, 7.05250)(12.00000, 6.84435)(12.05000, 6.63744)(12.10000, 6.47176)(12.15000, 6.30736)(12.20000, 6.19728)(12.25000, 6.08887)(12.30000, 6.01685)(12.35000, 5.94595)(12.40000, 5.89982)(12.45000, 5.85445)(12.50000, 5.82141)(12.55000, 17.42584)(12.60000, 16.57865)(12.65000, 15.86993)(12.70000, 15.19192)(12.75000, 14.62222)(12.80000, 14.07653)(12.85000, 13.61479)(12.90000, 13.17168)(12.95000, 12.77743)(13.00000, 12.39403)(13.05000, 12.01977)(13.10000, 11.64754)(13.15000, 11.30196)(13.20000, 10.96229)(13.25000, 10.67861)(13.30000, 10.40735)(13.35000, 10.20004)(13.40000, 17.85813)(13.45000, 17.06802)(13.50000, 16.28020)(13.55000, 15.65425)(13.60000, 15.03045)(13.65000, 14.52227)(13.70000, 14.01563)(13.75000, 13.57506)(13.80000, 13.13537)(13.85000, 12.73112)(13.90000, 18.09498)(13.95000, 17.24979)(14.00000, 16.52149)(14.05000, 15.85154)(14.10000, 15.26720)(14.15000, 14.72561)(14.20000, 14.22713)(14.25000, 13.75018)(14.30000, 13.30918)(14.35000, 12.88613)(14.40000, 12.50150)(14.45000, 12.13605)(14.50000, 11.82247)(14.55000, 11.53478)(14.60000, 11.29543)(14.65000, 11.08022)(14.70000, 10.89386)(14.75000, 10.72190)(14.80000, 19.96282)(14.85000, 19.01532)(14.90000, 18.13568)(14.95000, 17.36430)(15.00000, 16.64803)(15.05000, 16.01970)(15.10000, 15.42500)(15.15000, 14.88398)(15.20000, 14.36974)(15.25000, 20.20828)(15.30000, 19.23325)(15.35000, 18.33562)(15.40000, 17.55075)(15.45000, 16.82449)(15.50000, 16.18363)(15.55000, 15.57753)(15.60000, 15.02209)(15.65000, 14.49915)(15.70000, 14.02355)(15.75000, 13.58825)(15.80000, 13.21168)(15.85000, 12.86761)(15.90000, 12.57088)(15.95000, 12.29529)(16.00000, 12.05053)(16.05000, 11.80648)(16.10000, 11.56346)(16.15000, 11.30868)(16.20000, 11.03676)(16.25000, 10.75871)(16.30000, 10.47172)(16.35000, 10.18305)(16.40000, 9.89195)(16.45000, 9.61315)(16.50000, 9.35228)(16.55000, 9.11305)(16.60000, 8.90533)(16.65000, 8.72022)(16.70000, 8.56806)(16.75000, 8.43578)(16.80000, 8.33248)(16.85000, 8.24165)(16.90000, 8.16900)(16.95000, 8.10806)(17.00000, 8.06421)(17.05000, 8.02641)(17.10000, 7.99744)(17.15000, 7.97172)(17.20000, 7.95074)(17.25000, 7.92809)(17.30000, 7.90302)(17.35000, 7.87458)(17.40000, 7.84122)(17.45000, 7.80002)(17.50000, 21.07206)(17.55000, 19.99114)(17.60000, 19.02524)(17.65000, 18.15380)(17.70000, 17.37768)(17.75000, 16.67983)(17.80000, 16.05593)(17.85000, 15.49277)(17.90000, 14.98992)(17.95000, 14.53661)(18.00000, 14.11249)(18.05000, 13.71233)(18.10000, 13.31867)(18.15000, 12.93034)(18.20000, 12.54843)(18.25000, 12.17179)(18.30000, 11.80305)(18.35000, 11.44080)(18.40000, 11.10724)(18.45000, 10.79724)(18.50000, 10.52289)(18.55000, 10.27781)(18.60000, 10.06984)(18.65000, 9.89234)(18.70000, 9.74468)(18.75000, 9.62151)(18.80000, 9.52006)(18.85000, 9.43645)(18.90000, 9.36511)(18.95000, 9.30386)(19.00000, 9.24671)(19.05000, 9.19293)(19.10000, 9.13403)(19.15000, 9.07091)(19.20000, 8.99520)(19.25000, 8.90914)(19.30000, 8.80757)(19.35000, 8.69326)(19.40000, 8.56207)(19.45000, 8.41699)(19.50000, 8.25744)(19.55000, 8.08600)(19.60000, 7.90435)(19.65000, 7.71432)(19.70000, 7.51903)(19.75000, 7.31944)(19.80000, 7.12308)(19.85000, 6.92939)(19.90000, 6.74035)(19.95000, 6.55515)(20.00000, 6.37976)
\psline[linecolor=black, plotstyle=curve, linewidth=0.4mm, showpoints=false, linestyle=dashed, linecolor=black, dotstyle=o, dotscale=1.2 1.2, linewidth=0.4mm](-9.58361, 0.00000)(-9.53361, 0.00000)(-9.48361, 0.00000)(-9.43361, 0.00000)(-9.38361, 0.00000)(-9.33361, 0.00000)(-9.28361, 0.00000)(-9.23361, 0.00000)(-9.18361, 0.00000)(-9.13361, 0.00000)(-9.08361, 0.00000)(-9.03361, 0.00000)(-8.98361, 0.00000)(-8.93361, 0.00000)(-8.88361, 0.00000)(-8.83361, 0.00000)(-8.78361, 0.00000)(-8.73361, 0.00000)(-8.68361, 0.00000)(-8.63361, 0.00000)(-8.58361, 0.00000)(-8.53361, 0.00000)(-8.48361, 0.00000)(-8.43361, 0.00000)(-8.38361, 0.00000)(-8.33361, 0.00000)(-8.28361, 0.00000)(-8.23361, 0.00000)(-8.18361, 0.00000)(-8.13361, 0.00000)(-8.08361, 0.00000)(-8.03361, 0.00000)(-7.98361, 0.00000)(-7.93361, 0.00000)(-7.88361, 0.00000)(-7.83361, 0.00000)(-7.78361, 0.00000)(-7.73361, 0.00000)(-7.68361, 0.00000)(-7.63361, 0.00000)(-7.58361, 0.00000)(-7.53361, 0.00000)(-7.48361, 0.00000)(-7.43361, 0.00000)(-7.38361, 0.00000)(-7.33361, 0.00000)(-7.28361, 0.00000)(-7.23361, 0.00000)(-7.18361, 0.00000)(-7.13361, 0.00000)(-7.08361, 0.00000)(-7.03361, 0.00000)(-6.98361, 0.00000)(-6.93361, 0.00000)(-6.88361, 0.00000)(-6.83361, 0.00000)(-6.78361, 0.00000)(-6.73361, 0.00000)(-6.68361, 0.00000)(-6.63361, 0.00000)(-6.58361, 1.72261)(-6.53361, 1.64482)(-6.48361, 1.58096)(-6.43361, 1.52672)(-6.38361, 1.47951)(-6.33361, 1.43768)(-6.28361, 1.40010)(-6.23361, 1.36594)(-6.18361, 1.33461)(-6.13361, 1.30566)(-6.08361, 1.27874)(-6.03361, 1.25356)(-5.98361, 1.22991)(-5.93361, 1.20759)(-5.88361, 1.18645)(-5.83361, 1.16637)(-5.78361, 1.14723)(-5.73361, 1.12895)(-5.68361, 1.11145)(-5.63361, 1.09465)(-5.58361, 1.07850)(-5.53361, 2.01985)(-5.48361, 1.92996)(-5.43361, 1.85525)(-5.38361, 1.79125)(-5.33361, 1.73521)(-5.28361, 1.68531)(-5.23361, 1.64031)(-5.18361, 1.59929)(-5.13361, 1.56158)(-5.08361, 1.52667)(-5.03361, 1.49415)(-4.98361, 1.46369)(-4.93361, 1.43504)(-4.88361, 1.40798)(-4.83361, 1.38233)(-4.78361, 1.35795)(-4.73361, 1.33470)(-4.68361, 1.31247)(-4.63361, 1.29118)(-4.58361, 2.33632)(-4.53361, 2.23180)(-4.48361, 2.14436)(-4.43361, 2.06912)(-4.38361, 2.00301)(-4.33361, 1.94400)(-4.28361, 1.89067)(-4.23361, 1.84199)(-4.18361, 1.79719)(-4.13361, 1.75566)(-4.08361, 1.71694)(-4.03361, 1.68066)(-3.98361, 1.64650)(-3.93361, 1.61423)(-3.88361, 1.58363)(-3.83361, 1.55453)(-3.78361, 1.52677)(-3.73361, 1.50024)(-3.68361, 1.47481)(-3.63361, 1.45039)(-3.58361, 1.42690)(-3.53361, 1.40427)(-3.48361, 1.38243)(-3.43361, 2.77406)(-3.38361, 2.64854)(-3.33361, 2.54286)(-3.28361, 2.45150)(-3.23361, 2.37097)(-3.18361, 2.29891)(-3.13361, 2.23366)(-3.08361, 2.17401)(-3.03361, 2.11903)(-2.98361, 2.06803)(-2.93361, 2.02044)(-2.88361, 1.97581)(-2.83361, 1.93379)(-2.78361, 1.89406)(-2.73361, 1.85637)(-2.68361, 1.82052)(-2.63361, 1.78632)(-2.58361, 1.75362)(-2.53361, 1.72228)(-2.48361, 1.69219)(-2.43361, 3.22236)(-2.38361, 3.07293)(-2.33361, 2.94688)(-2.28361, 2.83777)(-2.23361, 2.74149)(-2.18361, 2.65529)(-2.13361, 2.57719)(-2.08361, 2.50577)(-2.03361, 2.43993)(-1.98361, 2.37885)(-1.93361, 2.32184)(-1.88361, 2.26838)(-1.83361, 2.21804)(-1.78361, 2.17045)(-1.73361, 2.12531)(-1.68361, 2.08237)(-1.63361, 3.59928)(-1.58361, 3.43125)(-1.53361, 3.28891)(-1.48361, 3.16533)(-1.43361, 3.05606)(-1.38361, 2.95806)(-1.33361, 2.86916)(-1.28361, 2.78777)(-1.23361, 2.71270)(-1.18361, 2.64299)(-1.13361, 2.57790)(-1.08361, 2.51684)(-1.03361, 2.45932)(-0.98361, 2.40493)(-0.93361, 2.35333)(-0.88361, 2.30424)(-0.83361, 4.29001)(-0.78361, 4.05470)(-0.73361, 3.86176)(-0.68361, 3.69809)(-0.63361, 3.55585)(-0.58361, 3.42998)(-0.53361, 3.31704)(-0.48361, 3.21456)(-0.43361, 3.12072)(-0.38361, 3.03414)(-0.33361, 2.95373)(-0.28361, 2.87866)(-0.23361, 2.80824)(-0.18361, 2.74189)(-0.13361, 2.67916)(-0.08361, 2.61966)(-0.03361, 2.56305)(0.01639, 2.50906)(0.06639, 4.67506)(0.11639, 4.43494)(0.16639, 4.23365)(0.21639, 4.06023)(0.26639, 3.90779)(0.31639, 3.77173)(0.36639, 3.64879)(0.41639, 3.53662)(0.46639, 3.43343)(0.51639, 3.33787)(0.56639, 3.24885)(0.61639, 3.16550)(0.66639, 3.08712)(0.71639, 3.01314)(0.76639, 2.94307)(0.81639, 2.87650)(0.86639, 5.30537)(0.91639, 5.01570)(0.96639, 4.77522)(1.01639, 4.56949)(1.06639, 4.38960)(1.11639, 4.22969)(1.16639, 4.08570)(1.21639, 3.95469)(1.26639, 3.83447)(1.31639, 3.72335)(1.36639, 3.62002)(1.41639, 3.52343)(1.46639, 3.43273)(1.51639, 3.34722)(1.56639, 3.26633)(1.61639, 5.82848)(1.66639, 5.50711)(1.71639, 5.23981)(1.76639, 5.01083)(1.81639, 4.81043)(1.86639, 4.63218)(1.91639, 4.47160)(1.96639, 4.32545)(2.01639, 4.19129)(2.06639, 4.06727)(2.11639, 3.95194)(2.16639, 3.84412)(2.21639, 3.74287)(2.26639, 3.64742)(2.31639, 3.55712)(2.36639, 3.47143)(2.41639, 3.38988)(2.46639, 3.31208)(2.51639, 3.23769)(2.56639, 3.16641)(2.61639, 6.75232)(2.66639, 6.35398)(2.71639, 6.02654)(2.76639, 5.74835)(2.81639, 5.50640)(2.86639, 5.29224)(2.91639, 5.10006)(2.96639, 4.92570)(3.01639, 4.76610)(3.06639, 4.61891)(3.11639, 4.48230)(3.16639, 4.35482)(3.21639, 4.23531)(3.26639, 4.12280)(3.31639, 4.01651)(3.36639, 3.91575)(3.41639, 3.81998)(3.46639, 3.72871)(3.51639, 3.64151)(3.56639, 3.55804)(3.61639, 3.47798)(3.66639, 3.40105)(3.71639, 3.32701)(3.76639, 3.25564)(3.81639, 7.56334)(3.86639, 7.13343)(3.91639, 6.77470)(3.96639, 6.46676)(4.01639, 6.19690)(4.06639, 5.95663)(4.11639, 5.74005)(4.16639, 5.54284)(4.21639, 5.36178)(4.26639, 5.19439)(4.31639, 5.03871)(4.36639, 4.89319)(4.41639, 4.75657)(4.46639, 4.62778)(4.51639, 4.50598)(4.56639, 4.39042)(4.61639, 4.28048)(4.66639, 8.10065)(4.71639, 7.65745)(4.76639, 7.28302)(4.81639, 6.95873)(4.86639, 6.67264)(4.91639, 6.41662)(4.96639, 6.18489)(5.01639, 5.97319)(5.06639, 5.77828)(5.11639, 5.59768)(5.16639, 5.42939)(5.21639, 5.27182)(5.26639, 5.12366)(5.31639, 4.98384)(5.36639, 4.85145)(5.41639, 8.81874)(5.46639, 8.32692)(5.51639, 7.91224)(5.56639, 7.55363)(5.61639, 7.23763)(5.66639, 6.95512)(5.71639, 6.69961)(5.76639, 6.46635)(5.81639, 6.25173)(5.86639, 6.05296)(5.91639, 5.86783)(5.96639, 5.69457)(6.01639, 5.53173)(6.06639, 5.37811)(6.11639, 5.23270)(6.16639, 5.09466)(6.21639, 4.96327)(6.26639, 4.83790)(6.31639, 4.71801)(6.36639, 4.60315)(6.41639, 9.88487)(6.46639, 9.31439)(6.51639, 8.83591)(6.56639, 8.42374)(6.61639, 8.06162)(6.66639, 7.73864)(6.71639, 10.64954)(6.76639, 9.97723)(6.81639, 9.42568)(6.86639, 8.95791)(6.91639, 8.55169)(6.96639, 8.19261)(7.01639, 7.87081)(7.06639, 7.57922)(7.11639, 7.31260)(7.16639, 7.06699)(7.21639, 6.83929)(7.26639, 6.62704)(7.31639, 6.42826)(7.36639, 6.24131)(7.41639, 6.06487)(7.46639, 11.00676)(7.51639, 10.35783)(7.56639, 9.81500)(7.61639, 9.34828)(7.66639, 8.93885)(7.71639, 8.57412)(7.76639, 8.24522)(7.81639, 7.94570)(7.86639, 7.67071)(7.91639, 7.41651)(7.96639, 7.18014)(8.01639, 6.95925)(8.06639, 6.75192)(8.11639, 6.55656)(8.16639, 6.37186)(8.21639, 6.19670)(8.26639, 6.03013)(8.31639, 5.87135)(8.36639, 5.71964)(8.41639, 5.57439)(8.46639, 5.43508)(8.51639, 5.30123)(8.56639, 12.16412)(8.61639, 11.44116)(8.66639, 10.83623)(8.71639, 10.31606)(8.76639, 9.85971)(8.81639, 9.45316)(8.86639, 9.08657)(8.91639, 8.75274)(8.96639, 8.44626)(9.01639, 8.16297)(9.06639, 7.89958)(9.11639, 7.65346)(9.16639, 7.42246)(9.21639, 7.20483)(9.26639, 12.81867)(9.31639, 12.06489)(9.36639, 11.43176)(9.41639, 10.88584)(9.46639, 10.40594)(9.51639, 9.97774)(9.56639, 9.59114)(9.61639, 9.23873)(9.66639, 8.91493)(9.71639, 8.61542)(9.76639, 8.33679)(9.81639, 8.07630)(9.86639, 7.83172)(9.91639, 7.60120)(9.96639, 7.38320)(10.01639, 7.17643)(10.06639, 6.97978)(10.11639, 6.79229)(10.16639, 6.61315)(10.21639, 14.02082)(10.26639, 13.17517)(10.31639, 12.46833)(10.36639, 11.86102)(10.41639, 11.32857)(10.46639, 10.85450)(10.51639, 10.42721)(10.56639, 10.03828)(10.61639, 9.68135)(10.66639, 9.35154)(10.71639, 9.04499)(10.76639, 8.75864)(10.81639, 8.48995)(10.86639, 8.23688)(10.91639, 7.99770)(10.96639, 7.77095)(11.01639, 7.55540)(11.06639, 7.34998)(11.11639, 15.11257)(11.16639, 14.19036)(11.21639, 13.42094)(11.26639, 12.76075)(11.31639, 12.18253)(11.36639, 11.66812)(11.41639, 11.20479)(11.46639, 10.78329)(11.51639, 10.39665)(11.56639, 10.03953)(11.61639, 9.70773)(11.66639, 9.39788)(11.71639, 9.10724)(11.76639, 8.83356)(11.81639, 8.57497)(11.86639, 8.32987)(11.91639, 8.09692)(11.96639, 7.87497)(12.01639, 7.66302)(12.06639, 7.46020)(12.11639, 7.26576)(12.16639, 7.07903)(12.21639, 6.89941)(12.26639, 6.72639)(12.31639, 6.55948)(12.36639, 6.39828)(12.41639, 6.24239)(12.46639, 6.09149)(12.51639, 17.06308)(12.56639, 15.98142)(12.61639, 15.08657)(12.66639, 14.32334)(12.71639, 13.65788)(12.76639, 13.06794)(12.81639, 12.53806)(12.86639, 12.05713)(12.91639, 11.61684)(12.96639, 11.21084)(13.01639, 10.83416)(13.06639, 10.48284)(13.11639, 10.15366)(13.16639, 9.84400)(13.21639, 9.55165)(13.26639, 9.27479)(13.31639, 9.01184)(13.36639, 8.76147)(13.41639, 17.53517)(13.46639, 16.49057)(13.51639, 15.61206)(13.56639, 14.85395)(13.61639, 14.18714)(13.66639, 13.59196)(13.71639, 13.05448)(13.76639, 12.56447)(13.81639, 12.11422)(13.86639, 18.77414)(13.91639, 17.57351)(13.96639, 16.58154)(14.01639, 15.73628)(14.06639, 14.99984)(14.11639, 14.34733)(14.16639, 13.76154)(14.21639, 13.23007)(14.26639, 12.74367)(14.31639, 12.29529)(14.36639, 11.87939)(14.41639, 11.49159)(14.46639, 11.12830)(14.51639, 10.78662)(14.56639, 10.46410)(14.61639, 10.15871)(14.66639, 9.86872)(14.71639, 9.59264)(14.76639, 9.32919)(14.81639, 19.32507)(14.86639, 18.15860)(14.91639, 17.17997)(14.96639, 16.33696)(15.01639, 15.59648)(15.06639, 14.93625)(15.11639, 14.34055)(15.16639, 13.79786)(15.21639, 13.29951)(15.26639, 19.63257)(15.31639, 18.47235)(15.36639, 17.49371)(15.41639, 16.64739)(15.46639, 15.90181)(15.51639, 15.23549)(15.56639, 14.63318)(15.61639, 14.08364)(15.66639, 13.57834)(15.71639, 13.11069)(15.76639, 12.67544)(15.81639, 12.26840)(15.86639, 11.88613)(15.91639, 11.52577)(15.96639, 11.18495)(16.01639, 10.86165)(16.06639, 10.55416)(16.11639, 10.26099)(16.16639, 9.98087)(16.21639, 9.71269)(16.26639, 9.45547)(16.31639, 9.20834)(16.36639, 8.97054)(16.41639, 8.74139)(16.46639, 8.52028)(16.51639, 8.30667)(16.56639, 8.10006)(16.61639, 7.90001)(16.66639, 7.70611)(16.71639, 7.51800)(16.76639, 7.33533)(16.81639, 7.15780)(16.86639, 6.98513)(16.91639, 6.81707)(16.96639, 6.65336)(17.01639, 6.49379)(17.06639, 6.33815)(17.11639, 6.18626)(17.16639, 6.03794)(17.21639, 5.89302)(17.26639, 5.75136)(17.31639, 5.61280)(17.36639, 5.47721)(17.41639, 5.34448)(17.46639, 5.21447)(17.51639, 22.33007)(17.56639, 21.01219)(17.61639, 19.89934)(17.66639, 18.93620)(17.71639, 18.08721)(17.76639, 17.32814)(17.81639, 16.64174)(17.86639, 16.01530)(17.91639, 15.43917)(17.96639, 14.90586)(18.01639, 14.40944)(18.06639, 13.94513)(18.11639, 13.50902)(18.16639, 13.09788)(18.21639, 12.70900)(18.26639, 12.34009)(18.31639, 11.98920)(18.36639, 11.65465)(18.41639, 11.33497)(18.46639, 11.02891)(18.51639, 10.73535)(18.56639, 10.45330)(18.61639, 10.18190)(18.66639, 9.92037)(18.71639, 9.66802)(18.76639, 9.42422)(18.81639, 9.18841)(18.86639, 8.96008)(18.91639, 8.73878)(18.96639, 8.52408)(19.01639, 8.31560)(19.06639, 8.11299)(19.11639, 7.91593)(19.16639, 7.72411)(19.21639, 7.53728)(19.26639, 7.35517)(19.31639, 7.17755)(19.36639, 7.00421)(19.41639, 6.83494)(19.46639, 6.66956)(19.51639, 6.50790)(19.56639, 6.34978)(19.61639, 6.19506)(19.66639, 6.04359)(19.71639, 5.89524)(19.76639, 5.74988)(19.81639, 5.60740)(19.86639, 5.46768)(19.91639, 5.33062)(19.96639, 5.19612)(20.01639, 5.06408)(20.06639, 4.93441)(20.11639, 4.80704)(20.16639, 4.68188)(20.21639, 4.55886)(20.26639, 4.43790)(20.31639, 4.31894)(20.36639, 4.20191)(20.41639, 4.08675)(20.46639, 3.97340)
\endpsclip
\psframe[linecolor=black, fillstyle=solid, fillcolor=white, shadowcolor=lightgray, shadowsize=1mm, shadow=true](-7.16981, 17.32955)(6.41509, 22.72727)
\rput[l](-4.62264, 21.02273){\footnotesize{Empirical, max. 8 iter.}}
\psline[linecolor=blue, linestyle=solid, linewidth=0.3mm](-6.60377, 21.02273)(-5.47170, 21.02273)
\rput[l](-4.62264, 19.03409){\footnotesize{Analytical}}
\psline[linecolor=black, linestyle=dashed, linewidth=0.3mm](-6.60377, 19.03409)(-5.47170, 19.03409)
}\end{pspicture}
\endgroup
 

%% file: Complexity_scaling_NR.tex
\begingroup
\unitlength=1mm
\psset{xunit=1.37500mm, yunit=2.83333mm, linewidth=0.1mm}
\psset{arrowsize=2pt 3, arrowlength=1.4, arrowinset=.4}\psset{axesstyle=frame}
\begin{pspicture}(-0.90909, -3.64706)(50.00000, 14.00000)
\rput(-1.45455, -1.76471){%
\psaxes[subticks=0, labels=all, xsubticks=1, ysubticks=1, Ox=10, Oy=2, Dx=5, Dy=2]{-}(10.00000, 2.00000)(10.00000, 2.00000)(50.00000, 14.00000)%
\multips(15.00000, 2.00000)(5.00000, 0.0){7}{\psline[linecolor=black, linestyle=dotted, linewidth=0.2mm](0, 0)(0, 12.00000)}
\multips(10.00000, 4.00000)(0, 2.00000){5}{\psline[linecolor=black, linestyle=dotted, linewidth=0.2mm](0, 0)(40.00000, 0)}
\rput[b](30.00000, -1.88235){\small{Number of MCS schemes $N_R$}}
\rput[t]{90}(0.54545, 8.00000){\small{$E\{\mathcal{C}(\gamma,\Delta\gamma)\}$ [bit-iter pcu]}}
\psclip{\psframe(10.00000, 2.00000)(50.00000, 14.00000)}
\psline[linecolor=blue, plotstyle=curve, linewidth=0.4mm, showpoints=false, linestyle=solid, linecolor=blue, dotstyle=o, dotscale=1.2 1.2, linewidth=0.4mm](10.00000, 5.49278)(15.00000, 6.62074)(20.00000, 7.38212)(25.00000, 7.93857)(30.00000, 8.36731)(35.00000, 8.71011)(40.00000, 8.99178)(45.00000, 9.22813)(50.00000, 9.42981)
\psline[linecolor=blue, plotstyle=curve, linewidth=0.4mm, showpoints=true, linestyle=none, linecolor=blue, dotstyle=o, dotscale=1.2 1.2, linewidth=0.4mm](10.00000, 5.19547)(15.00000, 6.44653)(20.00000, 7.26759)(25.00000, 7.85165)(30.00000, 8.29981)(35.00000, 8.66387)(40.00000, 8.94599)(45.00000, 9.19443)(50.00000, 9.39876)
\psline[linecolor=red, plotstyle=curve, linewidth=0.4mm, showpoints=false, linestyle=solid, linecolor=red, dotstyle=triangle, dotscale=1.2 1.2, linewidth=0.4mm](10.00000, 4.22301)(15.00000, 5.03719)(20.00000, 5.55603)(25.00000, 5.91721)(30.00000, 6.18415)(35.00000, 6.39000)(40.00000, 6.55387)(45.00000, 6.68757)(50.00000, 6.79883)
\psline[linecolor=red, plotstyle=curve, linewidth=0.4mm, showpoints=true, linestyle=none, linecolor=red, dotstyle=triangle, dotscale=1.2 1.2, linewidth=0.4mm](10.00000, 4.04961)(15.00000, 4.95165)(20.00000, 5.49832)(25.00000, 5.87838)(30.00000, 6.15626)(35.00000, 6.36834)(40.00000, 6.54581)(45.00000, 6.67933)(50.00000, 6.79438)
\psline[linecolor=black, plotstyle=curve, linewidth=0.4mm, showpoints=false, linestyle=solid, linecolor=black, dotstyle=diamond, dotscale=1.2 1.2, linewidth=0.4mm](10.00000, 3.18889)(15.00000, 3.78825)(20.00000, 4.15660)(25.00000, 4.40568)(30.00000, 4.58551)(35.00000, 4.72156)(40.00000, 4.82813)(45.00000, 4.91390)(50.00000, 4.98444)
\psline[linecolor=black, plotstyle=curve, linewidth=0.4mm, showpoints=true, linestyle=none, linecolor=black, dotstyle=diamond, dotscale=1.2 1.2, linewidth=0.4mm](10.00000, 3.10032)(15.00000, 3.75277)(20.00000, 4.14852)(25.00000, 4.40521)(30.00000, 4.59143)(35.00000, 4.73344)(40.00000, 4.83741)(45.00000, 4.92764)(50.00000, 4.99910)
\endpsclip
\psframe[linecolor=black, fillstyle=solid, fillcolor=white, shadowcolor=lightgray, shadowsize=1mm, shadow=true](13.63636, 9.90588)(34.00000, 14.35294)
\rput[l](20.18182, 13.29412){\scriptsize{$\Delta\gamma=\unit[0]{dB}$}}
\psline[linecolor=blue, linestyle=solid, linewidth=0.3mm](15.09091, 13.29412)(18.00000, 13.29412)
\psline[linecolor=blue, linestyle=solid, linewidth=0.3mm](15.09091, 13.29412)(18.00000, 13.29412)
\psdots[linecolor=blue, linestyle=solid, linewidth=0.3mm, dotstyle=o, dotscale=1.2 1.2, linecolor=blue](16.54545, 13.29412)
\rput[l](20.18182, 12.12941){\scriptsize{$\Delta\gamma=\unit[0.4]{dB}$}}
\psline[linecolor=red, linestyle=solid, linewidth=0.3mm](15.09091, 12.12941)(18.00000, 12.12941)
\psline[linecolor=red, linestyle=solid, linewidth=0.3mm](15.09091, 12.12941)(18.00000, 12.12941)
\psdots[linecolor=red, linestyle=solid, linewidth=0.3mm, dotstyle=triangle, dotscale=1.2 1.2, linecolor=red](16.54545, 12.12941)
\rput[l](20.18182, 10.96471){\scriptsize{$\Delta\gamma=\unit[0.9]{dB}$}}
\psline[linecolor=black, linestyle=solid, linewidth=0.3mm](15.09091, 10.96471)(18.00000, 10.96471)
\psline[linecolor=black, linestyle=solid, linewidth=0.3mm](15.09091, 10.96471)(18.00000, 10.96471)
\psdots[linecolor=black, linestyle=solid, linewidth=0.3mm, dotstyle=diamond, dotscale=1.2 1.2, linecolor=black](16.54545, 10.96471)
}\end{pspicture}
\endgroup
 

%% file: Complexity_scaling_Var_NR.tex
\begingroup
\unitlength=1mm
\psset{xunit=1.37500mm, yunit=2.26667mm, linewidth=0.1mm}
\psset{arrowsize=2pt 3, arrowlength=1.4, arrowinset=.4}\psset{axesstyle=frame}
\begin{pspicture}(-0.90909, -4.05882)(50.00000, 18.00000)
\rput(-1.45455, -2.20588){%
\psaxes[subticks=0, labels=all, xsubticks=1, ysubticks=1, Ox=10, Oy=3, Dx=5, Dy=3]{-}(10.00000, 3.00000)(10.00000, 3.00000)(50.00000, 18.00000)%
\multips(15.00000, 3.00000)(5.00000, 0.0){7}{\psline[linecolor=black, linestyle=dotted, linewidth=0.2mm](0, 0)(0, 15.00000)}
\multips(10.00000, 6.00000)(0, 3.00000){4}{\psline[linecolor=black, linestyle=dotted, linewidth=0.2mm](0, 0)(40.00000, 0)}
\rput[b](30.00000, -1.85294){\small{Number of MCS schemes $N_R$}}
\rput[t]{90}(0.54545, 10.50000){\small{$10\log_{10} Var\{\mathcal{C}(\gamma,\Delta\gamma)\}$}}
\psclip{\psframe(10.00000, 3.00000)(50.00000, 18.00000)}
\psline[linecolor=blue, plotstyle=curve, linewidth=0.4mm, showpoints=false, linestyle=solid, linecolor=blue, dotstyle=o, dotscale=1.2 1.2, linewidth=0.4mm](10.00000, 11.19109)(15.00000, 11.40184)(20.00000, 11.62959)(25.00000, 11.83553)(30.00000, 12.01638)(35.00000, 12.17509)(40.00000, 12.31187)(45.00000, 12.43352)(50.00000, 12.54058)
\psline[linecolor=blue, plotstyle=curve, linewidth=0.4mm, showpoints=true, linestyle=none, linecolor=blue, dotstyle=o, dotscale=1.2 1.2, linewidth=0.4mm](10.00000, 10.66572)(15.00000, 11.12027)(20.00000, 11.45092)(25.00000, 11.71364)(30.00000, 11.91846)(35.00000, 12.09337)(40.00000, 12.24847)(45.00000, 12.37885)(50.00000, 12.49529)
\psline[linecolor=red, plotstyle=curve, linewidth=0.4mm, showpoints=false, linestyle=solid, linecolor=red, dotstyle=triangle, dotscale=1.2 1.2, linewidth=0.4mm](10.00000, 8.15925)(15.00000, 8.27828)(20.00000, 8.47412)(25.00000, 8.66526)(30.00000, 8.83398)(35.00000, 8.97944)(40.00000, 9.10428)(45.00000, 9.21188)(50.00000, 9.30480)
\psline[linecolor=red, plotstyle=curve, linewidth=0.4mm, showpoints=true, linestyle=none, linecolor=red, dotstyle=triangle, dotscale=1.2 1.2, linewidth=0.4mm](10.00000, 7.72333)(15.00000, 8.07620)(20.00000, 8.34731)(25.00000, 8.58111)(30.00000, 8.79315)(35.00000, 8.95459)(40.00000, 9.07753)(45.00000, 9.19314)(50.00000, 9.29923)
\psline[linecolor=black, plotstyle=curve, linewidth=0.4mm, showpoints=false, linestyle=solid, linecolor=black, dotstyle=diamond, dotscale=1.2 1.2, linewidth=0.4mm](10.00000, 5.32077)(15.00000, 5.38199)(20.00000, 5.56783)(25.00000, 5.75710)(30.00000, 5.92375)(35.00000, 6.06560)(40.00000, 6.18554)(45.00000, 6.28746)(50.00000, 6.37475)
\psline[linecolor=black, plotstyle=curve, linewidth=0.4mm, showpoints=true, linestyle=none, linecolor=black, dotstyle=diamond, dotscale=1.2 1.2, linewidth=0.4mm](10.00000, 4.95910)(15.00000, 5.23649)(20.00000, 5.51143)(25.00000, 5.73943)(30.00000, 5.93688)(35.00000, 6.08302)(40.00000, 6.21362)(45.00000, 6.32831)(50.00000, 6.39653)
\endpsclip
\psframe[linecolor=black, fillstyle=solid, fillcolor=white, shadowcolor=lightgray, shadowsize=1mm, shadow=true](13.63636, 12.88235)(34.00000, 18.44118)
\rput[l](20.18182, 17.11765){\scriptsize{$\Delta\gamma=\unit[0]{dB}$}}
\psline[linecolor=blue, linestyle=solid, linewidth=0.3mm](15.09091, 17.11765)(18.00000, 17.11765)
\psline[linecolor=blue, linestyle=solid, linewidth=0.3mm](15.09091, 17.11765)(18.00000, 17.11765)
\psdots[linecolor=blue, linestyle=solid, linewidth=0.3mm, dotstyle=o, dotscale=1.2 1.2, linecolor=blue](16.54545, 17.11765)
\rput[l](20.18182, 15.66176){\scriptsize{$\Delta\gamma=\unit[0.4]{dB}$}}
\psline[linecolor=red, linestyle=solid, linewidth=0.3mm](15.09091, 15.66176)(18.00000, 15.66176)
\psline[linecolor=red, linestyle=solid, linewidth=0.3mm](15.09091, 15.66176)(18.00000, 15.66176)
\psdots[linecolor=red, linestyle=solid, linewidth=0.3mm, dotstyle=triangle, dotscale=1.2 1.2, linecolor=red](16.54545, 15.66176)
\rput[l](20.18182, 14.20588){\scriptsize{$\Delta\gamma=\unit[0.9]{dB}$}}
\psline[linecolor=black, linestyle=solid, linewidth=0.3mm](15.09091, 14.20588)(18.00000, 14.20588)
\psline[linecolor=black, linestyle=solid, linewidth=0.3mm](15.09091, 14.20588)(18.00000, 14.20588)
\psdots[linecolor=black, linestyle=solid, linewidth=0.3mm, dotstyle=diamond, dotscale=1.2 1.2, linecolor=black](16.54545, 14.20588)
}\end{pspicture}
\endgroup
 

%% file: Complexity_outage_NR.tex
\begingroup
\unitlength=1mm
\psset{xunit=1.37500mm, yunit=2.26667mm, linewidth=0.1mm}
\psset{arrowsize=2pt 3, arrowlength=1.4, arrowinset=.4}\psset{axesstyle=frame}
\begin{pspicture}(-0.90909, -2.05882)(50.00000, 20.00000)
\rput(-1.45455, -2.20588){%
\psaxes[subticks=0, labels=all, xsubticks=1, ysubticks=1, Ox=10, Oy=5, Dx=5, Dy=5]{-}(10.00000, 5.00000)(10.00000, 5.00000)(50.00000, 20.00000)%
\multips(15.00000, 5.00000)(5.00000, 0.0){7}{\psline[linecolor=black, linestyle=dotted, linewidth=0.2mm](0, 0)(0, 15.00000)}
\multips(10.00000, 10.00000)(0, 5.00000){2}{\psline[linecolor=black, linestyle=dotted, linewidth=0.2mm](0, 0)(40.00000, 0)}
\rput[b](30.00000, 0.14706){\small{Number of MCS schemes $N_R$}}
\rput[t]{90}(0.54545, 12.50000){\small{$\mathcal{C}_\text{out}(\hat\epsilon,1)$ [bit-iter pcu]}}
\psclip{\psframe(10.00000, 5.00000)(50.00000, 20.00000)}
\psline[linecolor=blue, plotstyle=curve, linewidth=0.4mm, showpoints=false, linestyle=solid, linecolor=blue, dotstyle=o, dotscale=1.2 1.2, linewidth=0.4mm](10.00000, 10.06653)(15.00000, 11.46498)(20.00000, 12.39617)(25.00000, 13.05849)(30.00000, 13.56638)(35.00000, 13.98832)(40.00000, 14.34083)(45.00000, 14.63795)(50.00000, 14.89953)
\psline[linecolor=blue, plotstyle=curve, linewidth=0.4mm, showpoints=true, linestyle=none, linecolor=blue, dotstyle=o, dotscale=1.2 1.2, linewidth=0.4mm](10.00000, 10.05597)(15.00000, 11.45800)(20.00000, 12.40200)(25.00000, 13.05512)(30.00000, 13.55997)(35.00000, 13.99751)(40.00000, 14.33602)(45.00000, 14.63843)(50.00000, 14.89570)
\psline[linecolor=red, plotstyle=curve, linewidth=0.4mm, showpoints=false, linestyle=solid, linecolor=red, dotstyle=triangle, dotscale=1.2 1.2, linewidth=0.4mm](10.00000, 7.53693)(15.00000, 8.45193)(20.00000, 9.03944)(25.00000, 9.44958)(30.00000, 9.76324)(35.00000, 10.02079)(40.00000, 10.22881)(45.00000, 10.40182)(50.00000, 10.54982)
\psline[linecolor=red, plotstyle=curve, linewidth=0.4mm, showpoints=true, linestyle=none, linecolor=red, dotstyle=triangle, dotscale=1.2 1.2, linewidth=0.4mm](10.00000, 7.56373)(15.00000, 8.48402)(20.00000, 9.05198)(25.00000, 9.46165)(30.00000, 9.78403)(35.00000, 10.03368)(40.00000, 10.24339)(45.00000, 10.41935)(50.00000, 10.56773)
\psline[linecolor=black, plotstyle=curve, linewidth=0.4mm, showpoints=false, linestyle=solid, linecolor=black, dotstyle=diamond, dotscale=1.2 1.2, linewidth=0.4mm](10.00000, 5.62169)(15.00000, 6.23026)(20.00000, 6.63420)(25.00000, 6.91412)(30.00000, 7.13139)(35.00000, 7.30844)(40.00000, 7.45049)(45.00000, 7.56663)(50.00000, 7.66321)
\psline[linecolor=black, plotstyle=curve, linewidth=0.4mm, showpoints=true, linestyle=none, linecolor=black, dotstyle=diamond, dotscale=1.2 1.2, linewidth=0.4mm](10.00000, 5.67936)(15.00000, 6.27321)(20.00000, 6.67339)(25.00000, 6.96060)(30.00000, 7.18042)(35.00000, 7.35452)(40.00000, 7.48736)(45.00000, 7.60894)(50.00000, 7.69649)
\endpsclip
\psframe[linecolor=black, fillstyle=solid, fillcolor=white, shadowcolor=lightgray, shadowsize=1mm, shadow=true](13.63636, 14.88235)(34.00000, 20.44118)
\rput[l](20.18182, 19.11765){\scriptsize{$\Delta\gamma=\unit[0]{dB}$}}
\psline[linecolor=blue, linestyle=solid, linewidth=0.3mm](15.09091, 19.11765)(18.00000, 19.11765)
\psline[linecolor=blue, linestyle=solid, linewidth=0.3mm](15.09091, 19.11765)(18.00000, 19.11765)
\psdots[linecolor=blue, linestyle=solid, linewidth=0.3mm, dotstyle=o, dotscale=1.2 1.2, linecolor=blue](16.54545, 19.11765)
\rput[l](20.18182, 17.66176){\scriptsize{$\Delta\gamma=\unit[0.4]{dB}$}}
\psline[linecolor=red, linestyle=solid, linewidth=0.3mm](15.09091, 17.66176)(18.00000, 17.66176)
\psline[linecolor=red, linestyle=solid, linewidth=0.3mm](15.09091, 17.66176)(18.00000, 17.66176)
\psdots[linecolor=red, linestyle=solid, linewidth=0.3mm, dotstyle=triangle, dotscale=1.2 1.2, linecolor=red](16.54545, 17.66176)
\rput[l](20.18182, 16.20588){\scriptsize{$\Delta\gamma=\unit[0.9]{dB}$}}
\psline[linecolor=black, linestyle=solid, linewidth=0.3mm](15.09091, 16.20588)(18.00000, 16.20588)
\psline[linecolor=black, linestyle=solid, linewidth=0.3mm](15.09091, 16.20588)(18.00000, 16.20588)
\psdots[linecolor=black, linestyle=solid, linewidth=0.3mm, dotstyle=diamond, dotscale=1.2 1.2, linecolor=black](16.54545, 16.20588)
}\end{pspicture}
\endgroup
 

%% file: Rate_scaling_NR.tex
\begingroup
\unitlength=1mm
\psset{xunit=1.37500mm, yunit=48.56449mm, linewidth=0.1mm}
\psset{arrowsize=2pt 3, arrowlength=1.4, arrowinset=.4}\psset{axesstyle=frame}
\begin{pspicture}(-0.90909, 1.97054)(50.00000, 3.00010)
\rput(-1.45455, -0.10296){%
\psaxes[subticks=0, labels=all, xsubticks=1, ysubticks=1, Ox=10, Oy=2.3, Dx=5, Dy=0.1]{-}(10.00000, 2.30000)(10.00000, 2.30000)(50.00000, 3.00010)%
\multips(15.00000, 2.30000)(5.00000, 0.0){7}{\psline[linecolor=black, linestyle=dotted, linewidth=0.2mm](0, 0)(0, 0.70010)}
\multips(10.00000, 2.40000)(0, 0.10000){6}{\psline[linecolor=black, linestyle=dotted, linewidth=0.2mm](0, 0)(40.00000, 0)}
\rput[b](30.00000, 2.07350){\small{Number of MCS schemes $N_R$}}
\rput[t]{90}(0.54545, 2.65005){\small{$\text{E}\{R(\Delta\gamma)\}$ [bit pcu]}}
\psclip{\psframe(10.00000, 2.30000)(50.00000, 3.00010)}
\psline[linecolor=blue, plotstyle=curve, linewidth=0.4mm, showpoints=false, linestyle=solid, linecolor=blue, dotstyle=o, dotscale=1.2 1.2, linewidth=0.4mm](10.00000, 2.51148)(15.00000, 2.62933)(20.00000, 2.68625)(25.00000, 2.71975)(30.00000, 2.74183)(35.00000, 2.75747)(40.00000, 2.76913)(45.00000, 2.77815)(50.00000, 2.78535)
\psline[linecolor=blue, plotstyle=curve, linewidth=0.4mm, showpoints=true, linestyle=none, linecolor=blue, dotstyle=o, dotscale=1.2 1.2, linewidth=0.4mm](10.00000, 2.51185)(15.00000, 2.62945)(20.00000, 2.68797)(25.00000, 2.71952)(30.00000, 2.74234)(35.00000, 2.76054)(40.00000, 2.76901)(45.00000, 2.77935)(50.00000, 2.78598)
\psline[linecolor=red, plotstyle=curve, linewidth=0.4mm, showpoints=false, linestyle=solid, linecolor=red, dotstyle=triangle, dotscale=1.2 1.2, linewidth=0.4mm](10.00000, 2.41272)(15.00000, 2.52849)(20.00000, 2.58446)(25.00000, 2.61743)(30.00000, 2.63916)(35.00000, 2.65455)(40.00000, 2.66603)(45.00000, 2.67492)(50.00000, 2.68200)
\psline[linecolor=red, plotstyle=curve, linewidth=0.4mm, showpoints=true, linestyle=none, linecolor=red, dotstyle=triangle, dotscale=1.2 1.2, linewidth=0.4mm](10.00000, 2.41234)(15.00000, 2.52948)(20.00000, 2.58372)(25.00000, 2.61698)(30.00000, 2.63713)(35.00000, 2.65264)(40.00000, 2.66826)(45.00000, 2.67549)(50.00000, 2.68257)
\psline[linecolor=black, plotstyle=curve, linewidth=0.4mm, showpoints=false, linestyle=solid, linecolor=black, dotstyle=diamond, dotscale=1.2 1.2, linewidth=0.4mm](10.00000, 2.29176)(15.00000, 2.40483)(20.00000, 2.45957)(25.00000, 2.49183)(30.00000, 2.51310)(35.00000, 2.52818)(40.00000, 2.53942)(45.00000, 2.54813)(50.00000, 2.55507)
\psline[linecolor=black, plotstyle=curve, linewidth=0.4mm, showpoints=true, linestyle=none, linecolor=black, dotstyle=diamond, dotscale=1.2 1.2, linewidth=0.4mm](10.00000, 2.29265)(15.00000, 2.40405)(20.00000, 2.46094)(25.00000, 2.49145)(30.00000, 2.51293)(35.00000, 2.53015)(40.00000, 2.53844)(45.00000, 2.54890)(50.00000, 2.55536)
\endpsclip
\psframe[linecolor=black, fillstyle=solid, fillcolor=white, shadowcolor=lightgray, shadowsize=1mm, shadow=true](13.63636, 2.76124)(34.00000, 3.02069)
\rput[l](20.18182, 2.95892){\scriptsize{$\Delta\gamma=\unit[0.1]{dB}$}}
\psline[linecolor=blue, linestyle=solid, linewidth=0.3mm](15.09091, 2.95892)(18.00000, 2.95892)
\psline[linecolor=blue, linestyle=solid, linewidth=0.3mm](15.09091, 2.95892)(18.00000, 2.95892)
\psdots[linecolor=blue, linestyle=solid, linewidth=0.3mm, dotstyle=o, dotscale=1.2 1.2, linecolor=blue](16.54545, 2.95892)
\rput[l](20.18182, 2.89097){\scriptsize{$\Delta\gamma=\unit[0.4]{dB}$}}
\psline[linecolor=red, linestyle=solid, linewidth=0.3mm](15.09091, 2.89097)(18.00000, 2.89097)
\psline[linecolor=red, linestyle=solid, linewidth=0.3mm](15.09091, 2.89097)(18.00000, 2.89097)
\psdots[linecolor=red, linestyle=solid, linewidth=0.3mm, dotstyle=triangle, dotscale=1.2 1.2, linecolor=red](16.54545, 2.89097)
\rput[l](20.18182, 2.82302){\scriptsize{$\Delta\gamma=\unit[0.9]{dB}$}}
\psline[linecolor=black, linestyle=solid, linewidth=0.3mm](15.09091, 2.82302)(18.00000, 2.82302)
\psline[linecolor=black, linestyle=solid, linewidth=0.3mm](15.09091, 2.82302)(18.00000, 2.82302)
\psdots[linecolor=black, linestyle=solid, linewidth=0.3mm, dotstyle=diamond, dotscale=1.2 1.2, linecolor=black](16.54545, 2.82302)
}\end{pspicture}
\endgroup
 

%% file: Complexity_outage_network_Nusers_with_pl.tex
\begingroup
\unitlength=1mm
\psset{xunit=31.42857mm, yunit=3.41667mm, linewidth=0.1mm}
\psset{arrowsize=2pt 3, arrowlength=1.4, arrowinset=.4}\psset{axesstyle=frame}
\begin{pspicture}(-0.44545, -4.09756)(2.10000, 12.00000)
\rput(-0.06364, -1.46341){%
\psaxes[subticks=10, xlogBase=10, logLines=x, labels=all, xsubticks=10, ysubticks=1, Ox=0, Oy=0, Dx=1, Dy=3]{-}(0.00000, 0.00000)(0.00000, 0.00000)(2.10000, 12.00000)%
\multips(0.00000, 3.00000)(0, 3.00000){3}{\psline[linecolor=black, linestyle=dotted, linewidth=0.2mm](0, 0)(2.10000, 0)}
\rput[b](1.05000, -2.63415){\small{$N_\text{c}$}}
\rput[t]{90}(-0.38182, 6.00000){\small{$\mathcal{C}_\text{out}(\hat\epsilon, N_\text{c})/N_\text{c}$ [bit-iter pcu]}}
\psclip{\psframe(0.00000, 0.00000)(2.10000, 12.00000)}
\psline[linecolor=blue, plotstyle=curve, linewidth=0.4mm, showpoints=false, linestyle=solid, linecolor=blue, dotstyle=o, dotscale=1.2 1.2, linewidth=0.4mm](0.00000, 9.07758)(0.30103, 6.25610)(0.47712, 5.50161)(0.60206, 5.07722)(0.69897, 4.80037)(1.00000, 4.16775)(1.30103, 3.78025)(1.47712, 3.62946)(1.60206, 3.54702)(1.69897, 3.49437)(2.00000, 3.37826)
\psline[linecolor=blue, plotstyle=curve, linewidth=0.4mm, showpoints=true, linestyle=none, linecolor=blue, dotstyle=o, dotscale=1.2 1.2, linewidth=0.4mm](0.00000, 9.07695)(0.30103, 6.38316)(0.47712, 5.47748)(0.60206, 4.98803)(0.69897, 4.68853)(1.00000, 4.02545)(1.30103, 3.62982)(1.47712, 3.48769)(1.60206, 3.38480)(1.69897, 3.34646)(2.00000, 3.28897)
\psline[linecolor=blue, plotstyle=curve, linewidth=0.4mm, showpoints=false, linestyle=solid, linecolor=blue, dotstyle=+, dotscale=1.2 1.2, linewidth=0.4mm](2.06070, 3.28090)(2.06108, 3.28090)(2.06145, 3.28090)(2.06183, 3.28090)(2.06221, 3.28090)(2.06258, 3.28090)(2.06296, 3.28090)(2.06333, 3.28090)(2.06371, 3.28090)(2.06408, 3.28090)(2.06446, 3.28090)(2.06483, 3.28090)(2.06521, 3.28090)(2.06558, 3.28090)(2.06595, 3.28090)(2.06633, 3.28090)(2.06670, 3.28090)(2.06707, 3.28090)(2.06744, 3.28090)(2.06781, 3.28090)(2.06819, 3.28090)(2.06856, 3.28090)(2.06893, 3.28090)(2.06930, 3.28090)(2.06967, 3.28090)(2.07004, 3.28090)(2.07041, 3.28090)(2.07078, 3.28090)(2.07115, 3.28090)(2.07151, 3.28090)(2.07188, 3.28090)(2.07225, 3.28090)(2.07262, 3.28090)(2.07298, 3.28090)(2.07335, 3.28090)(2.07372, 3.28090)(2.07408, 3.28090)(2.07445, 3.28090)(2.07482, 3.28090)(2.07518, 3.28090)(2.07555, 3.28090)(2.07591, 3.28090)(2.07628, 3.28090)(2.07664, 3.28090)(2.07700, 3.28090)(2.07737, 3.28090)(2.07773, 3.28090)(2.07809, 3.28090)(2.07846, 3.28090)(2.07882, 3.28090)(2.07918, 3.28090)(2.07954, 3.28090)(2.07990, 3.28090)(2.08027, 3.28090)(2.08063, 3.28090)(2.08099, 3.28090)(2.08135, 3.28090)(2.08171, 3.28090)(2.08207, 3.28090)(2.08243, 3.28090)(2.08279, 3.28090)(2.08314, 3.28090)(2.08350, 3.28090)(2.08386, 3.28090)(2.08422, 3.28090)(2.08458, 3.28090)(2.08493, 3.28090)(2.08529, 3.28090)(2.08565, 3.28090)(2.08600, 3.28090)(2.08636, 3.28090)(2.08672, 3.28090)(2.08707, 3.28090)(2.08743, 3.28090)(2.08778, 3.28090)(2.08814, 3.28090)(2.08849, 3.28090)(2.08884, 3.28090)(2.08920, 3.28090)(2.08955, 3.28090)(2.08991, 3.28090)(2.09026, 3.28090)(2.09061, 3.28090)(2.09096, 3.28090)(2.09132, 3.28090)(2.09167, 3.28090)(2.09202, 3.28090)(2.09237, 3.28090)(2.09272, 3.28090)(2.09307, 3.28090)(2.09342, 3.28090)(2.09377, 3.28090)(2.09412, 3.28090)(2.09447, 3.28090)(2.09482, 3.28090)(2.09517, 3.28090)(2.09552, 3.28090)(2.09587, 3.28090)(2.09621, 3.28090)(2.09656, 3.28090)(2.09691, 3.28090)
\psline[linecolor=red, plotstyle=curve, linewidth=0.4mm, showpoints=false, linestyle=solid, linecolor=red, dotstyle=triangle, dotscale=1.2 1.2, linewidth=0.4mm](0.00000, 6.63840)(0.30103, 4.62270)(0.47712, 4.08941)(0.60206, 3.78944)(0.69897, 3.59376)(1.00000, 3.14661)(1.30103, 2.87271)(1.47712, 2.76613)(1.60206, 2.70786)(1.69897, 2.67065)(2.00000, 2.58858)
\psline[linecolor=red, plotstyle=curve, linewidth=0.4mm, showpoints=true, linestyle=none, linecolor=red, dotstyle=triangle, dotscale=1.2 1.2, linewidth=0.4mm](0.00000, 6.65756)(0.30103, 4.76591)(0.47712, 4.10467)(0.60206, 3.75738)(0.69897, 3.53835)(1.00000, 3.05409)(1.30103, 2.76774)(1.47712, 2.65650)(1.60206, 2.59230)(1.69897, 2.56636)(2.00000, 2.47523)
\psline[linecolor=red, plotstyle=curve, linewidth=0.4mm, showpoints=false, linestyle=solid, linecolor=red, dotstyle=+, dotscale=1.2 1.2, linewidth=0.4mm](2.06070, 2.51976)(2.06108, 2.51976)(2.06145, 2.51976)(2.06183, 2.51976)(2.06221, 2.51976)(2.06258, 2.51976)(2.06296, 2.51976)(2.06333, 2.51976)(2.06371, 2.51976)(2.06408, 2.51976)(2.06446, 2.51976)(2.06483, 2.51976)(2.06521, 2.51976)(2.06558, 2.51976)(2.06595, 2.51976)(2.06633, 2.51976)(2.06670, 2.51976)(2.06707, 2.51976)(2.06744, 2.51976)(2.06781, 2.51976)(2.06819, 2.51976)(2.06856, 2.51976)(2.06893, 2.51976)(2.06930, 2.51976)(2.06967, 2.51976)(2.07004, 2.51976)(2.07041, 2.51976)(2.07078, 2.51976)(2.07115, 2.51976)(2.07151, 2.51976)(2.07188, 2.51976)(2.07225, 2.51976)(2.07262, 2.51976)(2.07298, 2.51976)(2.07335, 2.51976)(2.07372, 2.51976)(2.07408, 2.51976)(2.07445, 2.51976)(2.07482, 2.51976)(2.07518, 2.51976)(2.07555, 2.51976)(2.07591, 2.51976)(2.07628, 2.51976)(2.07664, 2.51976)(2.07700, 2.51976)(2.07737, 2.51976)(2.07773, 2.51976)(2.07809, 2.51976)(2.07846, 2.51976)(2.07882, 2.51976)(2.07918, 2.51976)(2.07954, 2.51976)(2.07990, 2.51976)(2.08027, 2.51976)(2.08063, 2.51976)(2.08099, 2.51976)(2.08135, 2.51976)(2.08171, 2.51976)(2.08207, 2.51976)(2.08243, 2.51976)(2.08279, 2.51976)(2.08314, 2.51976)(2.08350, 2.51976)(2.08386, 2.51976)(2.08422, 2.51976)(2.08458, 2.51976)(2.08493, 2.51976)(2.08529, 2.51976)(2.08565, 2.51976)(2.08600, 2.51976)(2.08636, 2.51976)(2.08672, 2.51976)(2.08707, 2.51976)(2.08743, 2.51976)(2.08778, 2.51976)(2.08814, 2.51976)(2.08849, 2.51976)(2.08884, 2.51976)(2.08920, 2.51976)(2.08955, 2.51976)(2.08991, 2.51976)(2.09026, 2.51976)(2.09061, 2.51976)(2.09096, 2.51976)(2.09132, 2.51976)(2.09167, 2.51976)(2.09202, 2.51976)(2.09237, 2.51976)(2.09272, 2.51976)(2.09307, 2.51976)(2.09342, 2.51976)(2.09377, 2.51976)(2.09412, 2.51976)(2.09447, 2.51976)(2.09482, 2.51976)(2.09517, 2.51976)(2.09552, 2.51976)(2.09587, 2.51976)(2.09621, 2.51976)(2.09656, 2.51976)(2.09691, 2.51976)
\psline[linecolor=black, plotstyle=curve, linewidth=0.4mm, showpoints=false, linestyle=solid, linecolor=black, dotstyle=diamond, dotscale=1.2 1.2, linewidth=0.4mm](0.00000, 4.88205)(0.30103, 3.45617)(0.47712, 3.06719)(0.60206, 2.84839)(0.69897, 2.70566)(1.00000, 2.37951)(1.30103, 2.17974)(1.47712, 2.10200)(1.60206, 2.05950)(1.69897, 2.03235)(2.00000, 1.97249)
\psline[linecolor=black, plotstyle=curve, linewidth=0.4mm, showpoints=true, linestyle=none, linecolor=black, dotstyle=diamond, dotscale=1.2 1.2, linewidth=0.4mm](0.00000, 4.91756)(0.30103, 3.57953)(0.47712, 3.09302)(0.60206, 2.84328)(0.69897, 2.67952)(1.00000, 2.31831)(1.30103, 2.11010)(1.47712, 2.02646)(1.60206, 1.98239)(1.69897, 1.96025)(2.00000, 1.87007)
\psline[linecolor=black, plotstyle=curve, linewidth=0.4mm, showpoints=false, linestyle=solid, linecolor=black, dotstyle=+, dotscale=1.2 1.2, linewidth=0.4mm](2.06070, 1.92230)(2.06108, 1.92230)(2.06145, 1.92230)(2.06183, 1.92230)(2.06221, 1.92230)(2.06258, 1.92230)(2.06296, 1.92230)(2.06333, 1.92230)(2.06371, 1.92230)(2.06408, 1.92230)(2.06446, 1.92230)(2.06483, 1.92230)(2.06521, 1.92230)(2.06558, 1.92230)(2.06595, 1.92230)(2.06633, 1.92230)(2.06670, 1.92230)(2.06707, 1.92230)(2.06744, 1.92230)(2.06781, 1.92230)(2.06819, 1.92230)(2.06856, 1.92230)(2.06893, 1.92230)(2.06930, 1.92230)(2.06967, 1.92230)(2.07004, 1.92230)(2.07041, 1.92230)(2.07078, 1.92230)(2.07115, 1.92230)(2.07151, 1.92230)(2.07188, 1.92230)(2.07225, 1.92230)(2.07262, 1.92230)(2.07298, 1.92230)(2.07335, 1.92230)(2.07372, 1.92230)(2.07408, 1.92230)(2.07445, 1.92230)(2.07482, 1.92230)(2.07518, 1.92230)(2.07555, 1.92230)(2.07591, 1.92230)(2.07628, 1.92230)(2.07664, 1.92230)(2.07700, 1.92230)(2.07737, 1.92230)(2.07773, 1.92230)(2.07809, 1.92230)(2.07846, 1.92230)(2.07882, 1.92230)(2.07918, 1.92230)(2.07954, 1.92230)(2.07990, 1.92230)(2.08027, 1.92230)(2.08063, 1.92230)(2.08099, 1.92230)(2.08135, 1.92230)(2.08171, 1.92230)(2.08207, 1.92230)(2.08243, 1.92230)(2.08279, 1.92230)(2.08314, 1.92230)(2.08350, 1.92230)(2.08386, 1.92230)(2.08422, 1.92230)(2.08458, 1.92230)(2.08493, 1.92230)(2.08529, 1.92230)(2.08565, 1.92230)(2.08600, 1.92230)(2.08636, 1.92230)(2.08672, 1.92230)(2.08707, 1.92230)(2.08743, 1.92230)(2.08778, 1.92230)(2.08814, 1.92230)(2.08849, 1.92230)(2.08884, 1.92230)(2.08920, 1.92230)(2.08955, 1.92230)(2.08991, 1.92230)(2.09026, 1.92230)(2.09061, 1.92230)(2.09096, 1.92230)(2.09132, 1.92230)(2.09167, 1.92230)(2.09202, 1.92230)(2.09237, 1.92230)(2.09272, 1.92230)(2.09307, 1.92230)(2.09342, 1.92230)(2.09377, 1.92230)(2.09412, 1.92230)(2.09447, 1.92230)(2.09482, 1.92230)(2.09517, 1.92230)(2.09552, 1.92230)(2.09587, 1.92230)(2.09621, 1.92230)(2.09656, 1.92230)(2.09691, 1.92230)
\endpsclip
\psframe[linecolor=black, fillstyle=solid, fillcolor=white, shadowcolor=lightgray, shadowsize=1mm, shadow=true](0.79545, 8.45854)(1.65455, 12.14634)
\rput[l](1.08182, 11.26829){\scriptsize{$\Delta\gamma=\unit[0]{dB}$}}
\psline[linecolor=blue, linestyle=solid, linewidth=0.3mm](0.85909, 11.26829)(0.98636, 11.26829)
\psline[linecolor=blue, linestyle=solid, linewidth=0.3mm](0.85909, 11.26829)(0.98636, 11.26829)
\psdots[linecolor=blue, linestyle=solid, linewidth=0.3mm, dotstyle=o, dotscale=1.2 1.2, linecolor=blue](0.92273, 11.26829)
\rput[l](1.08182, 10.30244){\scriptsize{$\Delta\gamma=\unit[0.4]{dB}$}}
\psline[linecolor=red, linestyle=solid, linewidth=0.3mm](0.85909, 10.30244)(0.98636, 10.30244)
\psline[linecolor=red, linestyle=solid, linewidth=0.3mm](0.85909, 10.30244)(0.98636, 10.30244)
\psdots[linecolor=red, linestyle=solid, linewidth=0.3mm, dotstyle=triangle, dotscale=1.2 1.2, linecolor=red](0.92273, 10.30244)
\rput[l](1.08182, 9.33659){\scriptsize{$\Delta\gamma=\unit[0.9]{dB}$}}
\psline[linecolor=black, linestyle=solid, linewidth=0.3mm](0.85909, 9.33659)(0.98636, 9.33659)
\psline[linecolor=black, linestyle=solid, linewidth=0.3mm](0.85909, 9.33659)(0.98636, 9.33659)
\psdots[linecolor=black, linestyle=solid, linewidth=0.3mm, dotstyle=diamond, dotscale=1.2 1.2, linecolor=black](0.92273, 9.33659)
}\end{pspicture}
\endgroup
 

%% file: Complexity_outage_network_Nusers_rel_with_pl.tex
\begingroup
\unitlength=1mm
\psset{xunit=31.42857mm, yunit=58.48787mm, linewidth=0.1mm}
\psset{arrowsize=2pt 3, arrowlength=1.4, arrowinset=.4}\psset{axesstyle=frame}
\begin{pspicture}(-0.44545, 0.06063)(2.10000, 1.00100)
\rput(-0.06364, -0.08549){%
\psaxes[subticks=10, xlogBase=10, logLines=x, labels=all, xsubticks=10, ysubticks=1, Ox=0, Oy=0.3, Dx=1, Dy=0.1]{-}(0.00000, 0.30000)(0.00000, 0.30000)(2.10000, 1.00100)%
\multips(0.00000, 0.40000)(0, 0.10000){6}{\psline[linecolor=black, linestyle=dotted, linewidth=0.2mm](0, 0)(2.10000, 0)}
\rput[b](1.05000, 0.14612){\small{$N_\text{c}$}}
\rput[t]{90}(-0.38182, 0.65050){\small{$\mathcal{C}_\text{out}(\hat\epsilon, N_\text{c})/N_\text{c}\mathcal{C}_\text{out}(\hat\epsilon)$}}
\psclip{\psframe(0.00000, 0.30000)(2.10000, 1.00100)}
\psline[linecolor=blue, plotstyle=curve, linewidth=0.4mm, showpoints=false, linestyle=solid, linecolor=blue, dotstyle=o, dotscale=1.2 1.2, linewidth=0.4mm](0.00000, 1.00000)(0.30103, 0.68918)(0.47712, 0.60607)(0.60206, 0.55931)(0.69897, 0.52882)(1.00000, 0.45913)(1.30103, 0.41644)(1.47712, 0.39983)(1.60206, 0.39075)(1.69897, 0.38495)(2.00000, 0.37215)
\psline[linecolor=blue, plotstyle=curve, linewidth=0.4mm, showpoints=true, linestyle=none, linecolor=blue, dotstyle=o, dotscale=1.2 1.2, linewidth=0.4mm](0.00000, 1.00000)(0.30103, 0.70323)(0.47712, 0.60345)(0.60206, 0.54953)(0.69897, 0.51653)(1.00000, 0.44348)(1.30103, 0.39989)(1.47712, 0.38424)(1.60206, 0.37290)(1.69897, 0.36868)(2.00000, 0.36234)
\psline[linecolor=blue, plotstyle=curve, linewidth=0.4mm, showpoints=false, linestyle=solid, linecolor=blue, dotstyle=+, dotscale=1.2 1.2, linewidth=0.4mm](2.06070, 0.36143)(2.06108, 0.36143)(2.06145, 0.36143)(2.06183, 0.36143)(2.06221, 0.36143)(2.06258, 0.36143)(2.06296, 0.36143)(2.06333, 0.36143)(2.06371, 0.36143)(2.06408, 0.36143)(2.06446, 0.36143)(2.06483, 0.36143)(2.06521, 0.36143)(2.06558, 0.36143)(2.06595, 0.36143)(2.06633, 0.36143)(2.06670, 0.36143)(2.06707, 0.36143)(2.06744, 0.36143)(2.06781, 0.36143)(2.06819, 0.36143)(2.06856, 0.36143)(2.06893, 0.36143)(2.06930, 0.36143)(2.06967, 0.36143)(2.07004, 0.36143)(2.07041, 0.36143)(2.07078, 0.36143)(2.07115, 0.36143)(2.07151, 0.36143)(2.07188, 0.36143)(2.07225, 0.36143)(2.07262, 0.36143)(2.07298, 0.36143)(2.07335, 0.36143)(2.07372, 0.36143)(2.07408, 0.36143)(2.07445, 0.36143)(2.07482, 0.36143)(2.07518, 0.36143)(2.07555, 0.36143)(2.07591, 0.36143)(2.07628, 0.36143)(2.07664, 0.36143)(2.07700, 0.36143)(2.07737, 0.36143)(2.07773, 0.36143)(2.07809, 0.36143)(2.07846, 0.36143)(2.07882, 0.36143)(2.07918, 0.36143)(2.07954, 0.36143)(2.07990, 0.36143)(2.08027, 0.36143)(2.08063, 0.36143)(2.08099, 0.36143)(2.08135, 0.36143)(2.08171, 0.36143)(2.08207, 0.36143)(2.08243, 0.36143)(2.08279, 0.36143)(2.08314, 0.36143)(2.08350, 0.36143)(2.08386, 0.36143)(2.08422, 0.36143)(2.08458, 0.36143)(2.08493, 0.36143)(2.08529, 0.36143)(2.08565, 0.36143)(2.08600, 0.36143)(2.08636, 0.36143)(2.08672, 0.36143)(2.08707, 0.36143)(2.08743, 0.36143)(2.08778, 0.36143)(2.08814, 0.36143)(2.08849, 0.36143)(2.08884, 0.36143)(2.08920, 0.36143)(2.08955, 0.36143)(2.08991, 0.36143)(2.09026, 0.36143)(2.09061, 0.36143)(2.09096, 0.36143)(2.09132, 0.36143)(2.09167, 0.36143)(2.09202, 0.36143)(2.09237, 0.36143)(2.09272, 0.36143)(2.09307, 0.36143)(2.09342, 0.36143)(2.09377, 0.36143)(2.09412, 0.36143)(2.09447, 0.36143)(2.09482, 0.36143)(2.09517, 0.36143)(2.09552, 0.36143)(2.09587, 0.36143)(2.09621, 0.36143)(2.09656, 0.36143)(2.09691, 0.36143)
\psline[linecolor=red, plotstyle=curve, linewidth=0.4mm, showpoints=false, linestyle=solid, linecolor=red, dotstyle=triangle, dotscale=1.2 1.2, linewidth=0.4mm](0.00000, 1.00000)(0.30103, 0.69636)(0.47712, 0.61602)(0.60206, 0.57084)(0.69897, 0.54136)(1.00000, 0.47400)(1.30103, 0.43274)(1.47712, 0.41669)(1.60206, 0.40791)(1.69897, 0.40230)(2.00000, 0.38994)
\psline[linecolor=red, plotstyle=curve, linewidth=0.4mm, showpoints=true, linestyle=none, linecolor=red, dotstyle=triangle, dotscale=1.2 1.2, linewidth=0.4mm](0.00000, 1.00000)(0.30103, 0.71586)(0.47712, 0.61654)(0.60206, 0.56438)(0.69897, 0.53148)(1.00000, 0.45874)(1.30103, 0.41573)(1.47712, 0.39902)(1.60206, 0.38938)(1.69897, 0.38548)(2.00000, 0.37179)
\psline[linecolor=red, plotstyle=curve, linewidth=0.4mm, showpoints=false, linestyle=solid, linecolor=red, dotstyle=+, dotscale=1.2 1.2, linewidth=0.4mm](2.06070, 0.37957)(2.06108, 0.37957)(2.06145, 0.37957)(2.06183, 0.37957)(2.06221, 0.37957)(2.06258, 0.37957)(2.06296, 0.37957)(2.06333, 0.37957)(2.06371, 0.37957)(2.06408, 0.37957)(2.06446, 0.37957)(2.06483, 0.37957)(2.06521, 0.37957)(2.06558, 0.37957)(2.06595, 0.37957)(2.06633, 0.37957)(2.06670, 0.37957)(2.06707, 0.37957)(2.06744, 0.37957)(2.06781, 0.37957)(2.06819, 0.37957)(2.06856, 0.37957)(2.06893, 0.37957)(2.06930, 0.37957)(2.06967, 0.37957)(2.07004, 0.37957)(2.07041, 0.37957)(2.07078, 0.37957)(2.07115, 0.37957)(2.07151, 0.37957)(2.07188, 0.37957)(2.07225, 0.37957)(2.07262, 0.37957)(2.07298, 0.37957)(2.07335, 0.37957)(2.07372, 0.37957)(2.07408, 0.37957)(2.07445, 0.37957)(2.07482, 0.37957)(2.07518, 0.37957)(2.07555, 0.37957)(2.07591, 0.37957)(2.07628, 0.37957)(2.07664, 0.37957)(2.07700, 0.37957)(2.07737, 0.37957)(2.07773, 0.37957)(2.07809, 0.37957)(2.07846, 0.37957)(2.07882, 0.37957)(2.07918, 0.37957)(2.07954, 0.37957)(2.07990, 0.37957)(2.08027, 0.37957)(2.08063, 0.37957)(2.08099, 0.37957)(2.08135, 0.37957)(2.08171, 0.37957)(2.08207, 0.37957)(2.08243, 0.37957)(2.08279, 0.37957)(2.08314, 0.37957)(2.08350, 0.37957)(2.08386, 0.37957)(2.08422, 0.37957)(2.08458, 0.37957)(2.08493, 0.37957)(2.08529, 0.37957)(2.08565, 0.37957)(2.08600, 0.37957)(2.08636, 0.37957)(2.08672, 0.37957)(2.08707, 0.37957)(2.08743, 0.37957)(2.08778, 0.37957)(2.08814, 0.37957)(2.08849, 0.37957)(2.08884, 0.37957)(2.08920, 0.37957)(2.08955, 0.37957)(2.08991, 0.37957)(2.09026, 0.37957)(2.09061, 0.37957)(2.09096, 0.37957)(2.09132, 0.37957)(2.09167, 0.37957)(2.09202, 0.37957)(2.09237, 0.37957)(2.09272, 0.37957)(2.09307, 0.37957)(2.09342, 0.37957)(2.09377, 0.37957)(2.09412, 0.37957)(2.09447, 0.37957)(2.09482, 0.37957)(2.09517, 0.37957)(2.09552, 0.37957)(2.09587, 0.37957)(2.09621, 0.37957)(2.09656, 0.37957)(2.09691, 0.37957)
\psline[linecolor=black, plotstyle=curve, linewidth=0.4mm, showpoints=false, linestyle=solid, linecolor=black, dotstyle=diamond, dotscale=1.2 1.2, linewidth=0.4mm](0.00000, 1.00000)(0.30103, 0.70793)(0.47712, 0.62826)(0.60206, 0.58344)(0.69897, 0.55421)(1.00000, 0.48740)(1.30103, 0.44648)(1.47712, 0.43056)(1.60206, 0.42185)(1.69897, 0.41629)(2.00000, 0.40403)
\psline[linecolor=black, plotstyle=curve, linewidth=0.4mm, showpoints=true, linestyle=none, linecolor=black, dotstyle=diamond, dotscale=1.2 1.2, linewidth=0.4mm](0.00000, 1.00000)(0.30103, 0.72791)(0.47712, 0.62897)(0.60206, 0.57819)(0.69897, 0.54489)(1.00000, 0.47143)(1.30103, 0.42909)(1.47712, 0.41209)(1.60206, 0.40312)(1.69897, 0.39862)(2.00000, 0.38028)
\psline[linecolor=black, plotstyle=curve, linewidth=0.4mm, showpoints=false, linestyle=solid, linecolor=black, dotstyle=+, dotscale=1.2 1.2, linewidth=0.4mm](2.06070, 0.39375)(2.06108, 0.39375)(2.06145, 0.39375)(2.06183, 0.39375)(2.06221, 0.39375)(2.06258, 0.39375)(2.06296, 0.39375)(2.06333, 0.39375)(2.06371, 0.39375)(2.06408, 0.39375)(2.06446, 0.39375)(2.06483, 0.39375)(2.06521, 0.39375)(2.06558, 0.39375)(2.06595, 0.39375)(2.06633, 0.39375)(2.06670, 0.39375)(2.06707, 0.39375)(2.06744, 0.39375)(2.06781, 0.39375)(2.06819, 0.39375)(2.06856, 0.39375)(2.06893, 0.39375)(2.06930, 0.39375)(2.06967, 0.39375)(2.07004, 0.39375)(2.07041, 0.39375)(2.07078, 0.39375)(2.07115, 0.39375)(2.07151, 0.39375)(2.07188, 0.39375)(2.07225, 0.39375)(2.07262, 0.39375)(2.07298, 0.39375)(2.07335, 0.39375)(2.07372, 0.39375)(2.07408, 0.39375)(2.07445, 0.39375)(2.07482, 0.39375)(2.07518, 0.39375)(2.07555, 0.39375)(2.07591, 0.39375)(2.07628, 0.39375)(2.07664, 0.39375)(2.07700, 0.39375)(2.07737, 0.39375)(2.07773, 0.39375)(2.07809, 0.39375)(2.07846, 0.39375)(2.07882, 0.39375)(2.07918, 0.39375)(2.07954, 0.39375)(2.07990, 0.39375)(2.08027, 0.39375)(2.08063, 0.39375)(2.08099, 0.39375)(2.08135, 0.39375)(2.08171, 0.39375)(2.08207, 0.39375)(2.08243, 0.39375)(2.08279, 0.39375)(2.08314, 0.39375)(2.08350, 0.39375)(2.08386, 0.39375)(2.08422, 0.39375)(2.08458, 0.39375)(2.08493, 0.39375)(2.08529, 0.39375)(2.08565, 0.39375)(2.08600, 0.39375)(2.08636, 0.39375)(2.08672, 0.39375)(2.08707, 0.39375)(2.08743, 0.39375)(2.08778, 0.39375)(2.08814, 0.39375)(2.08849, 0.39375)(2.08884, 0.39375)(2.08920, 0.39375)(2.08955, 0.39375)(2.08991, 0.39375)(2.09026, 0.39375)(2.09061, 0.39375)(2.09096, 0.39375)(2.09132, 0.39375)(2.09167, 0.39375)(2.09202, 0.39375)(2.09237, 0.39375)(2.09272, 0.39375)(2.09307, 0.39375)(2.09342, 0.39375)(2.09377, 0.39375)(2.09412, 0.39375)(2.09447, 0.39375)(2.09482, 0.39375)(2.09517, 0.39375)(2.09552, 0.39375)(2.09587, 0.39375)(2.09621, 0.39375)(2.09656, 0.39375)(2.09691, 0.39375)
\endpsclip
\psframe[linecolor=black, fillstyle=solid, fillcolor=white, shadowcolor=lightgray, shadowsize=1mm, shadow=true](0.79545, 0.79412)(1.65455, 1.00955)
\rput[l](1.08182, 0.95826){\scriptsize{$\Delta\gamma=\unit[0]{dB}$}}
\psline[linecolor=blue, linestyle=solid, linewidth=0.3mm](0.85909, 0.95826)(0.98636, 0.95826)
\psline[linecolor=blue, linestyle=solid, linewidth=0.3mm](0.85909, 0.95826)(0.98636, 0.95826)
\psdots[linecolor=blue, linestyle=solid, linewidth=0.3mm, dotstyle=o, dotscale=1.2 1.2, linecolor=blue](0.92273, 0.95826)
\rput[l](1.08182, 0.90183){\scriptsize{$\Delta\gamma=\unit[0.4]{dB}$}}
\psline[linecolor=red, linestyle=solid, linewidth=0.3mm](0.85909, 0.90183)(0.98636, 0.90183)
\psline[linecolor=red, linestyle=solid, linewidth=0.3mm](0.85909, 0.90183)(0.98636, 0.90183)
\psdots[linecolor=red, linestyle=solid, linewidth=0.3mm, dotstyle=triangle, dotscale=1.2 1.2, linecolor=red](0.92273, 0.90183)
\rput[l](1.08182, 0.84541){\scriptsize{$\Delta\gamma=\unit[0.9]{dB}$}}
\psline[linecolor=black, linestyle=solid, linewidth=0.3mm](0.85909, 0.84541)(0.98636, 0.84541)
\psline[linecolor=black, linestyle=solid, linewidth=0.3mm](0.85909, 0.84541)(0.98636, 0.84541)
\psdots[linecolor=black, linestyle=solid, linewidth=0.3mm, dotstyle=diamond, dotscale=1.2 1.2, linecolor=black](0.92273, 0.84541)
}\end{pspicture}
\endgroup
 

%% file: AbsComplexity.tex
\begingroup
\unitlength=1mm
\psset{xunit=31.42857mm, yunit=2.05000mm, linewidth=0.1mm}
\psset{arrowsize=2pt 3, arrowlength=1.4, arrowinset=.4}\psset{axesstyle=frame}
\begin{pspicture}(-0.44545, -6.82927)(2.10000, 20.00000)
\rput(-0.06364, -2.43902){%
\psaxes[subticks=10, xlogBase=10, logLines=x, labels=all, xsubticks=10, ysubticks=1, Ox=0, Oy=0, Dx=1, Dy=5]{-}(0.00000, 0.00000)(0.00000, 0.00000)(2.10000, 20.00000)%
\multips(0.00000, 5.00000)(0, 5.00000){3}{\psline[linecolor=black, linestyle=dotted, linewidth=0.2mm](0, 0)(2.10000, 0)}
\rput[b](1.05000, -4.39024){\small{$N_\text{c}$}}
\rput[t]{90}(-0.38182, 10.00000){\small{$\mathcal{C}_\text{out}(\hat\epsilon, N_\text{c})/N_\text{c}$ [bit-iter pcu]}}
\psclip{\psframe(0.00000, 0.00000)(2.10000, 20.00000)}
\psline[linecolor=blue, plotstyle=curve, linewidth=0.4mm, showpoints=false, linestyle=solid, linecolor=blue, dotstyle=o, dotscale=1.2 1.2, linewidth=0.4mm](0.00000, 13.27602)(0.30103, 10.46928)(0.47712, 9.41970)(0.60206, 8.82932)(0.69897, 8.44420)(1.00000, 7.56414)(1.30103, 7.02509)(1.47712, 6.81532)(1.60206, 6.70064)(1.69897, 6.62740)(2.00000, 6.46588)
\psline[linecolor=blue, plotstyle=curve, linewidth=0.4mm, showpoints=false, linestyle=solid, linecolor=blue, dotstyle=+, dotscale=1.2 1.2, linewidth=0.4mm](2.06070, 6.33044)(2.06108, 6.33044)(2.06145, 6.33044)(2.06183, 6.33044)(2.06221, 6.33044)(2.06258, 6.33044)(2.06296, 6.33044)(2.06333, 6.33044)(2.06371, 6.33044)(2.06408, 6.33044)(2.06446, 6.33044)(2.06483, 6.33044)(2.06521, 6.33044)(2.06558, 6.33044)(2.06595, 6.33044)(2.06633, 6.33044)(2.06670, 6.33044)(2.06707, 6.33044)(2.06744, 6.33044)(2.06781, 6.33044)(2.06819, 6.33044)(2.06856, 6.33044)(2.06893, 6.33044)(2.06930, 6.33044)(2.06967, 6.33044)(2.07004, 6.33044)(2.07041, 6.33044)(2.07078, 6.33044)(2.07115, 6.33044)(2.07151, 6.33044)(2.07188, 6.33044)(2.07225, 6.33044)(2.07262, 6.33044)(2.07298, 6.33044)(2.07335, 6.33044)(2.07372, 6.33044)(2.07408, 6.33044)(2.07445, 6.33044)(2.07482, 6.33044)(2.07518, 6.33044)(2.07555, 6.33044)(2.07591, 6.33044)(2.07628, 6.33044)(2.07664, 6.33044)(2.07700, 6.33044)(2.07737, 6.33044)(2.07773, 6.33044)(2.07809, 6.33044)(2.07846, 6.33044)(2.07882, 6.33044)(2.07918, 6.33044)(2.07954, 6.33044)(2.07990, 6.33044)(2.08027, 6.33044)(2.08063, 6.33044)(2.08099, 6.33044)(2.08135, 6.33044)(2.08171, 6.33044)(2.08207, 6.33044)(2.08243, 6.33044)(2.08279, 6.33044)(2.08314, 6.33044)(2.08350, 6.33044)(2.08386, 6.33044)(2.08422, 6.33044)(2.08458, 6.33044)(2.08493, 6.33044)(2.08529, 6.33044)(2.08565, 6.33044)(2.08600, 6.33044)(2.08636, 6.33044)(2.08672, 6.33044)(2.08707, 6.33044)(2.08743, 6.33044)(2.08778, 6.33044)(2.08814, 6.33044)(2.08849, 6.33044)(2.08884, 6.33044)(2.08920, 6.33044)(2.08955, 6.33044)(2.08991, 6.33044)(2.09026, 6.33044)(2.09061, 6.33044)(2.09096, 6.33044)(2.09132, 6.33044)(2.09167, 6.33044)(2.09202, 6.33044)(2.09237, 6.33044)(2.09272, 6.33044)(2.09307, 6.33044)(2.09342, 6.33044)(2.09377, 6.33044)(2.09412, 6.33044)(2.09447, 6.33044)(2.09482, 6.33044)(2.09517, 6.33044)(2.09552, 6.33044)(2.09587, 6.33044)(2.09621, 6.33044)(2.09656, 6.33044)(2.09691, 6.33044)
\psline[linecolor=red, plotstyle=curve, linewidth=0.4mm, showpoints=false, linestyle=solid, linecolor=red, dotstyle=triangle, dotscale=1.2 1.2, linewidth=0.4mm](0.00000, 9.57195)(0.30103, 7.64266)(0.47712, 6.91805)(0.60206, 6.51046)(0.69897, 6.24458)(1.00000, 5.63700)(1.30103, 5.26485)(1.47712, 5.12003)(1.60206, 5.04086)(1.69897, 4.99029)(2.00000, 4.87878)
\psline[linecolor=red, plotstyle=curve, linewidth=0.4mm, showpoints=false, linestyle=solid, linecolor=red, dotstyle=+, dotscale=1.2 1.2, linewidth=0.4mm](2.06070, 4.78527)(2.06108, 4.78527)(2.06145, 4.78527)(2.06183, 4.78527)(2.06221, 4.78527)(2.06258, 4.78527)(2.06296, 4.78527)(2.06333, 4.78527)(2.06371, 4.78527)(2.06408, 4.78527)(2.06446, 4.78527)(2.06483, 4.78527)(2.06521, 4.78527)(2.06558, 4.78527)(2.06595, 4.78527)(2.06633, 4.78527)(2.06670, 4.78527)(2.06707, 4.78527)(2.06744, 4.78527)(2.06781, 4.78527)(2.06819, 4.78527)(2.06856, 4.78527)(2.06893, 4.78527)(2.06930, 4.78527)(2.06967, 4.78527)(2.07004, 4.78527)(2.07041, 4.78527)(2.07078, 4.78527)(2.07115, 4.78527)(2.07151, 4.78527)(2.07188, 4.78527)(2.07225, 4.78527)(2.07262, 4.78527)(2.07298, 4.78527)(2.07335, 4.78527)(2.07372, 4.78527)(2.07408, 4.78527)(2.07445, 4.78527)(2.07482, 4.78527)(2.07518, 4.78527)(2.07555, 4.78527)(2.07591, 4.78527)(2.07628, 4.78527)(2.07664, 4.78527)(2.07700, 4.78527)(2.07737, 4.78527)(2.07773, 4.78527)(2.07809, 4.78527)(2.07846, 4.78527)(2.07882, 4.78527)(2.07918, 4.78527)(2.07954, 4.78527)(2.07990, 4.78527)(2.08027, 4.78527)(2.08063, 4.78527)(2.08099, 4.78527)(2.08135, 4.78527)(2.08171, 4.78527)(2.08207, 4.78527)(2.08243, 4.78527)(2.08279, 4.78527)(2.08314, 4.78527)(2.08350, 4.78527)(2.08386, 4.78527)(2.08422, 4.78527)(2.08458, 4.78527)(2.08493, 4.78527)(2.08529, 4.78527)(2.08565, 4.78527)(2.08600, 4.78527)(2.08636, 4.78527)(2.08672, 4.78527)(2.08707, 4.78527)(2.08743, 4.78527)(2.08778, 4.78527)(2.08814, 4.78527)(2.08849, 4.78527)(2.08884, 4.78527)(2.08920, 4.78527)(2.08955, 4.78527)(2.08991, 4.78527)(2.09026, 4.78527)(2.09061, 4.78527)(2.09096, 4.78527)(2.09132, 4.78527)(2.09167, 4.78527)(2.09202, 4.78527)(2.09237, 4.78527)(2.09272, 4.78527)(2.09307, 4.78527)(2.09342, 4.78527)(2.09377, 4.78527)(2.09412, 4.78527)(2.09447, 4.78527)(2.09482, 4.78527)(2.09517, 4.78527)(2.09552, 4.78527)(2.09587, 4.78527)(2.09621, 4.78527)(2.09656, 4.78527)(2.09691, 4.78527)
\psline[linecolor=black, plotstyle=curve, linewidth=0.4mm, showpoints=false, linestyle=solid, linecolor=black, dotstyle=diamond, dotscale=1.2 1.2, linewidth=0.4mm](0.00000, 7.01404)(0.30103, 5.62959)(0.47712, 5.11229)(0.60206, 4.82132)(0.69897, 4.63150)(1.00000, 4.19776)(1.30103, 3.93208)(1.47712, 3.82869)(1.60206, 3.77217)(1.69897, 3.73607)(2.00000, 3.65646)
\psline[linecolor=black, plotstyle=curve, linewidth=0.4mm, showpoints=false, linestyle=solid, linecolor=black, dotstyle=+, dotscale=1.2 1.2, linewidth=0.4mm](2.06070, 3.58971)(2.06108, 3.58971)(2.06145, 3.58971)(2.06183, 3.58971)(2.06221, 3.58971)(2.06258, 3.58971)(2.06296, 3.58971)(2.06333, 3.58971)(2.06371, 3.58971)(2.06408, 3.58971)(2.06446, 3.58971)(2.06483, 3.58971)(2.06521, 3.58971)(2.06558, 3.58971)(2.06595, 3.58971)(2.06633, 3.58971)(2.06670, 3.58971)(2.06707, 3.58971)(2.06744, 3.58971)(2.06781, 3.58971)(2.06819, 3.58971)(2.06856, 3.58971)(2.06893, 3.58971)(2.06930, 3.58971)(2.06967, 3.58971)(2.07004, 3.58971)(2.07041, 3.58971)(2.07078, 3.58971)(2.07115, 3.58971)(2.07151, 3.58971)(2.07188, 3.58971)(2.07225, 3.58971)(2.07262, 3.58971)(2.07298, 3.58971)(2.07335, 3.58971)(2.07372, 3.58971)(2.07408, 3.58971)(2.07445, 3.58971)(2.07482, 3.58971)(2.07518, 3.58971)(2.07555, 3.58971)(2.07591, 3.58971)(2.07628, 3.58971)(2.07664, 3.58971)(2.07700, 3.58971)(2.07737, 3.58971)(2.07773, 3.58971)(2.07809, 3.58971)(2.07846, 3.58971)(2.07882, 3.58971)(2.07918, 3.58971)(2.07954, 3.58971)(2.07990, 3.58971)(2.08027, 3.58971)(2.08063, 3.58971)(2.08099, 3.58971)(2.08135, 3.58971)(2.08171, 3.58971)(2.08207, 3.58971)(2.08243, 3.58971)(2.08279, 3.58971)(2.08314, 3.58971)(2.08350, 3.58971)(2.08386, 3.58971)(2.08422, 3.58971)(2.08458, 3.58971)(2.08493, 3.58971)(2.08529, 3.58971)(2.08565, 3.58971)(2.08600, 3.58971)(2.08636, 3.58971)(2.08672, 3.58971)(2.08707, 3.58971)(2.08743, 3.58971)(2.08778, 3.58971)(2.08814, 3.58971)(2.08849, 3.58971)(2.08884, 3.58971)(2.08920, 3.58971)(2.08955, 3.58971)(2.08991, 3.58971)(2.09026, 3.58971)(2.09061, 3.58971)(2.09096, 3.58971)(2.09132, 3.58971)(2.09167, 3.58971)(2.09202, 3.58971)(2.09237, 3.58971)(2.09272, 3.58971)(2.09307, 3.58971)(2.09342, 3.58971)(2.09377, 3.58971)(2.09412, 3.58971)(2.09447, 3.58971)(2.09482, 3.58971)(2.09517, 3.58971)(2.09552, 3.58971)(2.09587, 3.58971)(2.09621, 3.58971)(2.09656, 3.58971)(2.09691, 3.58971)
\psline[linecolor=blue, plotstyle=curve, linewidth=0.4mm, showpoints=true, linestyle=none, linecolor=blue, dotstyle=o, dotscale=1.2 1.2, linewidth=0.4mm](0.00000, 13.60500)(0.30103, 10.40500)(0.47712, 9.23500)(0.60206, 8.60500)(0.69897, 8.21500)(1.00000, 7.32000)(1.30103, 6.79000)(1.47712, 6.58500)(1.60206, 6.49000)(1.69897, 6.39500)(2.00000, 5.92000)
\psline[linecolor=red, plotstyle=curve, linewidth=0.4mm, showpoints=true, linestyle=none, linecolor=red, dotstyle=triangle, dotscale=1.2 1.2, linewidth=0.4mm](0.00000, 9.58500)(0.30103, 7.27500)(0.47712, 6.51000)(0.60206, 6.09500)(0.69897, 5.82000)(1.00000, 5.21000)(1.30103, 4.84500)(1.47712, 4.70500)(1.60206, 4.64250)(1.69897, 4.58000)(2.00000, 4.26750)
\psline[linecolor=black, plotstyle=curve, linewidth=0.4mm, showpoints=true, linestyle=none, linecolor=black, dotstyle=diamond, dotscale=1.2 1.2, linewidth=0.4mm](0.00000, 5.33500)(0.30103, 4.34000)(0.47712, 3.88000)(0.60206, 3.63500)(0.69897, 3.47000)(1.00000, 3.10500)(1.30103, 2.89000)(1.47712, 2.81000)(1.60206, 2.77000)(1.69897, 2.73000)(2.00000, 2.53000)
\endpsclip
\psframe[linecolor=black, fillstyle=solid, fillcolor=white, shadowcolor=lightgray, shadowsize=1mm, shadow=true](0.79545, 14.09756)(1.65455, 20.24390)
\rput[l](1.08182, 18.78049){\scriptsize{$\Delta \gamma=\unit[0]{dB}$}}
\psline[linecolor=blue, linestyle=solid, linewidth=0.3mm](0.85909, 18.78049)(0.98636, 18.78049)
\psline[linecolor=blue, linestyle=solid, linewidth=0.3mm](0.85909, 18.78049)(0.98636, 18.78049)
\psdots[linecolor=blue, linestyle=solid, linewidth=0.3mm, dotstyle=o, dotscale=1.2 1.2, linecolor=blue](0.92273, 18.78049)
\rput[l](1.08182, 17.17073){\scriptsize{$\Delta \gamma=\unit[0.4]{dB}$}}
\psline[linecolor=red, linestyle=solid, linewidth=0.3mm](0.85909, 17.17073)(0.98636, 17.17073)
\psline[linecolor=red, linestyle=solid, linewidth=0.3mm](0.85909, 17.17073)(0.98636, 17.17073)
\psdots[linecolor=red, linestyle=solid, linewidth=0.3mm, dotstyle=triangle, dotscale=1.2 1.2, linecolor=red](0.92273, 17.17073)
\rput[l](1.08182, 15.56098){\scriptsize{$\Delta \gamma=\unit[0.9]{dB}$}}
\psline[linecolor=black, linestyle=solid, linewidth=0.3mm](0.85909, 15.56098)(0.98636, 15.56098)
\psline[linecolor=black, linestyle=solid, linewidth=0.3mm](0.85909, 15.56098)(0.98636, 15.56098)
\psdots[linecolor=black, linestyle=solid, linewidth=0.3mm, dotstyle=diamond, dotscale=1.2 1.2, linecolor=black](0.92273, 15.56098)
}\end{pspicture}
\endgroup
 

%% file: AbsComplexity_PathLoss.tex
\begingroup
\unitlength=1mm
\psset{xunit=31.42857mm, yunit=2.73333mm, linewidth=0.1mm}
\psset{arrowsize=2pt 3, arrowlength=1.4, arrowinset=.4}\psset{axesstyle=frame}
\begin{pspicture}(-0.44545, -5.12195)(2.10000, 15.00000)
\rput(-0.06364, -1.82927){%
\psaxes[subticks=10, xlogBase=10, logLines=x, labels=all, xsubticks=10, ysubticks=1, Ox=0, Oy=0, Dx=1, Dy=5]{-}(0.00000, 0.00000)(0.00000, 0.00000)(2.10000, 15.00000)%
\multips(0.00000, 5.00000)(0, 5.00000){2}{\psline[linecolor=black, linestyle=dotted, linewidth=0.2mm](0, 0)(2.10000, 0)}
\rput[b](1.05000, -3.29268){\small{$N_\text{c}$}}
\rput[t]{90}(-0.38182, 7.50000){\small{$\mathcal{C}_\text{out}(\hat\epsilon, N_\text{c})/N_\text{c}$ [bit-iter pcu]}}
\psclip{\psframe(0.00000, 0.00000)(2.10000, 15.00000)}
\psline[linecolor=blue, plotstyle=curve, linewidth=0.4mm, showpoints=false, linestyle=solid, linecolor=blue, dotstyle=o, dotscale=1.2 1.2, linewidth=0.4mm](0.00000, 9.07526)(0.30103, 6.25610)(0.47712, 5.50161)(0.60206, 5.07722)(0.69897, 4.80037)(1.00000, 4.16775)(1.30103, 3.78025)(1.47712, 3.62946)(1.60206, 3.54702)(1.69897, 3.49437)(2.00000, 3.37826)
\psline[linecolor=blue, plotstyle=curve, linewidth=0.4mm, showpoints=false, linestyle=solid, linecolor=blue, dotstyle=+, dotscale=1.2 1.2, linewidth=0.4mm](2.06070, 3.28090)(2.06108, 3.28090)(2.06145, 3.28090)(2.06183, 3.28090)(2.06221, 3.28090)(2.06258, 3.28090)(2.06296, 3.28090)(2.06333, 3.28090)(2.06371, 3.28090)(2.06408, 3.28090)(2.06446, 3.28090)(2.06483, 3.28090)(2.06521, 3.28090)(2.06558, 3.28090)(2.06595, 3.28090)(2.06633, 3.28090)(2.06670, 3.28090)(2.06707, 3.28090)(2.06744, 3.28090)(2.06781, 3.28090)(2.06819, 3.28090)(2.06856, 3.28090)(2.06893, 3.28090)(2.06930, 3.28090)(2.06967, 3.28090)(2.07004, 3.28090)(2.07041, 3.28090)(2.07078, 3.28090)(2.07115, 3.28090)(2.07151, 3.28090)(2.07188, 3.28090)(2.07225, 3.28090)(2.07262, 3.28090)(2.07298, 3.28090)(2.07335, 3.28090)(2.07372, 3.28090)(2.07408, 3.28090)(2.07445, 3.28090)(2.07482, 3.28090)(2.07518, 3.28090)(2.07555, 3.28090)(2.07591, 3.28090)(2.07628, 3.28090)(2.07664, 3.28090)(2.07700, 3.28090)(2.07737, 3.28090)(2.07773, 3.28090)(2.07809, 3.28090)(2.07846, 3.28090)(2.07882, 3.28090)(2.07918, 3.28090)(2.07954, 3.28090)(2.07990, 3.28090)(2.08027, 3.28090)(2.08063, 3.28090)(2.08099, 3.28090)(2.08135, 3.28090)(2.08171, 3.28090)(2.08207, 3.28090)(2.08243, 3.28090)(2.08279, 3.28090)(2.08314, 3.28090)(2.08350, 3.28090)(2.08386, 3.28090)(2.08422, 3.28090)(2.08458, 3.28090)(2.08493, 3.28090)(2.08529, 3.28090)(2.08565, 3.28090)(2.08600, 3.28090)(2.08636, 3.28090)(2.08672, 3.28090)(2.08707, 3.28090)(2.08743, 3.28090)(2.08778, 3.28090)(2.08814, 3.28090)(2.08849, 3.28090)(2.08884, 3.28090)(2.08920, 3.28090)(2.08955, 3.28090)(2.08991, 3.28090)(2.09026, 3.28090)(2.09061, 3.28090)(2.09096, 3.28090)(2.09132, 3.28090)(2.09167, 3.28090)(2.09202, 3.28090)(2.09237, 3.28090)(2.09272, 3.28090)(2.09307, 3.28090)(2.09342, 3.28090)(2.09377, 3.28090)(2.09412, 3.28090)(2.09447, 3.28090)(2.09482, 3.28090)(2.09517, 3.28090)(2.09552, 3.28090)(2.09587, 3.28090)(2.09621, 3.28090)(2.09656, 3.28090)(2.09691, 3.28090)
\psline[linecolor=red, plotstyle=curve, linewidth=0.4mm, showpoints=false, linestyle=solid, linecolor=red, dotstyle=triangle, dotscale=1.2 1.2, linewidth=0.4mm](0.00000, 6.63759)(0.30103, 4.62270)(0.47712, 4.08941)(0.60206, 3.78944)(0.69897, 3.59376)(1.00000, 3.14661)(1.30103, 2.87271)(1.47712, 2.76613)(1.60206, 2.70786)(1.69897, 2.67065)(2.00000, 2.58858)
\psline[linecolor=red, plotstyle=curve, linewidth=0.4mm, showpoints=false, linestyle=solid, linecolor=red, dotstyle=+, dotscale=1.2 1.2, linewidth=0.4mm](2.06070, 2.51976)(2.06108, 2.51976)(2.06145, 2.51976)(2.06183, 2.51976)(2.06221, 2.51976)(2.06258, 2.51976)(2.06296, 2.51976)(2.06333, 2.51976)(2.06371, 2.51976)(2.06408, 2.51976)(2.06446, 2.51976)(2.06483, 2.51976)(2.06521, 2.51976)(2.06558, 2.51976)(2.06595, 2.51976)(2.06633, 2.51976)(2.06670, 2.51976)(2.06707, 2.51976)(2.06744, 2.51976)(2.06781, 2.51976)(2.06819, 2.51976)(2.06856, 2.51976)(2.06893, 2.51976)(2.06930, 2.51976)(2.06967, 2.51976)(2.07004, 2.51976)(2.07041, 2.51976)(2.07078, 2.51976)(2.07115, 2.51976)(2.07151, 2.51976)(2.07188, 2.51976)(2.07225, 2.51976)(2.07262, 2.51976)(2.07298, 2.51976)(2.07335, 2.51976)(2.07372, 2.51976)(2.07408, 2.51976)(2.07445, 2.51976)(2.07482, 2.51976)(2.07518, 2.51976)(2.07555, 2.51976)(2.07591, 2.51976)(2.07628, 2.51976)(2.07664, 2.51976)(2.07700, 2.51976)(2.07737, 2.51976)(2.07773, 2.51976)(2.07809, 2.51976)(2.07846, 2.51976)(2.07882, 2.51976)(2.07918, 2.51976)(2.07954, 2.51976)(2.07990, 2.51976)(2.08027, 2.51976)(2.08063, 2.51976)(2.08099, 2.51976)(2.08135, 2.51976)(2.08171, 2.51976)(2.08207, 2.51976)(2.08243, 2.51976)(2.08279, 2.51976)(2.08314, 2.51976)(2.08350, 2.51976)(2.08386, 2.51976)(2.08422, 2.51976)(2.08458, 2.51976)(2.08493, 2.51976)(2.08529, 2.51976)(2.08565, 2.51976)(2.08600, 2.51976)(2.08636, 2.51976)(2.08672, 2.51976)(2.08707, 2.51976)(2.08743, 2.51976)(2.08778, 2.51976)(2.08814, 2.51976)(2.08849, 2.51976)(2.08884, 2.51976)(2.08920, 2.51976)(2.08955, 2.51976)(2.08991, 2.51976)(2.09026, 2.51976)(2.09061, 2.51976)(2.09096, 2.51976)(2.09132, 2.51976)(2.09167, 2.51976)(2.09202, 2.51976)(2.09237, 2.51976)(2.09272, 2.51976)(2.09307, 2.51976)(2.09342, 2.51976)(2.09377, 2.51976)(2.09412, 2.51976)(2.09447, 2.51976)(2.09482, 2.51976)(2.09517, 2.51976)(2.09552, 2.51976)(2.09587, 2.51976)(2.09621, 2.51976)(2.09656, 2.51976)(2.09691, 2.51976)
\psline[linecolor=black, plotstyle=curve, linewidth=0.4mm, showpoints=false, linestyle=solid, linecolor=black, dotstyle=diamond, dotscale=1.2 1.2, linewidth=0.4mm](0.00000, 4.87669)(0.30103, 3.45617)(0.47712, 3.06719)(0.60206, 2.84839)(0.69897, 2.70566)(1.00000, 2.37951)(1.30103, 2.17974)(1.47712, 2.10200)(1.60206, 2.05950)(1.69897, 2.03235)(2.00000, 1.97249)
\psline[linecolor=black, plotstyle=curve, linewidth=0.4mm, showpoints=false, linestyle=solid, linecolor=black, dotstyle=+, dotscale=1.2 1.2, linewidth=0.4mm](2.06070, 1.92230)(2.06108, 1.92230)(2.06145, 1.92230)(2.06183, 1.92230)(2.06221, 1.92230)(2.06258, 1.92230)(2.06296, 1.92230)(2.06333, 1.92230)(2.06371, 1.92230)(2.06408, 1.92230)(2.06446, 1.92230)(2.06483, 1.92230)(2.06521, 1.92230)(2.06558, 1.92230)(2.06595, 1.92230)(2.06633, 1.92230)(2.06670, 1.92230)(2.06707, 1.92230)(2.06744, 1.92230)(2.06781, 1.92230)(2.06819, 1.92230)(2.06856, 1.92230)(2.06893, 1.92230)(2.06930, 1.92230)(2.06967, 1.92230)(2.07004, 1.92230)(2.07041, 1.92230)(2.07078, 1.92230)(2.07115, 1.92230)(2.07151, 1.92230)(2.07188, 1.92230)(2.07225, 1.92230)(2.07262, 1.92230)(2.07298, 1.92230)(2.07335, 1.92230)(2.07372, 1.92230)(2.07408, 1.92230)(2.07445, 1.92230)(2.07482, 1.92230)(2.07518, 1.92230)(2.07555, 1.92230)(2.07591, 1.92230)(2.07628, 1.92230)(2.07664, 1.92230)(2.07700, 1.92230)(2.07737, 1.92230)(2.07773, 1.92230)(2.07809, 1.92230)(2.07846, 1.92230)(2.07882, 1.92230)(2.07918, 1.92230)(2.07954, 1.92230)(2.07990, 1.92230)(2.08027, 1.92230)(2.08063, 1.92230)(2.08099, 1.92230)(2.08135, 1.92230)(2.08171, 1.92230)(2.08207, 1.92230)(2.08243, 1.92230)(2.08279, 1.92230)(2.08314, 1.92230)(2.08350, 1.92230)(2.08386, 1.92230)(2.08422, 1.92230)(2.08458, 1.92230)(2.08493, 1.92230)(2.08529, 1.92230)(2.08565, 1.92230)(2.08600, 1.92230)(2.08636, 1.92230)(2.08672, 1.92230)(2.08707, 1.92230)(2.08743, 1.92230)(2.08778, 1.92230)(2.08814, 1.92230)(2.08849, 1.92230)(2.08884, 1.92230)(2.08920, 1.92230)(2.08955, 1.92230)(2.08991, 1.92230)(2.09026, 1.92230)(2.09061, 1.92230)(2.09096, 1.92230)(2.09132, 1.92230)(2.09167, 1.92230)(2.09202, 1.92230)(2.09237, 1.92230)(2.09272, 1.92230)(2.09307, 1.92230)(2.09342, 1.92230)(2.09377, 1.92230)(2.09412, 1.92230)(2.09447, 1.92230)(2.09482, 1.92230)(2.09517, 1.92230)(2.09552, 1.92230)(2.09587, 1.92230)(2.09621, 1.92230)(2.09656, 1.92230)(2.09691, 1.92230)
\psline[linecolor=blue, plotstyle=curve, linewidth=0.4mm, showpoints=true, linestyle=none, linecolor=blue, dotstyle=o, dotscale=1.2 1.2, linewidth=0.4mm](0.00000, 8.84000)(0.30103, 6.20000)(0.47712, 5.28500)(0.60206, 4.78500)(0.69897, 4.48000)(1.00000, 3.81000)(1.30103, 3.43000)(1.47712, 3.28500)(1.60206, 3.22250)(1.69897, 3.16000)(2.00000, 2.84750)
\psline[linecolor=red, plotstyle=curve, linewidth=0.4mm, showpoints=true, linestyle=none, linecolor=red, dotstyle=triangle, dotscale=1.2 1.2, linewidth=0.4mm](0.00000, 6.16500)(0.30103, 4.35000)(0.47712, 3.71000)(0.60206, 3.38000)(0.69897, 3.17000)(1.00000, 2.71000)(1.30103, 2.45500)(1.47712, 2.35500)(1.60206, 2.31500)(1.69897, 2.27500)(2.00000, 2.07500)
\psline[linecolor=black, plotstyle=curve, linewidth=0.4mm, showpoints=true, linestyle=none, linecolor=black, dotstyle=diamond, dotscale=1.2 1.2, linewidth=0.4mm](0.00000, 3.63000)(0.30103, 2.60500)(0.47712, 2.23000)(0.60206, 2.03500)(0.69897, 1.91000)(1.00000, 1.65000)(1.30103, 1.50500)(1.47712, 1.45000)(1.60206, 1.42500)(1.69897, 1.40000)(2.00000, 1.27500)
\endpsclip
\psframe[linecolor=black, fillstyle=solid, fillcolor=white, shadowcolor=lightgray, shadowsize=1mm, shadow=true](0.79545, 10.57317)(1.65455, 15.18293)
\rput[l](1.08182, 14.08537){\scriptsize{$\Delta \gamma=\unit[0]{dB}$}}
\psline[linecolor=blue, linestyle=solid, linewidth=0.3mm](0.85909, 14.08537)(0.98636, 14.08537)
\psline[linecolor=blue, linestyle=solid, linewidth=0.3mm](0.85909, 14.08537)(0.98636, 14.08537)
\psdots[linecolor=blue, linestyle=solid, linewidth=0.3mm, dotstyle=o, dotscale=1.2 1.2, linecolor=blue](0.92273, 14.08537)
\rput[l](1.08182, 12.87805){\scriptsize{$\Delta \gamma=\unit[0.4]{dB}$}}
\psline[linecolor=red, linestyle=solid, linewidth=0.3mm](0.85909, 12.87805)(0.98636, 12.87805)
\psline[linecolor=red, linestyle=solid, linewidth=0.3mm](0.85909, 12.87805)(0.98636, 12.87805)
\psdots[linecolor=red, linestyle=solid, linewidth=0.3mm, dotstyle=triangle, dotscale=1.2 1.2, linecolor=red](0.92273, 12.87805)
\rput[l](1.08182, 11.67073){\scriptsize{$\Delta \gamma=\unit[0.9]{dB}$}}
\psline[linecolor=black, linestyle=solid, linewidth=0.3mm](0.85909, 11.67073)(0.98636, 11.67073)
\psline[linecolor=black, linestyle=solid, linewidth=0.3mm](0.85909, 11.67073)(0.98636, 11.67073)
\psdots[linecolor=black, linestyle=solid, linewidth=0.3mm, dotstyle=diamond, dotscale=1.2 1.2, linecolor=black](0.92273, 11.67073)
}\end{pspicture}
\endgroup
 

%% file: Complexity_gain_Nusers.tex
\begingroup
\unitlength=1mm
\psset{xunit=31.42857mm, yunit=16.39344mm, linewidth=0.1mm}
\psset{arrowsize=2pt 3, arrowlength=1.4, arrowinset=.4}\psset{axesstyle=frame}
\begin{pspicture}(-0.44545, 0.14600)(2.10000, 3.50100)
\rput(-0.06364, -0.30500){%
\psaxes[subticks=10, xlogBase=10, logLines=x, labels=all, xsubticks=10, ysubticks=1, Ox=0, Oy=1, Dx=1, Dy=0.5]{-}(0.00000, 1.00000)(0.00000, 1.00000)(2.10000, 3.50100)%
\multips(0.00000, 1.50000)(0, 0.50000){4}{\psline[linecolor=black, linestyle=dotted, linewidth=0.2mm](0, 0)(2.10000, 0)}
\rput[b](1.05000, 0.45100){\small{$N_\text{c}$}}
\rput[t]{90}(-0.38182, 2.25050){{\small{$g_\text{comp}$}}}
\psclip{\psframe(0.00000, 1.00000)(2.10000, 3.50100)}
\psline[linecolor=blue, plotstyle=curve, linewidth=0.4mm, showpoints=true, linestyle=solid, linecolor=blue, dotstyle=o, dotscale=1.2 1.2, linewidth=0.4mm](0.00000, 1.00000)(0.30103, 1.46408)(0.47712, 1.66960)(0.60206, 1.81274)(0.69897, 1.92013)(1.00000, 2.22075)(1.30103, 2.45630)(1.47712, 2.56205)(1.60206, 2.62381)(1.69897, 2.66483)(2.00000, 2.76000)
\psline[linecolor=blue, plotstyle=curve, linewidth=0.4mm, showpoints=false, linestyle=solid, linecolor=blue, dotstyle=+, dotscale=1.2 1.2, linewidth=0.4mm](2.06070, 2.84520)(2.06108, 2.84520)(2.06145, 2.84520)(2.06183, 2.84520)(2.06221, 2.84520)(2.06258, 2.84520)(2.06296, 2.84520)(2.06333, 2.84520)(2.06371, 2.84520)(2.06408, 2.84520)(2.06446, 2.84520)(2.06483, 2.84520)(2.06521, 2.84520)(2.06558, 2.84520)(2.06595, 2.84520)(2.06633, 2.84520)(2.06670, 2.84520)(2.06707, 2.84520)(2.06744, 2.84520)(2.06781, 2.84520)(2.06819, 2.84520)(2.06856, 2.84520)(2.06893, 2.84520)(2.06930, 2.84520)(2.06967, 2.84520)(2.07004, 2.84520)(2.07041, 2.84520)(2.07078, 2.84520)(2.07115, 2.84520)(2.07151, 2.84520)(2.07188, 2.84520)(2.07225, 2.84520)(2.07262, 2.84520)(2.07298, 2.84520)(2.07335, 2.84520)(2.07372, 2.84520)(2.07408, 2.84520)(2.07445, 2.84520)(2.07482, 2.84520)(2.07518, 2.84520)(2.07555, 2.84520)(2.07591, 2.84520)(2.07628, 2.84520)(2.07664, 2.84520)(2.07700, 2.84520)(2.07737, 2.84520)(2.07773, 2.84520)(2.07809, 2.84520)(2.07846, 2.84520)(2.07882, 2.84520)(2.07918, 2.84520)(2.07954, 2.84520)(2.07990, 2.84520)(2.08027, 2.84520)(2.08063, 2.84520)(2.08099, 2.84520)(2.08135, 2.84520)(2.08171, 2.84520)(2.08207, 2.84520)(2.08243, 2.84520)(2.08279, 2.84520)(2.08314, 2.84520)(2.08350, 2.84520)(2.08386, 2.84520)(2.08422, 2.84520)(2.08458, 2.84520)(2.08493, 2.84520)(2.08529, 2.84520)(2.08565, 2.84520)(2.08600, 2.84520)(2.08636, 2.84520)(2.08672, 2.84520)(2.08707, 2.84520)(2.08743, 2.84520)(2.08778, 2.84520)(2.08814, 2.84520)(2.08849, 2.84520)(2.08884, 2.84520)(2.08920, 2.84520)(2.08955, 2.84520)(2.08991, 2.84520)(2.09026, 2.84520)(2.09061, 2.84520)(2.09096, 2.84520)(2.09132, 2.84520)(2.09167, 2.84520)(2.09202, 2.84520)(2.09237, 2.84520)(2.09272, 2.84520)(2.09307, 2.84520)(2.09342, 2.84520)(2.09377, 2.84520)(2.09412, 2.84520)(2.09447, 2.84520)(2.09482, 2.84520)(2.09517, 2.84520)(2.09552, 2.84520)(2.09587, 2.84520)(2.09621, 2.84520)(2.09656, 2.84520)(2.09691, 2.84520)
\psline[linecolor=red, plotstyle=curve, linewidth=0.4mm, showpoints=true, linestyle=solid, linecolor=red, dotstyle=triangle, dotscale=1.2 1.2, linewidth=0.4mm](0.00000, 1.00000)(0.30103, 1.44953)(0.47712, 1.64313)(0.60206, 1.77659)(0.69897, 1.87600)(1.00000, 2.15102)(1.30103, 2.36323)(1.47712, 2.45758)(1.60206, 2.51242)(1.69897, 2.54874)(2.00000, 2.63268)
\psline[linecolor=red, plotstyle=curve, linewidth=0.4mm, showpoints=false, linestyle=solid, linecolor=red, dotstyle=+, dotscale=1.2 1.2, linewidth=0.4mm](2.06070, 2.70744)(2.06108, 2.70744)(2.06145, 2.70744)(2.06183, 2.70744)(2.06221, 2.70744)(2.06258, 2.70744)(2.06296, 2.70744)(2.06333, 2.70744)(2.06371, 2.70744)(2.06408, 2.70744)(2.06446, 2.70744)(2.06483, 2.70744)(2.06521, 2.70744)(2.06558, 2.70744)(2.06595, 2.70744)(2.06633, 2.70744)(2.06670, 2.70744)(2.06707, 2.70744)(2.06744, 2.70744)(2.06781, 2.70744)(2.06819, 2.70744)(2.06856, 2.70744)(2.06893, 2.70744)(2.06930, 2.70744)(2.06967, 2.70744)(2.07004, 2.70744)(2.07041, 2.70744)(2.07078, 2.70744)(2.07115, 2.70744)(2.07151, 2.70744)(2.07188, 2.70744)(2.07225, 2.70744)(2.07262, 2.70744)(2.07298, 2.70744)(2.07335, 2.70744)(2.07372, 2.70744)(2.07408, 2.70744)(2.07445, 2.70744)(2.07482, 2.70744)(2.07518, 2.70744)(2.07555, 2.70744)(2.07591, 2.70744)(2.07628, 2.70744)(2.07664, 2.70744)(2.07700, 2.70744)(2.07737, 2.70744)(2.07773, 2.70744)(2.07809, 2.70744)(2.07846, 2.70744)(2.07882, 2.70744)(2.07918, 2.70744)(2.07954, 2.70744)(2.07990, 2.70744)(2.08027, 2.70744)(2.08063, 2.70744)(2.08099, 2.70744)(2.08135, 2.70744)(2.08171, 2.70744)(2.08207, 2.70744)(2.08243, 2.70744)(2.08279, 2.70744)(2.08314, 2.70744)(2.08350, 2.70744)(2.08386, 2.70744)(2.08422, 2.70744)(2.08458, 2.70744)(2.08493, 2.70744)(2.08529, 2.70744)(2.08565, 2.70744)(2.08600, 2.70744)(2.08636, 2.70744)(2.08672, 2.70744)(2.08707, 2.70744)(2.08743, 2.70744)(2.08778, 2.70744)(2.08814, 2.70744)(2.08849, 2.70744)(2.08884, 2.70744)(2.08920, 2.70744)(2.08955, 2.70744)(2.08991, 2.70744)(2.09026, 2.70744)(2.09061, 2.70744)(2.09096, 2.70744)(2.09132, 2.70744)(2.09167, 2.70744)(2.09202, 2.70744)(2.09237, 2.70744)(2.09272, 2.70744)(2.09307, 2.70744)(2.09342, 2.70744)(2.09377, 2.70744)(2.09412, 2.70744)(2.09447, 2.70744)(2.09482, 2.70744)(2.09517, 2.70744)(2.09552, 2.70744)(2.09587, 2.70744)(2.09621, 2.70744)(2.09656, 2.70744)(2.09691, 2.70744)
\psline[linecolor=black, plotstyle=curve, linewidth=0.4mm, showpoints=true, linestyle=solid, linecolor=black, dotstyle=diamond, dotscale=1.2 1.2, linewidth=0.4mm](0.00000, 1.00000)(0.30103, 1.42675)(0.47712, 1.61131)(0.60206, 1.73776)(0.69897, 1.83152)(1.00000, 2.08908)(1.30103, 2.28599)(1.47712, 2.37303)(1.60206, 2.42348)(1.69897, 2.45683)(2.00000, 2.53374)
\psline[linecolor=black, plotstyle=curve, linewidth=0.4mm, showpoints=false, linestyle=solid, linecolor=black, dotstyle=+, dotscale=1.2 1.2, linewidth=0.4mm](2.06070, 2.60204)(2.06108, 2.60204)(2.06145, 2.60204)(2.06183, 2.60204)(2.06221, 2.60204)(2.06258, 2.60204)(2.06296, 2.60204)(2.06333, 2.60204)(2.06371, 2.60204)(2.06408, 2.60204)(2.06446, 2.60204)(2.06483, 2.60204)(2.06521, 2.60204)(2.06558, 2.60204)(2.06595, 2.60204)(2.06633, 2.60204)(2.06670, 2.60204)(2.06707, 2.60204)(2.06744, 2.60204)(2.06781, 2.60204)(2.06819, 2.60204)(2.06856, 2.60204)(2.06893, 2.60204)(2.06930, 2.60204)(2.06967, 2.60204)(2.07004, 2.60204)(2.07041, 2.60204)(2.07078, 2.60204)(2.07115, 2.60204)(2.07151, 2.60204)(2.07188, 2.60204)(2.07225, 2.60204)(2.07262, 2.60204)(2.07298, 2.60204)(2.07335, 2.60204)(2.07372, 2.60204)(2.07408, 2.60204)(2.07445, 2.60204)(2.07482, 2.60204)(2.07518, 2.60204)(2.07555, 2.60204)(2.07591, 2.60204)(2.07628, 2.60204)(2.07664, 2.60204)(2.07700, 2.60204)(2.07737, 2.60204)(2.07773, 2.60204)(2.07809, 2.60204)(2.07846, 2.60204)(2.07882, 2.60204)(2.07918, 2.60204)(2.07954, 2.60204)(2.07990, 2.60204)(2.08027, 2.60204)(2.08063, 2.60204)(2.08099, 2.60204)(2.08135, 2.60204)(2.08171, 2.60204)(2.08207, 2.60204)(2.08243, 2.60204)(2.08279, 2.60204)(2.08314, 2.60204)(2.08350, 2.60204)(2.08386, 2.60204)(2.08422, 2.60204)(2.08458, 2.60204)(2.08493, 2.60204)(2.08529, 2.60204)(2.08565, 2.60204)(2.08600, 2.60204)(2.08636, 2.60204)(2.08672, 2.60204)(2.08707, 2.60204)(2.08743, 2.60204)(2.08778, 2.60204)(2.08814, 2.60204)(2.08849, 2.60204)(2.08884, 2.60204)(2.08920, 2.60204)(2.08955, 2.60204)(2.08991, 2.60204)(2.09026, 2.60204)(2.09061, 2.60204)(2.09096, 2.60204)(2.09132, 2.60204)(2.09167, 2.60204)(2.09202, 2.60204)(2.09237, 2.60204)(2.09272, 2.60204)(2.09307, 2.60204)(2.09342, 2.60204)(2.09377, 2.60204)(2.09412, 2.60204)(2.09447, 2.60204)(2.09482, 2.60204)(2.09517, 2.60204)(2.09552, 2.60204)(2.09587, 2.60204)(2.09621, 2.60204)(2.09656, 2.60204)(2.09691, 2.60204)
\endpsclip
\psframe[linecolor=black, fillstyle=solid, fillcolor=white, shadowcolor=lightgray, shadowsize=1mm, shadow=true](0.79545, 2.76290)(1.65455, 3.53150)
\rput[l](1.08182, 3.34850){\scriptsize{$\Delta \gamma=\unit[0]{dB}$}}
\psline[linecolor=blue, linestyle=solid, linewidth=0.3mm](0.85909, 3.34850)(0.98636, 3.34850)
\psline[linecolor=blue, linestyle=solid, linewidth=0.3mm](0.85909, 3.34850)(0.98636, 3.34850)
\psdots[linecolor=blue, linestyle=solid, linewidth=0.3mm, dotstyle=o, dotscale=1.2 1.2, linecolor=blue](0.92273, 3.34850)
\rput[l](1.08182, 3.14720){\scriptsize{$\Delta \gamma=\unit[0.4]{dB}$}}
\psline[linecolor=red, linestyle=solid, linewidth=0.3mm](0.85909, 3.14720)(0.98636, 3.14720)
\psline[linecolor=red, linestyle=solid, linewidth=0.3mm](0.85909, 3.14720)(0.98636, 3.14720)
\psdots[linecolor=red, linestyle=solid, linewidth=0.3mm, dotstyle=triangle, dotscale=1.2 1.2, linecolor=red](0.92273, 3.14720)
\rput[l](1.08182, 2.94590){\scriptsize{$\Delta \gamma=\unit[0.9]{dB}$}}
\psline[linecolor=black, linestyle=solid, linewidth=0.3mm](0.85909, 2.94590)(0.98636, 2.94590)
\psline[linecolor=black, linestyle=solid, linewidth=0.3mm](0.85909, 2.94590)(0.98636, 2.94590)
\psdots[linecolor=black, linestyle=solid, linewidth=0.3mm, dotstyle=diamond, dotscale=1.2 1.2, linecolor=black](0.92273, 2.94590)
}\end{pspicture}
\endgroup
 

%% file: Complexity_gain_epsilon.tex
\begingroup
\unitlength=1mm
\psset{xunit=33.00000mm, yunit=10.24744mm, linewidth=0.1mm}
\psset{arrowsize=2pt 3, arrowlength=1.4, arrowinset=.4}\psset{axesstyle=frame}
\begin{pspicture}(-3.42424, 0.63380)(-1.00000, 6.00100)
\rput(-0.06061, -0.48793){%
\psaxes[subticks=0, labels=all, xsubticks=1, ysubticks=1, Ox=-3, Oy=2, Dx=0.5, Dy=1]{-}(-3.00000, 2.00000)(-3.00000, 2.00000)(-1.00000, 6.00100)%
\multips(-2.50000, 2.00000)(0.50000, 0.0){3}{\psline[linecolor=black, linestyle=dotted, linewidth=0.2mm](0, 0)(0, 4.00100)}
\multips(-3.00000, 3.00000)(0, 1.00000){3}{\psline[linecolor=black, linestyle=dotted, linewidth=0.2mm](0, 0)(2.00000, 0)}
\rput[b](-2.00000, 1.12173){\small{$\log_{10}\hat\epsilon_\text{comp}$}}
\rput[t]{90}(-3.36364, 4.00050){\small{$g_\text{comp}$}}
\psclip{\psframe(-3.00000, 2.00000)(-1.00000, 6.00100)}
\psline[linecolor=blue, plotstyle=curve, linewidth=0.4mm, showpoints=true, linestyle=solid, linecolor=blue, dotstyle=o, dotscale=1.2 1.2, linewidth=0.4mm](-3.00000, 4.36357)(-2.75000, 4.21837)(-2.50000, 4.07691)(-2.25000, 3.90158)(-2.00000, 3.67209)(-1.75000, 3.43743)(-1.50000, 3.20585)(-1.25000, 2.97493)(-1.00000, 2.84520)
\psline[linecolor=red, plotstyle=curve, linewidth=0.4mm, showpoints=true, linestyle=solid, linecolor=red, dotstyle=triangle, dotscale=1.2 1.2, linewidth=0.4mm](-3.00000, 3.95634)(-2.75000, 3.86677)(-2.50000, 3.74602)(-2.25000, 3.63087)(-2.00000, 3.48255)(-1.75000, 3.27430)(-1.50000, 3.03981)(-1.25000, 2.84111)(-1.00000, 2.70744)
\psline[linecolor=black, plotstyle=curve, linewidth=0.4mm, showpoints=true, linestyle=solid, linecolor=black, dotstyle=diamond, dotscale=1.2 1.2, linewidth=0.4mm](-3.00000, 3.76377)(-2.75000, 3.68349)(-2.50000, 3.58130)(-2.25000, 3.47683)(-2.00000, 3.35072)(-1.75000, 3.15702)(-1.50000, 2.93632)(-1.25000, 2.75265)(-1.00000, 2.60204)
\endpsclip
\psframe[linecolor=black, fillstyle=solid, fillcolor=white, shadowcolor=lightgray, shadowsize=1mm, shadow=true](-2.24242, 4.82022)(-1.42424, 6.04979)
\rput[l](-1.96970, 5.75704){\scriptsize{$\Delta \gamma=\unit[0]{dB}$}}
\psline[linecolor=blue, linestyle=solid, linewidth=0.3mm](-2.18182, 5.75704)(-2.06061, 5.75704)
\psline[linecolor=blue, linestyle=solid, linewidth=0.3mm](-2.18182, 5.75704)(-2.06061, 5.75704)
\psdots[linecolor=blue, linestyle=solid, linewidth=0.3mm, dotstyle=o, dotscale=1.2 1.2, linecolor=blue](-2.12121, 5.75704)
\rput[l](-1.96970, 5.43500){\scriptsize{$\Delta \gamma=\unit[0.4]{dB}$}}
\psline[linecolor=red, linestyle=solid, linewidth=0.3mm](-2.18182, 5.43500)(-2.06061, 5.43500)
\psline[linecolor=red, linestyle=solid, linewidth=0.3mm](-2.18182, 5.43500)(-2.06061, 5.43500)
\psdots[linecolor=red, linestyle=solid, linewidth=0.3mm, dotstyle=triangle, dotscale=1.2 1.2, linecolor=red](-2.12121, 5.43500)
\rput[l](-1.96970, 5.11297){\scriptsize{$\Delta \gamma=\unit[0.9]{dB}$}}
\psline[linecolor=black, linestyle=solid, linewidth=0.3mm](-2.18182, 5.11297)(-2.06061, 5.11297)
\psline[linecolor=black, linestyle=solid, linewidth=0.3mm](-2.18182, 5.11297)(-2.06061, 5.11297)
\psdots[linecolor=black, linestyle=solid, linewidth=0.3mm, dotstyle=diamond, dotscale=1.2 1.2, linecolor=black](-2.12121, 5.11297)
}\end{pspicture}
\endgroup
 

%% file: Computational_diversity.tex
\begingroup
\unitlength=1mm
\psset{xunit=34.00000mm, yunit=5.12436mm, linewidth=0.1mm}
\psset{arrowsize=2pt 3, arrowlength=1.4, arrowinset=.4}\psset{axesstyle=frame}
\begin{pspicture}(-3.35294, -2.73205)(-1.00000, 8.00100)
\rput(-0.05882, -0.97573){%
\psaxes[subticks=0, labels=all, xsubticks=1, ysubticks=1, Ox=-3, Oy=0, Dx=0.5, Dy=2]{-}(-3.00000, 0.00000)(-3.00000, 0.00000)(-1.00000, 8.00100)%
\multips(-2.50000, 0.00000)(0.50000, 0.0){3}{\psline[linecolor=black, linestyle=dotted, linewidth=0.2mm](0, 0)(0, 8.00100)}
\multips(-3.00000, 2.00000)(0, 2.00000){3}{\psline[linecolor=black, linestyle=dotted, linewidth=0.2mm](0, 0)(2.00000, 0)}
\rput[b](-2.00000, -1.75632){\small{$\log_{10}\hat\epsilon_\text{comp}$}}
\rput[t]{90}(-3.29412, 4.00050){\small{$d_\text{comp}$}}
\psclip{\psframe(-3.00000, 0.00000)(-1.00000, 8.00100)}
\psline[linecolor=blue, plotstyle=curve, linewidth=0.4mm, showpoints=true, linestyle=solid, linecolor=blue, dotstyle=o, dotscale=1.2 1.2, linewidth=0.4mm](-3.00000, 6.84073)(-2.75000, 6.08872)(-2.50000, 5.37238)(-2.25000, 4.57869)(-2.00000, 3.69084)(-1.75000, 2.87294)(-1.50000, 2.14969)(-1.25000, 1.52272)(-1.00000, 1.08623)
\psline[linecolor=red, plotstyle=curve, linewidth=0.4mm, showpoints=true, linestyle=solid, linecolor=red, dotstyle=triangle, dotscale=1.2 1.2, linewidth=0.4mm](-3.00000, 6.02919)(-2.75000, 5.51013)(-2.50000, 4.89153)(-2.25000, 4.30225)(-2.00000, 3.63422)(-1.75000, 2.86263)(-1.50000, 2.12266)(-1.25000, 1.53551)(-1.00000, 1.09550)
\psline[linecolor=black, plotstyle=curve, linewidth=0.4mm, showpoints=true, linestyle=solid, linecolor=black, dotstyle=diamond, dotscale=1.2 1.2, linewidth=0.4mm](-3.00000, 5.71763)(-2.75000, 5.24373)(-2.50000, 4.69716)(-2.25000, 4.14981)(-2.00000, 3.54921)(-1.75000, 2.81192)(-1.50000, 2.09649)(-1.25000, 1.53128)(-1.00000, 1.07724)
\endpsclip
\psframe[linecolor=black, fillstyle=solid, fillcolor=white, shadowcolor=lightgray, shadowsize=1mm, shadow=true](-2.26471, 5.63973)(-1.47059, 8.09857)
\rput[l](-2.00000, 7.51313){\scriptsize{$\Delta \gamma=\unit[0]{dB}$}}
\psline[linecolor=blue, linestyle=solid, linewidth=0.3mm](-2.20588, 7.51313)(-2.08824, 7.51313)
\psline[linecolor=blue, linestyle=solid, linewidth=0.3mm](-2.20588, 7.51313)(-2.08824, 7.51313)
\psdots[linecolor=blue, linestyle=solid, linewidth=0.3mm, dotstyle=o, dotscale=1.2 1.2, linecolor=blue](-2.14706, 7.51313)
\rput[l](-2.00000, 6.86915){\scriptsize{$\Delta \gamma=\unit[0.4]{dB}$}}
\psline[linecolor=red, linestyle=solid, linewidth=0.3mm](-2.20588, 6.86915)(-2.08824, 6.86915)
\psline[linecolor=red, linestyle=solid, linewidth=0.3mm](-2.20588, 6.86915)(-2.08824, 6.86915)
\psdots[linecolor=red, linestyle=solid, linewidth=0.3mm, dotstyle=triangle, dotscale=1.2 1.2, linecolor=red](-2.14706, 6.86915)
\rput[l](-2.00000, 6.22517){\scriptsize{$\Delta \gamma=\unit[0.9]{dB}$}}
\psline[linecolor=black, linestyle=solid, linewidth=0.3mm](-2.20588, 6.22517)(-2.08824, 6.22517)
\psline[linecolor=black, linestyle=solid, linewidth=0.3mm](-2.20588, 6.22517)(-2.08824, 6.22517)
\psdots[linecolor=black, linestyle=solid, linewidth=0.3mm, dotstyle=diamond, dotscale=1.2 1.2, linecolor=black](-2.14706, 6.22517)
}\end{pspicture}
\endgroup
 

%% file: Complexity_rate_tradeoff.tex
\begingroup
\unitlength=1mm
\psset{xunit=30.00000mm, yunit=243.59775mm, linewidth=0.1mm}
\psset{arrowsize=2pt 3, arrowlength=1.4, arrowinset=.4}\psset{axesstyle=frame}
\begin{pspicture}(-0.56667, -0.06568)(2.10000, 0.16010)
\rput(-0.06667, -0.02053){%
\psaxes[subticks=10, xlogBase=10, logLines=x, labels=all, xsubticks=10, ysubticks=1, Ox=0, Oy=0, Dx=1, Dy=0.02]{-}(0.00000, 0.00000)(0.00000, 0.00000)(2.10000, 0.16010)%
\multips(0.00000, 0.02000)(0, 0.02000){7}{\psline[linecolor=black, linestyle=dotted, linewidth=0.2mm](0, 0)(2.10000, 0)}
\rput[b](1.05000, -0.03516){\small{$N_\text{c}$}}
\rput[t]{90}(-0.50000, 0.08005){\small{$t_\text{comp}$}}
\psclip{\psframe(0.00000, 0.00000)(2.10000, 0.16010)}
\psline[linecolor=blue, plotstyle=curve, linewidth=0.4mm, showpoints=true, linestyle=solid, linecolor=blue, dotstyle=o, dotscale=1.2 1.2, linewidth=0.4mm](0.30103, 0.02278)(0.47712, 0.02661)(0.60206, 0.02940)(0.69897, 0.03155)(1.00000, 0.03790)(1.30103, 0.04322)(1.47712, 0.04572)(1.60206, 0.04722)(1.69897, 0.04822)(2.00000, 0.05060)
\psline[linecolor=blue, plotstyle=curve, linewidth=0.4mm, showpoints=false, linestyle=solid, linecolor=blue, dotstyle=+, dotscale=1.2 1.2, linewidth=0.4mm](2.06070, 0.05278)(2.06108, 0.05278)(2.06145, 0.05278)(2.06183, 0.05278)(2.06221, 0.05278)(2.06258, 0.05278)(2.06296, 0.05278)(2.06333, 0.05278)(2.06371, 0.05278)(2.06408, 0.05278)(2.06446, 0.05278)(2.06483, 0.05278)(2.06521, 0.05278)(2.06558, 0.05278)(2.06595, 0.05278)(2.06633, 0.05278)(2.06670, 0.05278)(2.06707, 0.05278)(2.06744, 0.05278)(2.06781, 0.05278)(2.06819, 0.05278)(2.06856, 0.05278)(2.06893, 0.05278)(2.06930, 0.05278)(2.06967, 0.05278)(2.07004, 0.05278)(2.07041, 0.05278)(2.07078, 0.05278)(2.07115, 0.05278)(2.07151, 0.05278)(2.07188, 0.05278)(2.07225, 0.05278)(2.07262, 0.05278)(2.07298, 0.05278)(2.07335, 0.05278)(2.07372, 0.05278)(2.07408, 0.05278)(2.07445, 0.05278)(2.07482, 0.05278)(2.07518, 0.05278)(2.07555, 0.05278)(2.07591, 0.05278)(2.07628, 0.05278)(2.07664, 0.05278)(2.07700, 0.05278)(2.07737, 0.05278)(2.07773, 0.05278)(2.07809, 0.05278)(2.07846, 0.05278)(2.07882, 0.05278)(2.07918, 0.05278)(2.07954, 0.05278)(2.07990, 0.05278)(2.08027, 0.05278)(2.08063, 0.05278)(2.08099, 0.05278)(2.08135, 0.05278)(2.08171, 0.05278)(2.08207, 0.05278)(2.08243, 0.05278)(2.08279, 0.05278)(2.08314, 0.05278)(2.08350, 0.05278)(2.08386, 0.05278)(2.08422, 0.05278)(2.08458, 0.05278)(2.08493, 0.05278)(2.08529, 0.05278)(2.08565, 0.05278)(2.08600, 0.05278)(2.08636, 0.05278)(2.08672, 0.05278)(2.08707, 0.05278)(2.08743, 0.05278)(2.08778, 0.05278)(2.08814, 0.05278)(2.08849, 0.05278)(2.08884, 0.05278)(2.08920, 0.05278)(2.08955, 0.05278)(2.08991, 0.05278)(2.09026, 0.05278)(2.09061, 0.05278)(2.09096, 0.05278)(2.09132, 0.05278)(2.09167, 0.05278)(2.09202, 0.05278)(2.09237, 0.05278)(2.09272, 0.05278)(2.09307, 0.05278)(2.09342, 0.05278)(2.09377, 0.05278)(2.09412, 0.05278)(2.09447, 0.05278)(2.09482, 0.05278)(2.09517, 0.05278)(2.09552, 0.05278)(2.09587, 0.05278)(2.09621, 0.05278)(2.09656, 0.05278)(2.09691, 0.05278)
\psline[linecolor=red, plotstyle=curve, linewidth=0.4mm, showpoints=true, linestyle=solid, linecolor=red, dotstyle=triangle, dotscale=1.2 1.2, linewidth=0.4mm](0.30103, 0.04053)(0.47712, 0.04657)(0.60206, 0.05083)(0.69897, 0.05406)(1.00000, 0.06323)(1.30103, 0.07057)(1.47712, 0.07391)(1.60206, 0.07587)(1.69897, 0.07717)(2.00000, 0.08022)
\psline[linecolor=red, plotstyle=curve, linewidth=0.4mm, showpoints=false, linestyle=solid, linecolor=red, dotstyle=+, dotscale=1.2 1.2, linewidth=0.4mm](2.06070, 0.08297)(2.06108, 0.08297)(2.06145, 0.08297)(2.06183, 0.08297)(2.06221, 0.08297)(2.06258, 0.08297)(2.06296, 0.08297)(2.06333, 0.08297)(2.06371, 0.08297)(2.06408, 0.08297)(2.06446, 0.08297)(2.06483, 0.08297)(2.06521, 0.08297)(2.06558, 0.08297)(2.06595, 0.08297)(2.06633, 0.08297)(2.06670, 0.08297)(2.06707, 0.08297)(2.06744, 0.08297)(2.06781, 0.08297)(2.06819, 0.08297)(2.06856, 0.08297)(2.06893, 0.08297)(2.06930, 0.08297)(2.06967, 0.08297)(2.07004, 0.08297)(2.07041, 0.08297)(2.07078, 0.08297)(2.07115, 0.08297)(2.07151, 0.08297)(2.07188, 0.08297)(2.07225, 0.08297)(2.07262, 0.08297)(2.07298, 0.08297)(2.07335, 0.08297)(2.07372, 0.08297)(2.07408, 0.08297)(2.07445, 0.08297)(2.07482, 0.08297)(2.07518, 0.08297)(2.07555, 0.08297)(2.07591, 0.08297)(2.07628, 0.08297)(2.07664, 0.08297)(2.07700, 0.08297)(2.07737, 0.08297)(2.07773, 0.08297)(2.07809, 0.08297)(2.07846, 0.08297)(2.07882, 0.08297)(2.07918, 0.08297)(2.07954, 0.08297)(2.07990, 0.08297)(2.08027, 0.08297)(2.08063, 0.08297)(2.08099, 0.08297)(2.08135, 0.08297)(2.08171, 0.08297)(2.08207, 0.08297)(2.08243, 0.08297)(2.08279, 0.08297)(2.08314, 0.08297)(2.08350, 0.08297)(2.08386, 0.08297)(2.08422, 0.08297)(2.08458, 0.08297)(2.08493, 0.08297)(2.08529, 0.08297)(2.08565, 0.08297)(2.08600, 0.08297)(2.08636, 0.08297)(2.08672, 0.08297)(2.08707, 0.08297)(2.08743, 0.08297)(2.08778, 0.08297)(2.08814, 0.08297)(2.08849, 0.08297)(2.08884, 0.08297)(2.08920, 0.08297)(2.08955, 0.08297)(2.08991, 0.08297)(2.09026, 0.08297)(2.09061, 0.08297)(2.09096, 0.08297)(2.09132, 0.08297)(2.09167, 0.08297)(2.09202, 0.08297)(2.09237, 0.08297)(2.09272, 0.08297)(2.09307, 0.08297)(2.09342, 0.08297)(2.09377, 0.08297)(2.09412, 0.08297)(2.09447, 0.08297)(2.09482, 0.08297)(2.09517, 0.08297)(2.09552, 0.08297)(2.09587, 0.08297)(2.09621, 0.08297)(2.09656, 0.08297)(2.09691, 0.08297)
\psline[linecolor=black, plotstyle=curve, linewidth=0.4mm, showpoints=true, linestyle=solid, linecolor=black, dotstyle=diamond, dotscale=1.2 1.2, linewidth=0.4mm](0.30103, 0.06003)(0.47712, 0.06858)(0.60206, 0.07455)(0.69897, 0.07905)(1.00000, 0.09166)(1.30103, 0.10160)(1.47712, 0.10607)(1.60206, 0.10868)(1.69897, 0.11042)(2.00000, 0.11446)
\psline[linecolor=black, plotstyle=curve, linewidth=0.4mm, showpoints=false, linestyle=solid, linecolor=black, dotstyle=+, dotscale=1.2 1.2, linewidth=0.4mm](2.06070, 0.11809)(2.06108, 0.11809)(2.06145, 0.11809)(2.06183, 0.11809)(2.06221, 0.11809)(2.06258, 0.11809)(2.06296, 0.11809)(2.06333, 0.11809)(2.06371, 0.11809)(2.06408, 0.11809)(2.06446, 0.11809)(2.06483, 0.11809)(2.06521, 0.11809)(2.06558, 0.11809)(2.06595, 0.11809)(2.06633, 0.11809)(2.06670, 0.11809)(2.06707, 0.11809)(2.06744, 0.11809)(2.06781, 0.11809)(2.06819, 0.11809)(2.06856, 0.11809)(2.06893, 0.11809)(2.06930, 0.11809)(2.06967, 0.11809)(2.07004, 0.11809)(2.07041, 0.11809)(2.07078, 0.11809)(2.07115, 0.11809)(2.07151, 0.11809)(2.07188, 0.11809)(2.07225, 0.11809)(2.07262, 0.11809)(2.07298, 0.11809)(2.07335, 0.11809)(2.07372, 0.11809)(2.07408, 0.11809)(2.07445, 0.11809)(2.07482, 0.11809)(2.07518, 0.11809)(2.07555, 0.11809)(2.07591, 0.11809)(2.07628, 0.11809)(2.07664, 0.11809)(2.07700, 0.11809)(2.07737, 0.11809)(2.07773, 0.11809)(2.07809, 0.11809)(2.07846, 0.11809)(2.07882, 0.11809)(2.07918, 0.11809)(2.07954, 0.11809)(2.07990, 0.11809)(2.08027, 0.11809)(2.08063, 0.11809)(2.08099, 0.11809)(2.08135, 0.11809)(2.08171, 0.11809)(2.08207, 0.11809)(2.08243, 0.11809)(2.08279, 0.11809)(2.08314, 0.11809)(2.08350, 0.11809)(2.08386, 0.11809)(2.08422, 0.11809)(2.08458, 0.11809)(2.08493, 0.11809)(2.08529, 0.11809)(2.08565, 0.11809)(2.08600, 0.11809)(2.08636, 0.11809)(2.08672, 0.11809)(2.08707, 0.11809)(2.08743, 0.11809)(2.08778, 0.11809)(2.08814, 0.11809)(2.08849, 0.11809)(2.08884, 0.11809)(2.08920, 0.11809)(2.08955, 0.11809)(2.08991, 0.11809)(2.09026, 0.11809)(2.09061, 0.11809)(2.09096, 0.11809)(2.09132, 0.11809)(2.09167, 0.11809)(2.09202, 0.11809)(2.09237, 0.11809)(2.09272, 0.11809)(2.09307, 0.11809)(2.09342, 0.11809)(2.09377, 0.11809)(2.09412, 0.11809)(2.09447, 0.11809)(2.09482, 0.11809)(2.09517, 0.11809)(2.09552, 0.11809)(2.09587, 0.11809)(2.09621, 0.11809)(2.09656, 0.11809)(2.09691, 0.11809)
\endpsclip
\psframe[linecolor=black, fillstyle=solid, fillcolor=white, shadowcolor=lightgray, shadowsize=1mm, shadow=true](0.83333, 0.11864)(1.73333, 0.17036)
\rput[l](1.13333, 0.15805){\scriptsize{$\Delta \gamma=\unit[0]{dB}$}}
\psline[linecolor=blue, linestyle=solid, linewidth=0.3mm](0.90000, 0.15805)(1.03333, 0.15805)
\psline[linecolor=blue, linestyle=solid, linewidth=0.3mm](0.90000, 0.15805)(1.03333, 0.15805)
\psdots[linecolor=blue, linestyle=solid, linewidth=0.3mm, dotstyle=o, dotscale=1.2 1.2, linecolor=blue](0.96667, 0.15805)
\rput[l](1.13333, 0.14450){\scriptsize{$\Delta \gamma=\unit[0.4]{dB}$}}
\psline[linecolor=red, linestyle=solid, linewidth=0.3mm](0.90000, 0.14450)(1.03333, 0.14450)
\psline[linecolor=red, linestyle=solid, linewidth=0.3mm](0.90000, 0.14450)(1.03333, 0.14450)
\psdots[linecolor=red, linestyle=solid, linewidth=0.3mm, dotstyle=triangle, dotscale=1.2 1.2, linecolor=red](0.96667, 0.14450)
\rput[l](1.13333, 0.13095){\scriptsize{$\Delta \gamma=\unit[0.9]{dB}$}}
\psline[linecolor=black, linestyle=solid, linewidth=0.3mm](0.90000, 0.13095)(1.03333, 0.13095)
\psline[linecolor=black, linestyle=solid, linewidth=0.3mm](0.90000, 0.13095)(1.03333, 0.13095)
\psdots[linecolor=black, linestyle=solid, linewidth=0.3mm, dotstyle=diamond, dotscale=1.2 1.2, linecolor=black](0.96667, 0.13095)
}\end{pspicture}
\endgroup
 